\documentclass[a4paper,11pt]{article}
\pdfoutput=1 
\usepackage{jheppub}
\usepackage{amsmath}
\usepackage{amsfonts}  
\usepackage{amssymb}
\usepackage{slashed}
\usepackage{stmaryrd}
\usepackage{bm}
\usepackage{graphicx}%
\usepackage[T1]{fontenc} 
\usepackage{mathrsfs}
\usepackage{yfonts}
\usepackage{pifont}

\DeclareMathAlphabet{\mathpzc}{OT1}{pzc}{m}{it}

\usepackage{graphicx}
\usepackage{dcolumn}
\usepackage{epsfig}

\newcommand{\beq}{\begin{equation}} 
\newcommand{\eeq}{\end{equation}} 
\newcommand{\bega}{\begin{eqnarray}} 
\newcommand{\ega}{\end{eqnarray}} 

\newcommand{\dhd}{{\textstyle d}
\lower.03ex\hbox{\kern-0.38em$^{\scriptstyle-}$}\kern-0.05em{}}
\newcommand{\dbar}{{\textstyle \delta}
\lower.03ex\hbox{\kern-0.38em$^{\scriptstyle-}$}\kern-0.05em{}}
\newcommand{\half}{{1\over 2}}

\newcommand{\bu}{{\bullet}}
\newcommand{\ie}{i\epsilon}
\newcommand{\vro}{\varrho}

\newcommand{\baru}{{\bar u}}

\newcommand{\barv}{{\bar v}}

\newcommand{\bsi}{{\bar \psi}}

\newcommand{\barA}{{\bar A}}
\newcommand{\barB}{{\bar B}}
\newcommand{\barC}{{\bar C}}
\newcommand{\barD}{{\bar D}}
 
\newcommand{\barF}{{\bar F}}

\newcommand{\Bsi}{{\bar \Psi}}

\newcommand{\cala}{{\cal A}}
\newcommand{\calb}{{\cal B}}

\newcommand{\cald}{{\cal D}}  
\newcommand{\calf}{{\cal F}}

\newcommand{\kal}{{\cal L}}

\newcommand{\calp}{{\cal P}}

\newcommand{\hatW}{{\hat W}}

\newcommand{\hacalo}{{\hat {\cal O}}}

\newcommand{\tilA}{{\tilde A}}
\newcommand{\tilB}{{\tilde B}}
\newcommand{\tilC}{{\tilde C}}

\newcommand{\tilF}{{\tilde F}}

\newcommand{\tilS}{{\tilde S}}
\newcommand{\tilT}{{\tilde T}}

\newcommand{\tigma}{\tilde {\sigma}} 
\newcommand{\tsi}{\tilde {\psi}} 
\newcommand{\tipsi}{\tilde {\psi}}

\newcommand{\matA}{\mathbb{A}}

\newcommand{\matD}{\mathbb{D}} 
\newcommand{\matF}{\mathbb{F}} 
\newcommand{\matG}{\mathbb{G}} 
\newcommand{\matP}{\mathbb{P}} 
\newcommand{\matO}{\mathbb{O}}

\newcommand{\scra}{\mathscr{A}}

\newcommand{\scras}{\mathscr{C}}
\newcommand{\scrad}{\mathscr{D}}

\newcommand{\frc}{\mathfrak{C}}

\newcommand{\pizf}{\mathpzc{F}}
\newcommand{\pizg}{\mathpzc{G}}

\newcommand{\pizw}{\mathpzc{W}}
\newcommand{\pizW}{\mathpzc{W}}

\newcommand{\ve}{\varepsilon}

\pdfoutput=1
\abstract{Typically, a production of a particle with a small transverse momentum in hadron-hadron collisions is described
by CSS-based TMD factorization at moderate Bjorken $x_B\sim 1$ and by $k_T$-factorization at small $x_B$. 
A uniform description valid for all $x_B$ is provided by rapidity-only TMD factorization developed in a series
of recent papers at the tree level. In this paper the  rapidity-only TMD factorization for particle production by
gluon fusion is extended to the one-loop level.
 }
\arxivnumber{}

\emailAdd{balitsky@jlab.org}

\begin{document}
\title{Rapidity-only TMD factorization at one loop}

\author{ Ian Balitsky}

\affiliation{%
Physics Dept., Old Dominion University, Norfolk, VA 23529 \&
 Theory Group, JLAB, 12000 Jefferson Ave, Newport News, VA 23606
}

\preprint{JLAB-THY-23-3741}
\maketitle

\flushbottom

\section{Introduction\label{aba:sec1}}

Rapidity factorization and rapidity evolution are main tools for study of QCD processes at small $x$ \cite{Kovchegov:2012mbw}. 
On the other hand, at moderate $x$  conventional methods are based on CSS equation  \cite{Collins:1984kg} 
and closely related SCET
approach (see Refs. \cite{Collins:2011zzd} and  \cite{Rothstein:2016bsq}  reviews).
However, with the advent of EIC accelerator the region of energies intermediate between low  and moderate $x$
needs to be investigated. One of the ideas is to extend rapidity factorization methods beyond the ``pure'' small-$x$ region. In a 
series of recent papers A. Tarasov, G.A. Chirilli and the author applied the method of rapidity-only factorization
to processes of particle production in hadron-hadron collisions in the so-called Sudakov region where transverse momentum
of produced particle(s) is much smaller than their invariant mass. The typical examples of such processes are 
the Drell-Yan process or Higgs production by gluon fusion.  At moderate $x$ such processes are studied by
CSS-based TMD factorization  \cite{Collins:2011zzd, Collins:2014jpa}
\begin{eqnarray}
&&\hspace{-1mm}
{d\sigma\over  d\eta d^2q_\perp}~=~
\sum_f\!\int\! d^2b_\perp e^{i(q,b)_\perp}
\cald_{f/A}(x_A,b_\perp,\eta_a)\cald_{f/B}(x_B,b_\perp,\eta_2)
\nonumber\\
&&\hspace{14mm}
\times~
\sigma_{ff\rightarrow X}(\eta, \eta_a,\eta_b)~+~{\rm power ~corrections}~+~{\rm Y-terms}
\label{TMDf}
\end{eqnarray}
where $\eta=\half\ln{q^+\over q^-}$ is the rapidity, $\cald_{f/h}(x,z_\perp,\eta_i)$ is the 
TMD density of  a parton $f$  in hadron $h$ with rapidity cutoff $\eta_i$, and 
$\sigma_{ff\rightarrow X}(\eta, \eta_a,\eta_b)$ is the cross section of production of particle(s) $X$ 
of invariant mass $m_X^2=q^2\equiv Q^2\gg q_\perp^2$ in the scattering of two partons.  
The TMD parton densities are regularized with a combination of UV and rapidity cutoffs and 
the relevant Sudakov logarithms are obtain by solving double evolution with respect to $\mu_{\rm UV}$ and
the rapidity cutoffs $\eta_i$  \cite{Collins:2011zzd}.

It should be emphasized that the CSS approach and hence the formula (\ref{TMDf}) are valid at 
 $x_A\sim x_B\sim 1$. At small $x_A$ and/or $x_B$ one should resort to other factorization methods. 
 As I mentioned above, a  perspective approach is to apply methods based on rapidity-only factorization
 used  in small-$x$/BFKL physics.  In a series of papers \cite{Balitsky:2017flc,Balitsky:2017gis,Balitsky:2021fer,Balitsky:2022vnb} A. Tarasov and the author applied 
 rapidity-only factorization approach to get for the first time power corrections $\sim{q_\perp^2\over Q^2}$ 
 restoring EM gauge invariance of DY hadronic tensor both at moderate and small $x$. 
 Also, in recent papers \cite{Balitsky:2019ayf,Balitsky:2022vnb} G.A. Chirilli and the author calculated the rapidity-only 
 evolution of TMD operators, again both at moderate and small $x$. In the present paper I calculate coefficient function multiplying two TMD distributions at the one-loop level. This completes the task of performing the rapidity-only factorization 
at the one-loop accuracy.

Apart from requirement $Q^2=x_Ax_Bs\gg q_\perp^2$, in this paper it is  assumed  that
\begin{equation}
{Q^2\over q_\perp^2} ~\gg~ {q_\perp^2\over m_N^2}
\label{region}
\end{equation}
 The region (\ref{region})
can be understood in terms of rescaling $s\rightarrow\zeta s_0, ~\zeta\rightarrow \infty$ with $q_\perp^2$ fixed:
\begin{equation}
s~\sim~\zeta s_0, ~~~~q_\perp~\sim ~O(\zeta^0)
\label{rescale}
\end{equation}
It should be emphasized that we will not use the small-$x$ approximation $s\gg Q^2$ 
so our formulas are correct both at $x\ll 1$ and $x\sim 1$ provided that the  condition (\ref{region}) is satisfied.
Thus, at $x\sim 1$ our rapidity-evolution formulas should be equivalent to usual CSS approach, although the exact
relation between our rapidity evolution and CSS double evolution in rapidity and UV cutoff remains to be established, 
see the discussion in the Conclusions section.

The  
rapidity evolution of TMDs $\cald_{f/A}(x_A,b_\perp,\eta_a),\cald_{f/B}(x_B,b_\perp,\eta_b)$ 
should match the one-loop rapidity evolution of the coefficient function $\sigma_{ff\rightarrow H}(\eta, \eta_a,\eta_b)$
so that the cutoffs $\eta_a$ and $\eta_b$ disappear from physical amplitude.  
I will demonstrate that the result for the coefficient function in Eq. (\ref{TMDf}) for rapidity-only gluon TMD  factorization is proportional to
($\gamma\equiv\gamma_E\simeq 0.577$)
\bega
&&\hspace{-1mm}
\exp\Big\{ {\alpha_sN_c\over 2\pi}\Big[(\ln {b_\perp^2s\over 4}+\eta_a+\eta_b)^2
-2(\ln x_A-\eta_a+\gamma)(\ln x_B-\eta_b+\gamma)+\pi^2\Big]\Big\}
\label{aim}
\ega
and check that $\eta_a, \eta_b$ dependence matches the rapidity-only TMD evolution obtained in Refs. \cite{Balitsky:2019ayf,Balitsky:2022vnb}.

The paper is organized as follows. In Sect. \ref{sec:funt} I define hadronic tensor and TMD operators 
for particle production by gluon fusion.  In Sect. \ref{sec:fint} I discuss separation of functional integral
for particle production in three integrals according to the rapidity of the fields involved. 
In Sect. \ref{sec:coef} I set up the calculation of the coefficient function in front of TMD operators by computing
 diagrams in  two background fields. Sections \ref{sect:virtual}, \ref{sect:real}, \ref{sect:sumdiagrams},
 and \ref{sec:sgsubtract} 
 are devoted to calculation of these diagrams at the one-loop level. The result of the calculation and 
 check of matching to evolution of TMD operators are presented in Sect. \ref{sec:coefresult}. 
 The necessary technical and sidelined results are presented in the Appendix.

\section{TMD factorization for particle production by gluon fusion \label{sec:funt}}
Let us consider production of an (imaginary) scalar particle $\Phi$ by gluon fusion in proton-proton scattering. 
The particle is connected to gluons by the vertex
\begin{equation}
{\kal}_\Phi~=~g_\Phi\!\int\! d^4x~\Phi(x)g^2F^2(x),~~~~~F^2(x)~\equiv~F^a_{\mu\nu}(x)F^{a\mu\nu}(x)
\end{equation}
%
\begin{figure}[htb]
\begin{center}
\vspace{-11mm}
\includegraphics[width=151mm]{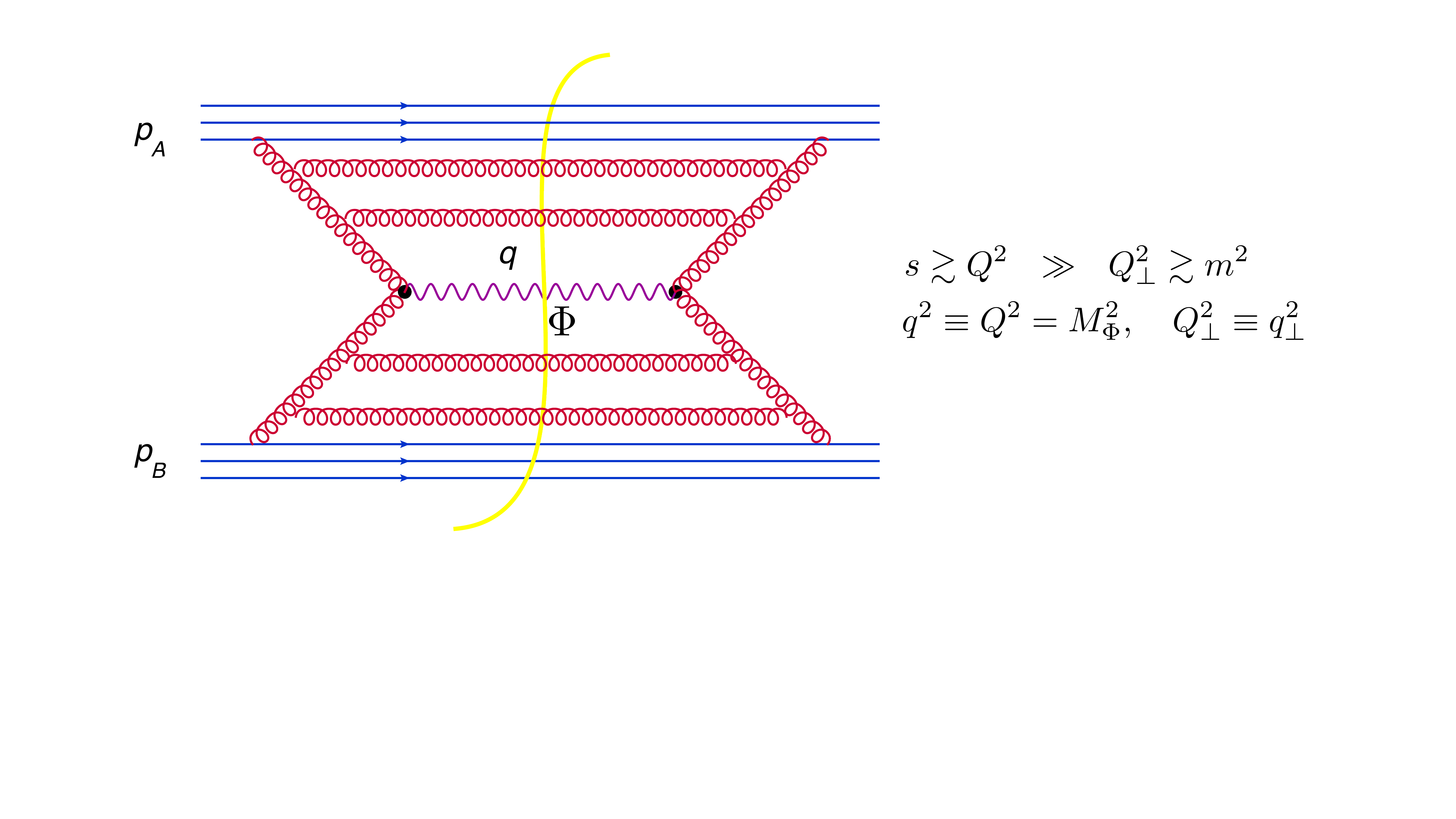}
\vspace{-44mm}
\end{center}
\caption{Particle production by gluon-gluon fusion \label{fig:1}}
\end{figure}
This is a ${m_H\over m_t}\ll 1$ approximation 
 for Higgs production via gluon fusion at the LHC.  The differential cross section of $\Phi$ production is defined by the ``hadronic tensor''  $W(p_A,p_B,q)$ 
\begin{eqnarray}
\hspace{-1mm}
W(p_A,p_B,q)~&\stackrel{\rm def}{=}&~{N_c^2-1\over 16}\sum_X\!\int\! d^4x~e^{-iqx}
\langle p_A,p_B|g^2F^2(x)|X\rangle\langle X|g^2F^2(0) |p_A,p_B\rangle
\nonumber\\
~&=&~{N_c^2-1\over 16}\!\int\! d^4x~e^{-iqx}
\langle p_A,p_B|g^4F^2(x)F^2(0) |p_A,p_B\rangle  
\label{W}
\end{eqnarray}
where the factor ${N_c^2-1\over 16}$ is added to simplify factorization formulas.
As usual,  $\sum_X$ denotes the sum over full set of ``out''  states.

We will study the hadronic tensor (\ref{W}) with non-zero momentum transfer in $t$-channel defined as a matrix element of the operator
\beq
\hatW(x_1,x_2)~\equiv~{N_c^2-1\over 16}g^4
F^a_{\mu\nu}(x_2)F^{\mu\nu;a}(x_2)F^b_{\lambda\rho}(x_1)F^{\lambda\rho;b}(x_1)
\label{operw}
\eeq
between initial and final states with slightly non-equal momenta
\begin{eqnarray}
&&\hspace{-1mm}
W(p_A,p_B,p'_A,p'_B; x_1,x_2)~=~\langle p'_A,p'_B|\hatW(x_1,x_2) |p_A,p_B\rangle  
\label{WC}
\end{eqnarray}
where
\begin{eqnarray}
&&\hspace{-1mm}
p_A~=~p_1+{m^2+l_\perp^2\over s}p_2-{l_\perp\over 2},~~~~~p'_A~=~p_1+{m^2+l_\perp^2\over s}p_2+{l_\perp\over 2},
\nonumber\\
&&\hspace{-1mm}
p_B~=~p_2+{m^2+l_\perp^2\over s}p_1+{l_\perp\over 2},~~~~~p'_B~=~p_2+{m^2+l_\perp^2\over s}p_1-{l_\perp\over 2}
\label{definishon}
\end{eqnarray}
Here $p_1$ and $p_2$ are light-like vectors close to $p_A$ and $p_B$, respectively.
\footnote{We assume that $t= -l_\perp^2\sim m_N^2$. If there is a longitudinal component of momentum transfer, 
one can redefine $p_1$ and $p_2$ in such a way  that with respect to new $p'_1$ and $p'_2$ the formulas are those of Eq. (\ref{definishon}). }  
We will use light-cone coordinates 
with respect to the frame where
 $p_1=\big({\sqrt{s}\over 2},0,0,{\sqrt{s}\over 2}\big)$ and $p_2=\big({\sqrt{s}\over 2},0,0,-{\sqrt{s}\over 2}\big)$ so
 that $p_1^+=p_2^-=\sqrt{s\over 2}$, $p_2^+=p_1^-=0$ and $p_{1_\perp}=p_{2_\perp}=0$. 
 
 The kinematical region (\ref{region}) in the coordinate space translates to
\begin{equation}
 x_\parallel^2\ll x_{12_\perp}^4m_N^2
\label{kinregion}
\end{equation}
where $x_\parallel^2\equiv 2x_{12}^+x_{12}^-$. 
Also, we must assume $x_{12_\perp}^2\leq m_N^{-2}$ so that the coupling constant $\alpha_s(x_\perp)$ is a valid perturbative parameter.

In the coordinate space,
TMD factorization  (\ref{TMDf})  for  hadronic tensor in  Eq. (\ref{WC}) should look like
\begin{eqnarray}
&&\hspace{-0mm}
{g^4\over 16}(N_c^2-1)\langle p'_A,p'_B| F^a_{\mu\nu}F^{a\mu\nu}(x_2)  F^b_{\lambda\rho}F^{b\lambda\rho}(x_1)|p_A,p_B\rangle
\nonumber\\
&&\hspace{-1mm}
=~\int\! dz_2^-dz_{2_\perp}dz_1^-dz_{1_\perp}dw_1^+dw_{1_\perp}dw_2^+dw_{2_\perp}
\frc(x_2,x_1;z_i^-,z_{i_\perp},w_i^+,w_{i_\perp};\sigma_p,\sigma_t)
\nonumber\\
&&\hspace{-1mm}
\times~\langle p'_A|\hacalo_{ij}^{\sigma_p}(z_2^-,z_{2_\perp};z_1^-,z_{1_\perp}) |p_A\rangle
\langle p'_B|\hacalo^{ij;\sigma_t}(w_2^+,w_{2_\perp};w_1^+,w_{1_\perp}) |p_B\rangle~+...
\label{facoord}
\end{eqnarray}
where the dots stand for power corrections $\sim{q_\perp^2\over Q^2}$ and
\begin{eqnarray}
&&\hspace{-2mm}
\hacalo_{ij}(z_2^+,z_{2_\perp};z_1^+,z_{1_\perp})
~=~\pizf^a_i(z_2)
[z_2-\infty^+,z_1-\infty^+]^{ab}\pizf^b_j(z_1)\Big|_{z_2^-=z_1^-=0}\,,
\label{kalo}
\\
&&\hspace{-2mm}
\hacalo_{ij}(z_2^-,z_{2_\perp};z_1^-,z_{1_\perp})
~=~
\pizf^a_i(z_2)
[z_2-\infty^-,z_1-\infty^-]^{ab}\pizf^b_j(z_1)\Big|_{z_2^+=z_1^+=0}\,,
\nonumber\\
&&\hspace{-2mm}
\pizf^{i,a}(z_\perp,z^+)~\equiv~gF^{- i,m}(z)[z^+,-\infty^+]_z^{ma}\Big|_{z^-=0},~~
\nonumber\\
&&\hspace{-2mm}
\pizf^{i,a}(z_\perp,z^-)~\equiv~gF^{+i,m}(z)[z^-,-\infty^-]^{ma}\Big|_{z^+=0}
\nonumber
\end{eqnarray}
are gluon TMD operators (the precise definitions of rapidity-only cutoffs $\sigma_a=e^{\eta_a}$ and 
$\sigma_b=e^{\eta_b}$ for gluon TMDs will be given later). Hereafter, we use the notation 
\begin{equation}
[x,y]~\equiv~{\rm P}e^{ig\!\int_0^1\! du~(x-y)^\mu A_\mu(ux+y-uy)}
\end{equation}
for the straight-line ordered gauge link between points $x$ and $y$, and 
space-saving notations
\begin{equation}
[x^+,y^+]_z~\equiv~[x^++z_\perp,y^++z_\perp],~~~~[x^-,y^-]_z~\equiv~[x^-+z_\perp,y^-+z_\perp]
\label{fla2.7}
\end{equation}

The coefficient function $\frc$ represents
a Fourier transform of $\sigma_{ff\rightarrow H}(\eta, \eta_1,\eta_2)$ in Eq. (\ref{TMDf}) with 
$\eta_i=\ln\sigma_i$. The normalization in the l.h.s. of Eq. (\ref{facoord}) is chosen in such a way that 
$\frc=1+{\alpha_sN_c\over 2\pi}\frc_1+O(\alpha_s^2)$.  The goal of this paper is to find 
the one-loop coefficient function $\frc_1(x_2,x_1;z_i^-,z_{i_\perp},w_i^+,w_{i_\perp};\sigma_a,\sigma_b)$ 
and check that the evolution of this coefficient function matches the evolutions of TMD operators.

\section{TMD factorization from functional integral \label{sec:fint}}

The hadronic tensor  (\ref{W})
 can be represented by double functional integral
\begin{eqnarray}
&&\hspace{-2mm}
W(p_A,p_B,p'_A,p'_B;x_2,x_1)~=~\sum_X
\langle p'_A,p'_B|g^2F^2(x_2)|X\rangle\langle X|g^2F^2(x_1) |p_A,p_B\rangle
\label{dablfun}\\
&&\hspace{-2mm}
=\lim_{t_i\rightarrow -\infty}^{t_f\rightarrow\infty}g^4
\!\int^{\tilA(t_f)=A(t_f)}\!\!  D\tilA_\mu DA_\mu \!\int^{\tsi(t_f)=\psi(t_f)}\! D\tilde{\bar\psi}D\tilde{\psi} D\bsi D\psi 
~e^{-iS_{\rm QCD}(\tilA,\tipsi)}e^{iS_{\rm QCD}(A,\psi)}
\nonumber\\
&&\hspace{-2mm}
\times~\Psi^\ast_{p'_A}(\vec{\tilA}(t_i),\tipsi(t_i))\Psi^\ast_{p'_B}(\vec{\tilA}(t_i),\tipsi(t_i))
\tilF^2(x_2)F^2(x_1)\Psi_{p_A}(\vec{A}(t_i),\psi(t_i))\Psi_{p_B}(\vec{A}(t_i),\psi(t_i))
\nonumber
\end{eqnarray}
Here the fields $A,\psi$ correspond to the amplitude  $\langle X|F^2(x_1) |p_A,p_B\rangle$, the fields $\tilA,\tipsi$ correspond to 
complex conjugate amplitude $\langle p'_A,p'_B|F^2(x_2)|X\rangle$
and  $\Psi_p(\vec{A}(t_i),\psi(t_i))$ denote the proton wave function at the initial time $t_i$. The boundary conditions
$\tilA(t_f)=A(t_f)$ and $\tsi(t_f)=\psi(t_f)$ reflect the sum over all states $X$, cf. Refs. \cite{Balitsky:1988fi,Balitsky:1990ck,Balitsky:1991yz}. We will also use the notation
\begin{equation}
\{x,y\}~\equiv~{\rm P}e^{ig\!\int_0^1\! du~(x-y)^\mu \tilA_\mu(ux+y-uy)}
\end{equation}
and similar notations like Eq. (\ref{fla2.7}) for gauge links in the left sector.

For calculations in the momentum space we will use Sudakov variables  related to  light-cone components $p^+,p^-,p_\perp$ by 
$\alpha\equiv p^+/\vro$ and $\beta\equiv p^-/\vro$ where $\vro\equiv\sqrt{s/2}$.
In terms of Sudakov variables 
 $p\cdot q~=~(\alpha_p\beta_q+\alpha_q\beta_p){s\over 2}-(p,q)_\perp$ where $(p,q)_\perp\equiv -p_iq^i$. 
Throughout the paper, the sum over the Latin indices $i$, $j$... runs over the two transverse 
components while the sum over Greek indices runs over the four components as usual. 
Also, since we  use Sudakov variables 
it is convenient to change the notations of gluon momentum fractions to
\beq
{\alpha_a}\equiv x_A,~~~{\beta_b}\equiv x_B
\label{xaxb}
\eeq
to avoid confusion with coordinates.

Following Refs. \cite{Balitsky:2017flc,Balitsky:2017gis}, to derive the factorization formula we separate gluon (and quark) fields in the functional integral (\ref{dablfun}) into three sectors: ``projectile'' fields $A_\mu, \psi_a$ 
with $|\beta|<\sigma_p\equiv\sigma_a$, 
`` target'' fields $B_\mu,\psi_b$ with $|\alpha|<\sigma_t\equiv\sigma_b$ and ``central rapidity'' fields $C_\mu,\psi$ with $|\alpha|>\sigma_t$ and $|\beta|>\sigma_p$. 
\begin{figure}[htb]
\begin{center}
\includegraphics[width=151mm]{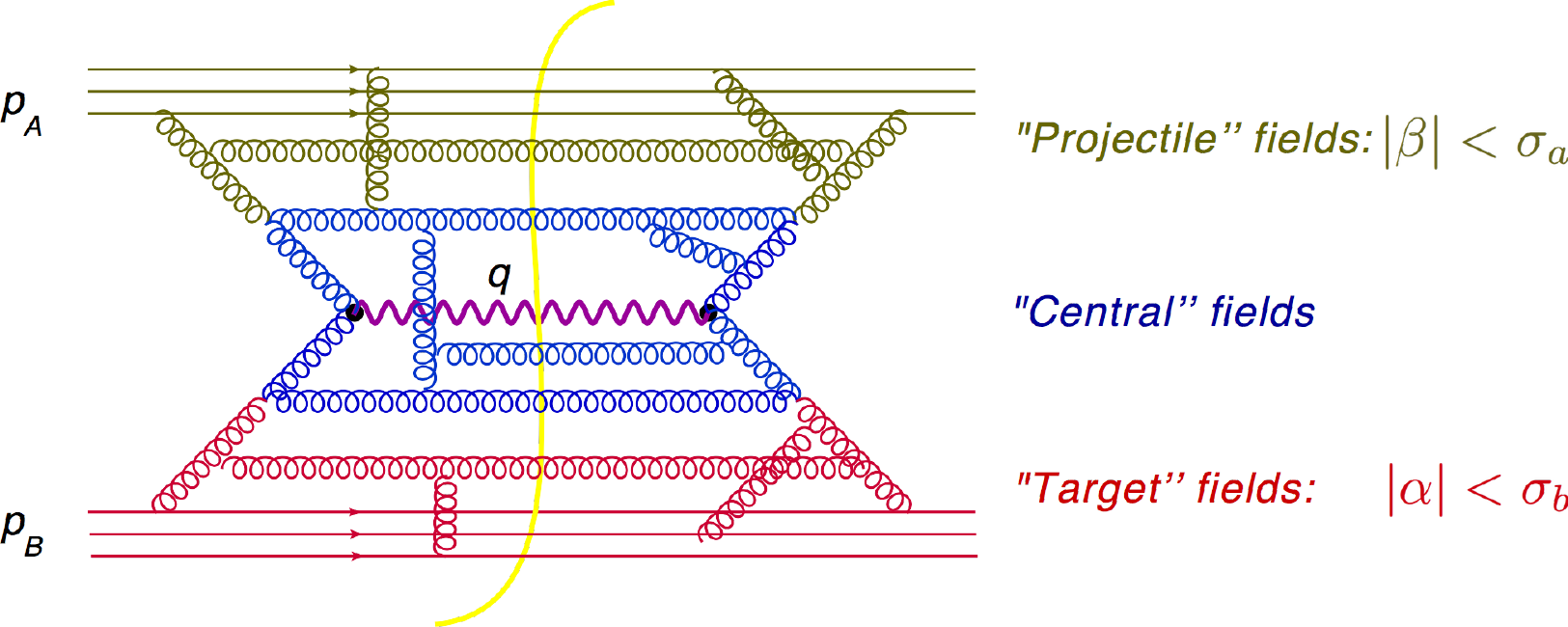}
\end{center}
\caption{Rapidity factorization for particle production \label{fig:2}}
\end{figure}
Let us specify the values of the TMD cutoffs $\sigma_p$ and $\sigma_t$ in our factorization. 
Needless to say, we should take $\sigma_t\ll {\alpha_a}$ and $\sigma_p\ll {\beta_b}$. Moreover,
as discussed in Ref. \cite{Balitsky:2022vnb}, power corrections to  rapidity evolution of TMDs are $\sim {Q_\perp^2\over{\beta_b}\sigma_ts}$ so we need to 
assume $\sigma_t{\beta_b}s\gg Q_\perp^2$, and similarly  $\sigma_p{\alpha_a}s\gg Q_\perp^2$ for the projectile. Next, as we shall see below, it is convenient to calculate coefficient function (\ref{ceodin}) at $m_N^2\gg \mu_\sigma^2\equiv\sigma_p\sigma_ts$ so finally we take
 the region 
of $\sigma_p$ and $\sigma_t$ as follows
\beq
{\alpha_a}\gg\sigma_t\gg{Q_\perp^2\over{\beta_b}s},~~~{\beta_b}\gg\sigma_p\gg{Q_\perp^2\over{\alpha_a}s},~~~~
m_N^2\gg\mu_\sigma^2\equiv\sigma_p\sigma_ts\gg{Q_\perp^4\over Q^2}
\label{condisigmas}
\eeq
Note that due to  Eq. (\ref{region})  we can choose $\mu_\sigma^2$ between $m_N^2$ and 
parametrically small ${Q_\perp^4\over Q^2}$.  In terms of rescaling (\ref{rescale}) this means that we can choose
$\sigma_p,\sigma_t\sim \zeta^{-{3\over 4}}\sim\big({Q_\perp\over Q}\big)^{3/2}$ so 
that 
\beq
\mu^2_\sigma\sim \zeta^{-1/2}~~~\Leftrightarrow~~~ 1\gg{\mu_\sigma^2\over Q_\perp^2}\gg {Q_\perp^2\over Q^2}\sim \zeta^{-1}
\label{musigma}
\eeq
and both conditions in Eq. (\ref{condisigmas}) 
are satisfied.

In this paper we are calculating logarithmical corrections so the power corrections due to the small parameters 
${O_\perp^2\over\sigma_p{\alpha'_a}s}$,
${O_\perp^2\over\sigma_t{\beta'_b}s}$ will be systematically neglected. 
The convenient notations for small
parameters are
\beq
\lambda\equiv{Q_\perp^2\over Q^2}\sim {1\over\zeta}\ll 1,~~~~\lambda_p\equiv{Q_\perp^2\over |{\alpha_a}|\sigma_p s}\sim\zeta^{-{1\over 4}}\ll 1,~~~~
\lambda_t\equiv{Q_\perp^2\over \sigma_t|{\beta_b}|s}\sim\zeta^{-{1\over 4}}\ll 1
\label{lambdas}
\eeq
In these notations the last condition in Eq. (\ref{condisigmas}) translates to
\beq
\lambda_p\lambda_t\gg\lambda
\label{condilambdas}
\eeq
Thus, in terms of rescaling (\ref{rescale}) we neglect
all power corrections $\sim 1/\zeta^{1/4}$ or smaller.

\begin{figure}[htb]
\begin{center}
\includegraphics[width=101mm]{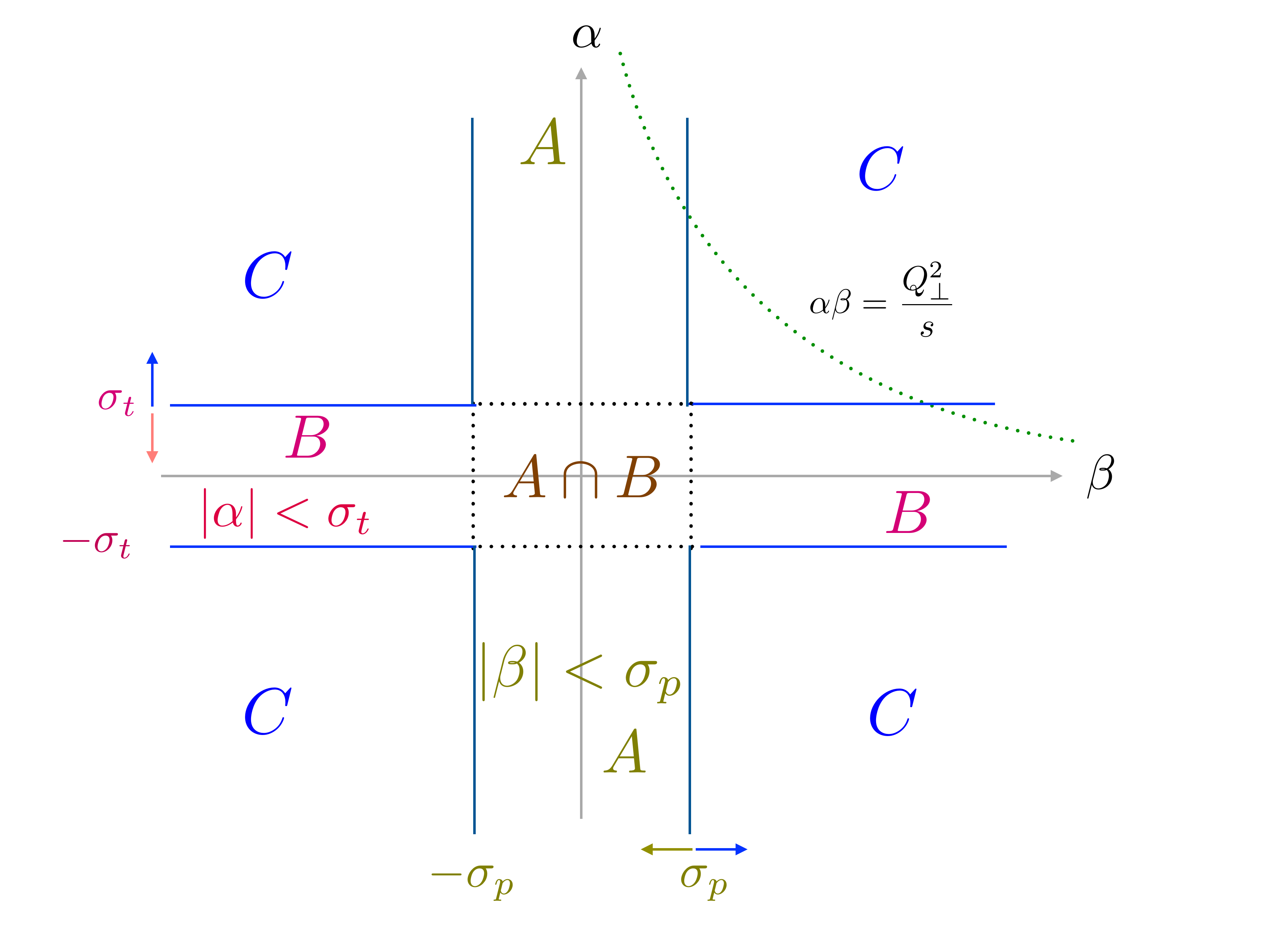}
\end{center}
\caption{Regions of factorization in the momentum space   \label{fig:albe1}}
\end{figure}

Note that while  central fields are well separated from projectile and target fields, the latter have an intersection 
when both $\alpha$ and $\beta$ are small, $\alpha\leq \sigma_t\sim\zeta^{-{3\over 4}}$ and
$\beta\leq \sigma_p\sim\zeta^{-{3\over 4}}$, 
see Fig. \ref{fig:albe1}.
 Depending on the scale of characteristic
transverse momenta they are called Glauber gluons (if $p_\perp\gg p^+,p^-\sim \zeta^{-{1\over 4}}$) or soft  gluons (if $p_\perp\sim p^+,p^-$).  
 We will denote both of them by notation $\scras=A\cap B$ and call them soft/Glauber (sG) gluons.  We will discuss later that 
 Glauber gluons do not contribute to factorization (\ref{TMDf}) \cite{Collins:1984kg,Collins:2011zzd} and soft gluons form a soft factor which 
 is a power correction with our rapidity cutoffs. 
 
To discuss interaction of central gluons with either $A$, $B$, or $\scras$ fields it is convenient do denote all the latter by notation
 $\scra=A\cup B$  which means all fields with $|\alpha|<\sigma_a$ and/or $|\beta|<\sigma_b$.

We get
\begin{eqnarray}
&&\hspace{-1mm}
W(p_A,p_B,p'_A,p'_B;x_1,x_2)~=~\lim_{t_i\rightarrow -\infty}g^4\!\int\! \scrad\Phi_\scra\scrad\Phi_C~
\Psi^\ast_{p'_A}(t_i)\Psi_{p_A}(t_i)
\nonumber\\
&&\hspace{-1mm}
\times~\Psi^\ast_{p'_B}(t_i)\Psi_{p_B}(t_i)(\tilF^\scra+\tilF^C)^2(x_2)(F^\scra+F^C)^2(x_1)
\label{W2}
\end{eqnarray}
where $F^\scra$ is the usual field tensor for $\scra$ field 
and $F^C_{\mu\nu}~\equiv~F_{\mu\nu}(\scra+C)-F_{\mu\nu}(\scra)$. 
Also, we use  shorthand notations 
\begin{eqnarray}
&&\hspace{-1mm}
\Psi_{p_A}(t_i)~\equiv~\Psi_{p_A}(\vec{A}(t_i),\psi(t_i)),~~~~~~~~~~\Psi^\ast_{p_A}(t_i)~\equiv~\Psi^\ast_{p_A}(\vec{\tilA}(t_i),\tipsi_a(t_i))
\nonumber\\
&&\hspace{-1mm}
\Psi_{p_B}(t_i)~\equiv~\Psi_{p_B}(\vec{B}(t_i),\psi_b(t_i)),~~~~~~~~~~
\Psi^\ast_{p_B}(t_i)~\equiv~\Psi^\ast_{p_B}(\vec{\tilB}(t_i),\tipsi_b(t_i)),
\end{eqnarray}
for projectile and target protons' wave functions, 
and 
\begin{eqnarray}
&&\hspace{-1mm}
\int\!\scrad\Phi_\scra~\equiv~
\int^{\tilde{\scra}(t_f)=\scra(t_f)}\! D\tilde{\scra}_\mu D\scra_\mu
\nonumber\\
&&\hspace{14mm}
\times~\int^{\tipsi_\scra(t_f)=\psi_\scra(t_f)} D\tilde{\psi}_\scra D\psi_\scra
e^{-iS_{\rm QCD}(\tilde{\scra},\tipsi_\scra)+iS_{\rm QCD}(\scra,\psi_\scra)},
\label{deFi}\\
&&\hspace{-1mm}
\int\!\scrad\Phi_C~\equiv~\int^{\tilC(t_f)=C(t_f)} \! D\tilC_\mu DC_\mu\int^{\tsi_C(t_f)=\psi_C(t_f)}\! D\psi_C D\tsi_C~
e^{-i\tilS_C+iS_C}
\label{defic}
\end{eqnarray}
for functional integrals. 
In the last line $S_C~\equiv~ S_{\rm QCD}(\scra+C)-S_{\rm QCD}(\scra)$.

Our goal is to integrate over central fields and get
the amplitude in the factorized form, as a (sum of) products of functional integrals over $A$ fields representing projectile matrix elements (TMDs) 
and functional integrals over $B$ fields representing target matrix elements. In the spirit of background-field method, we ``freeze'' projectile and target fields
 and get a sum of diagrams in these external fields. 
Since  $|\beta|<\sigma_p$ in the projectile fields and $|\alpha|<\sigma_t$  in the target fields, at the  tree-level 
one can set with power accuracy $\beta=0$ for the  projectile fields and $\alpha=0$ for the target fields - the corrections will
be $O\big({m^2\over\sigma_p s}\big)$ and  $O\big({m^2\over\sigma_t s}\big)$.
\footnote{Indeed, suppose an ``A'' gluon with momentum $k_a$ interacts with a ``C'' gluon with momentum $k_C$. The 
resulting propagator is $[({\alpha'_a}+\alpha_c)(\beta_a+\beta_c)s-(k_a+k_c)_\perp^2]^{-1}$ and one can neglect $\beta_a$
with $\beta_a/\beta_c\leq {m^2/s\over\sigma_p}$ accuracy. Similarly, for the target fields one gets the accuracy
$\alpha_b/\alpha_c\leq {m^2/s\over\sigma_t}$. }
Beyond the tree level,  the integration over $C$ fields will produce
the logarithms of the cutoffs $\sigma_p$ and $\sigma_t$ which cancel with the corresponding
logs in gluon TMDs of the projectile and the target, see the discussion in Sect. \ref{sec:coefresult}.

\section{Coefficient function from background-field diagrams \label{sec:coef}}
We will calculate the coefficient function $\frc$ in the first non-trivial order in perturbation theory: $\frc=1+{\alpha_sN_c\over 2\pi}\frc_1$. 
The desired formula looks like that
\begin{eqnarray}
&&\hspace{-2mm}
{1\over 16}(N_c^2-1)\langle p'_A,p'_B| g^2F^a_{\mu\nu}F^{a\mu\nu}(x_2)  g^2F^b_{\lambda\rho}F^{b\lambda\rho}(x_1)
|p_A,p_B\rangle
\nonumber\\
&&\hspace{-2mm}
=~
\!\int\!\scrad\Phi_\scra~\Psi^\ast_{p'_A}(t_i)\Psi_{p_A}(t_i)
\Psi^\ast_{p'_B}(t_i)\Psi_{p_B}(t_i)\Big[\hacalo_{ij}^{\sigma_p}(x_2^-,x_{2_\perp};z_1^-,x_{1_\perp})
\hacalo^{ij;\sigma_t}(x_2^+,x_{2_\perp};x_1^+,x_{1_\perp})
\nonumber\\
&&\hspace{11mm}
+~\int\! dz_1^-dz_{1_\perp}dz_2^-dz_{2_\perp}dw_1^+dw_{1_\perp}dw_2^+dw_{2_\perp}
{\alpha_sN_c\over 2\pi}\frc_1(x_1,x_2;z_i^-,z_{i_\perp},w_i^+,w_{i_\perp};\sigma_p,\sigma_t)
\nonumber\\
&&\hspace{22mm}
\times~\hacalo_{ij}^{\sigma_p}(z_2^-,z_{2_\perp};z_1^-,z_{1_\perp}) 
\hacalo^{ij;\sigma_t}(z_2^+,z_{2_\perp};z_1^+,z_{1_\perp}) ~+...\Big]
\label{facoord1}
\end{eqnarray}
where dots stand for higher orders in perturbation theory and/or power corrections.

The standard way to obtain a coefficient function is to rewrite Eq.  (\ref{facoord1}) as
an operator formula
\begin{eqnarray}
&&\hspace{-1mm}
\int\! dz_2^-dz_{2_\perp}dz_1^-dz_{1_\perp}dw_1^+dw_{1_\perp}dw_2^+dw_{2_\perp}
{\alpha_sN_c\over 2\pi}\frc_1(x_1,x_2;z_i^-,z_{i_\perp},w_i^+,w_{i_\perp};\sigma_p,\sigma_t)
\nonumber\\
&&\hspace{-1mm}
\times~\hacalo_{ij}^{\sigma_p}(z_2^-,z_{2_\perp};z_1^-,z_{1_\perp})
\hacalo^{ij;\sigma_t}(z_2^+,z_{2_\perp};z_1^+,z_{1_\perp})~+...
\label{facoord2}\\
&&\hspace{-1mm}
=~{g^4\over 16}(N_c^2-1)F^a_{\mu\nu}F^{a\mu\nu}(x_2)  F^b_{\lambda\rho}F^{b\lambda\rho}(x_1)
-~\hacalo_{ij}^{\sigma_p}(x_2^-,x_{2_\perp};x_1^-,x_{1_\perp})
\hacalo^{ij;\sigma_t}(x_2^+,x_{2_\perp};x_1^+,x_{1_\perp}),
\nonumber
\end{eqnarray}
calculate the l.h.s. and r.h.s. between two initial and two final gluon states and compare.  
However, the amplitudes with real gluons have infrared divergencies so we will consider
amplitudes with virtual gluon tails instead. As is well known,
a gauge-invariant  way to write down matrix elements ``between virtual gluons'' is to consider 
l.h.s. and r.h.s. of Eq. (\ref{facoord2}) in a suitable background field.

Following the analysis of rapidity factorization (\ref{W2}) in Refs. \cite{Balitsky:2017flc,Balitsky:2017gis}, we choose the background field as a result of interaction of  ``projectile''  field $\barA$ and ``target'' field $\barB$ where
the ``projectile''  field $\barA(z)$ depends only on $z_\perp,z_-$ (corresponding to $\beta_a$=0) and 
``target''  field $\barB(z)$ depends only on $z_\perp,z_+$ (corresponding to $\alpha_b$=0).
\footnote{See Ref. \cite{Vladimirov:2021hdn}  for similar approach}  As  demonstrated in 
 Ref.  \cite{Balitsky:2017flc}, in this case one can always choose the gauge where $\barA^-=\barB^+=0$. Moreover, since we are after logarithmical corrections to coefficient in front of the operators (\ref{kalo}), it is convenient to take $\barA^+=\barB^-=0$. 
\footnote{ The general case with  background fields $\barA^-,\barB^+\neq 0$ is relevant for obtaining power corrections
to Eq. (\ref{facoord}), see the discussion in Refs.  \cite{Balitsky:2017flc,Balitsky:2017gis}
}
Thus , we choose  ``projectile'' and ``target'' fields in the form
\begin{eqnarray}
\hspace{-1mm}
g\barA_i~=~U_i(x^-,x_\perp),~~A_+=A_-~=~0,&~~~~~~&g\barB_i~=~V_i(x^+,x_\perp),~~B_+=B_-~=~0
\nonumber\\
g\barF^{+ i}(A)~=~\partial^+U_i\equiv U^{+i}(x^-,x_\perp),&~~~~~~&g\barF^{+ i}(B)~=~\partial^-V_i\equiv V^{-i}(x^+,x_\perp)
\label{aibefildz}
\end{eqnarray}
We assume that the ``projectile'' and ``target'' fields $A(z_-,z_\perp)$ and $B(z_+,z_\perp)$ satisfy standard YM equations
\begin{equation}
(\partial_i-i[U_i,) U^{-i}~=~g^2\bsi_A\gamma^-\psi_A,~~~~~(\partial_i-i[V_i,) V^{+i}~=~g^2\bsi_B\gamma\gamma^+\psi_B
\end{equation}
and only ``good'' components of background quark fields  $\gamma^-\psi_A(z_-,z_\perp)$ and 
$\gamma^+\psi_B(z_+,z_\perp)$ do exist.
 
\begin{figure}[htb]
\begin{center}
\includegraphics[width=111mm]{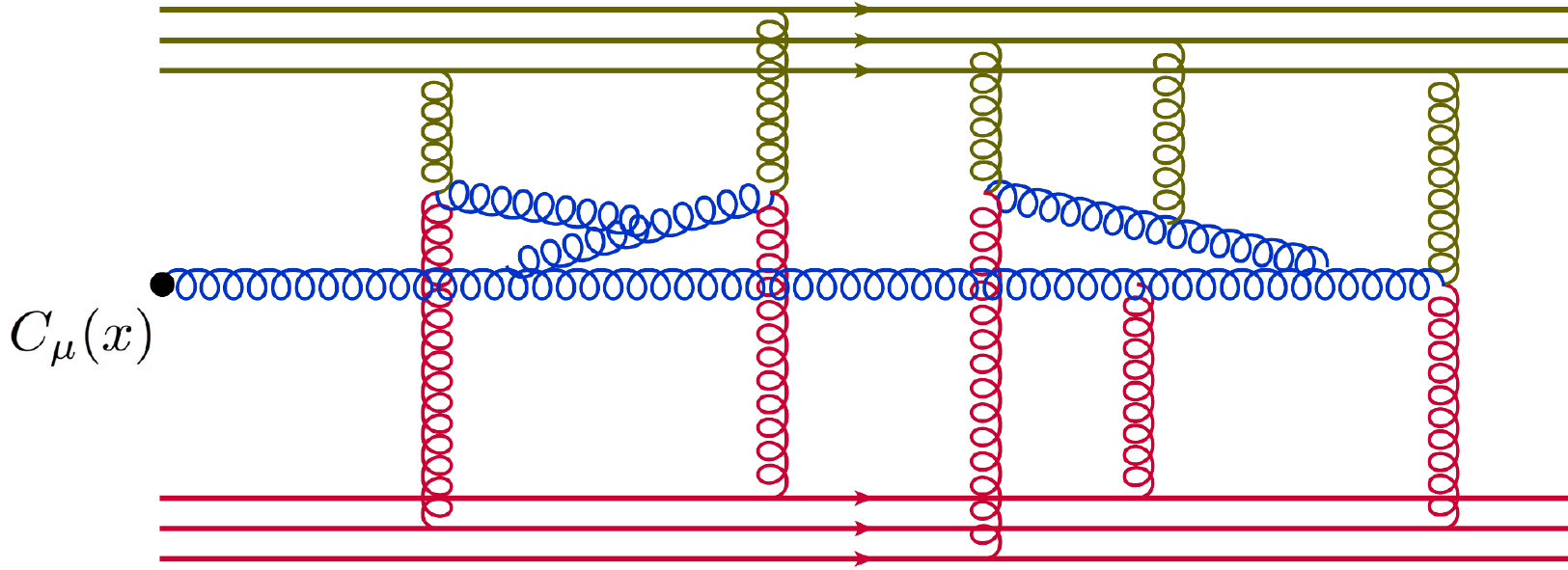}
\end{center}
\caption{Typical diagram for the classical field with projectile/target sources. 
The Green functions of the central fields are given by retarded propagators.   \label{fig:Ckl}}
\end{figure}

 The ``interaction'' field $\matA$ is defined as  is a classical field solving classical YM equations
\begin{equation}
\matD^\nu \matF^a_{\mu\nu}~=~\sum_fg\Bsi^f t^a\gamma_\mu\Psi^f,~~~~(\not\!P+m_f)\Psi^f~=~0
\label{kleqs}
\end{equation}
with boundary conditions
\footnote{It is convenient to fix redundant gauge transformations by requirements $\barA_i(-\infty_\bu,x_\perp)=0$ for the projectile and $\barB_i(-\infty_\ast,x_\perp)=0$ for the target, 
see the discussion in Ref. \cite{Belitsky:2002sm}.
}
\begin{eqnarray}
&&\hspace{-1mm}
\matA_\mu(x)\stackrel{x^+\rightarrow -\infty}{=}\barA_\mu(x^-,x_\perp),~~~~
\Psi(x)\stackrel{x^+\rightarrow -\infty}{=}\psi_A(x^-,x_\perp)
\nonumber\\
&&\hspace{-1mm}
\matA_\mu(x)\stackrel{x^-\rightarrow -\infty}{=}\barB_\mu(x^+,x_\perp),~~~~
\Psi(x)\stackrel{x^-\rightarrow -\infty}{=}\psi_B(x^+,x_\perp)
\label{inicondi}
\end{eqnarray}
reflecting the fact that at $t\rightarrow -\infty$ we have only incoming hadrons with ``A'' and ``B'' fields. 
An important property of the functional integral (\ref{W2}) is that since the projectile fields $\bar\tilA$ and $\barA$ 
should coincide at $t\rightarrow\infty$ (see Eq. (\ref{deFi}))
and since they do not depend on $x^+$, they coincide everywhere. Similar property is valid for the target fields 
 so we have the condition
\beq
\bar\tilA~=~\barA,~~~\bar\tilB~=~\barB
\label{kandishen}
\eeq
As proved
 in Refs. \cite{Balitsky:2017flc,Balitsky:2017gis} the solution of classical equations (\ref{kleqs}) has the same property:
 $\tilde\matA=\matA$. In terms of perturbative diagrams the solution of Eq. (\ref{kleqs}) 
  is given by the sum $\barA+\barB+\barC$ where
the ``correction'' field $\barC$ is given by the sum diagrams of the type shown in Fig. \ref{fig:Ckl} with {\it retarded} propagators.

The solution of YM equations (\ref{kleqs}) in general case  is yet unsolved problem,
 especially important for description of scattering of two heavy nuclei
in semiclassical approximation. Fortunately, for our case of particle production with ${q_\perp\over Q}\ll 1$ 
one can construct the ``correction'' field $\barC$ as a series in this small parameter.
The explicit form of the expansion of correction field $\barC$ in powers of this small parameter is presented in Refs. \cite{Balitsky:2017flc,Balitsky:2017gis}.  We will need only one term in this expansion shown in  Eq. (\ref{corefs}) below.

First, let us present the estimates of relative strength of different components of projectile and target fields
in our $Q^2\gg q_\perp^2$ kinematics from Ref. \cite{Balitsky:2017flc}
\begin{eqnarray}
&&\hspace{-1mm}
U^i, V^i~\sim~Q_\perp,~~~~~~~~U^{+i}=\partial^+U^i~\sim~Q_\perp\sqrt{s},~~~V^{-i}=\partial^-V^i~\sim~Q_\perp\sqrt{s}
\label{pc}\\
&&\hspace{-1mm}
U^{ij}=\partial^iU^j-\partial^jU_i-i[U^i,U^j]~\sim~Q_\perp^2,~~~V^{ij}=\partial^iV^j-\partial^jV_i-i[V^i,V^j]~\sim~Q_\perp^2
\nonumber
\end{eqnarray}
Note that Eq. (\ref{pc}) means that characteristic scales of projectile fields are such that extra $\partial^+$ brings $\sqrt{s}$ 
and extra $\barD_i$ is $\sim q_\perp$ so $\partial^+\gg\barD^i$ for the projectile fields. Similarly, for the target fields 
$\partial^-\gg\barD^i$. 
The characteristic longitudinal distances are $z^+\sim 1/\sqrt{s}$ for the projectile fields and $z^-\sim 1/\sqrt{s}$ for 
the target ones while the characteristic transverse distances  are $\sim 1/Q_\perp$  for both of them.
Also, in sorting out power corrections according to rescaling parameter $\zeta$ in Eq. (\ref{rescale}), we do not distinguish 
between ${m_N^2\over Q^2}$, ${q_\perp^2\over Q^2}$, ${m_N^2\over s}$, and  ${q_\perp^2\over s}$ and use the common notation 
$$
O\big({m_\perp^2\over s}\big)~\sim~O\Big({m_N^2\over Q^2}, {q_\perp^2\over Q^2},{m_N^2\over s},{q_\perp^2\over s}\Big)
$$
and similarly for other ratios.

The relevant terms in the expansion of correction fields $\barC$  are \cite{Balitsky:2017flc}
\begin{eqnarray}
&&\hspace{-1mm}
\barC^{-}(x)~=~{1\over 2\vro}\int\! dz (x|{1\over\alpha+\ie}|z)[U_k(z^-,z_\perp),V^k(z^+,z_\perp)]
\nonumber\\
&&\hspace{11mm}
=~-{i\over 2g}\!\int_{-\infty}^{x^+}\!dx'^+\!\int_{-\infty}^{x^-}\!dx'^-(x-x')^- 
[U_{~k}^{+}(x'^-,x_\perp),V^{- k}(x'^+,x_\perp)]~\sim~{m_\perp^2\over\sqrt{s}},    
\nonumber\\
&&\hspace{-1mm}
\barC^{+}(x)~=~-{1\over 2\vro}\int\! dz (x|{1\over\beta+\ie}|z)[U_k(z^-,z_\perp),V^k(z^+,z_\perp)]
\nonumber\\
&&\hspace{11mm}
=~{i\over 2g}\!\int_{-\infty}^{x^-}\!dx'^-\!\int_{-\infty}^{x^+}\!dx'^+(x-x')^+
[U_{~k}^{+}(x'^-,x_\perp),V^{- k}(x'^+,x_\perp)]~\sim~{m_\perp^2\over\sqrt{s}}
\nonumber\\
&&\hspace{-1mm}
\barC^{i}(x)~=~{-i\over 2g}\!\int_{-\infty}^{x^-}\!dx'^-\!\int_{-\infty}^{x^+}\!dx'^+
\Big([U^j(x'^-,x_\perp),V_{ij}(x'^+,x_\perp)]+\partial^j[U_i(x'^-,x_\perp),V_j(x'^+,x_\perp)]
\nonumber\\
&&\hspace{11mm}
+~[V^j(x'^+,x_\perp),U_{ij}(x'^-,x_\perp)]+\partial^j[V_i(x'^+,x_\perp),U_j(x'^-,x_\perp)]\Big)~\sim~{m_\perp^3\over s}
\label{corfildz}
\end{eqnarray}
where we used
\beq
\hspace{-0mm}
U^i(x^-,x_\perp)=\int_{-\infty}^{x^-}\!d{x'}^- U^{+i}({x'}^-,x_\perp),~~~~
V^i(x^+,x_\perp)=\int_{-\infty}^{x^+}\!d{x'}^+ V^{+i}({x'}^+,x_\perp)
\eeq
 It should be noted that in expressions (\ref{corfildz}), (\ref{corefs}) we neglected terms
$\sim U_iU_jV_k$ and $U_iV_jV_k$ since they are proportional to $F_{\xi\eta}^3$ and hence cannot contribute to our coefficient function.

Thus, we need to calculate l.h.s. and r.h.s. of Eq. (\ref{facoord2}) in the background of the field 
\begin{equation}
\matA~\simeq~\barA+\barB+\barC
\end{equation}
given by Eqs. 
(\ref{aibefildz}) and (\ref{corefs}) 
\begin{eqnarray}
&&\hspace{-1mm}
\int\! dz_2^-dz_{2_\perp}dz_1^-dz_{1_\perp}dw_1^+dw_{1_\perp}dw_2^+dw_{2_\perp}
{\alpha_sN_c\over 2\pi}\frc_1(x_1,x_2;z_i^-,z_{i_\perp},w_i^+,w_{i_\perp};\sigma_p,\sigma_t)
\nonumber\\
&&\hspace{-1mm}
\times~\langle \hacalo_{ij}^{\sigma_p}(z_2^-,z_{2_\perp};z_1^-,z_{1_\perp}) 
\hacalo^{ij;\sigma_t}(z_2^+,z_{2_\perp};z_1^+,z_{1_\perp}) \rangle_{\matA}~+...
\nonumber\\
&&\hspace{-0mm}
=~{N_c^2-1\over 16}g^4\langle \tilF^a_{\mu\nu}\tilF^{a\mu\nu}(x_2)  F^b_{\lambda\rho}F^{b\lambda\rho}(x_1)\rangle_{\matA}
\nonumber\\
&&\hspace{-1mm}
-~\langle \hacalo_{ij}^{\sigma_p}(x_2^-,x_{2_\perp};x_1^-,x_{1_\perp}) 
\hacalo^{ij;\sigma_t}(x_2^+,x_{2_\perp};x_1^+,x_{1_\perp}) \rangle_{\matA}
\label{facoord3}
\end{eqnarray}

Since we are after first order of perturbation theory, operators in the l.h.s. of this equation can be replaced in the leading order by corresponding classical fields
\begin{eqnarray}
&&\hspace{-1mm}
\langle \hacalo^{ij,\sigma_p}(z_2^-,z_{2_\perp};z_1^-,z_{1_\perp})\rangle_\barA
~=~U^{+ i,a}(z_2^-,z_{2_\perp})U^{+ j,a}(z_1^-,z_{1_\perp})+O(\alpha_s)
\nonumber\\
&&\hspace{-1mm}
\langle \hacalo^{ij;\sigma_t}(z_2^+,z_{2_\perp};z_1^+,z_{1_\perp})\rangle_\barB
~=~V^{-i,a}(z_2^+,z_{2_\perp})V^{-j,a}(z_1^+,z_{1_\perp})+O(\alpha_s)
\end{eqnarray}
so the master formula (\ref{facoord3}) takes the form
\begin{eqnarray}
&&\hspace{-1mm}
\int\! dz_2^-dz_{2_\perp}dz_1^-dz_{1_\perp}dw_1^+dw_{1_\perp}dw_2^+dw_{2_\perp}
{\alpha_sN_c\over 2\pi}\frc_1(x_1,x_2;z_i^-,z_{i_\perp},w_i^+,w_{i_\perp};\sigma_p,\sigma_t)
\nonumber\\
&&\hspace{11mm}
\times~U^{- i,a}(z_2^+,z_{2_\perp})U^{- j,a}(z_1^+,z_{1_\perp})
V^{+i,a}(z_2^-,z_{2_\perp})V^{+j,a}(z_1^-,z_{1_\perp})
\nonumber\\
&&\hspace{-0mm}
=~{N_c^2-1\over 16}g^4\langle \tilF^a_{\mu\nu}\tilF^{a\mu\nu}(x_2)
F^b_{\lambda\rho}F^{b\lambda\rho}(x_1)\rangle_{\matA}
\nonumber\\
&&\hspace{11mm}
-~\langle \hacalo^{ij,\sigma_p}(x_2^-,x_{2_\perp};x_1^-,x_{1_\perp}) 
\hacalo^{ij;\sigma_t}(x_2^+,x_{2_\perp};x_1^+,x_{1_\perp}) \rangle_{\matA}
\label{master}
\end{eqnarray}
In what follows we will calculate the r.h.s. of this equation in the background field (\ref{aibefildz}), (\ref{corefs}) and get the coefficient function.

The double functional integral for  the r.h.s. of Eq. (\ref{master}) has the form 
\begin{eqnarray}
&&\hspace{-1mm}
{N_c^2-1\over 16}g^4\langle \tilF^a_{\mu\nu}\tilF^{a\mu\nu}(x_2)  F^b_{\lambda\rho}F^{b\lambda\rho}(x_1)-
\hacalo^{ij,\sigma_p}(x_2^-,x_{2_\perp};x_1^-,x_{1_\perp}) 
\hacalo^{ij;\sigma_t}(x_2^+,x_{2_\perp};x_1^+,x_{1_\perp})\rangle_{\matA}
\nonumber\\
&&\hspace{-0mm}
=~\!\int\! D\tilA DA D\tsi D\psi~e^{-iS_{\matA,\Psi}(\tilA,\tsi)+iS_{\matA,\Psi}(A,\psi)}
\label{funtegral1}\\
&&\hspace{-0mm}
\times~\Big(
{N_c^2-1\over 16} g^4\tilF^a_{\mu\nu}\tilF^{a\mu\nu}(x_2) F^b_{\lambda\rho}F^{b\lambda\rho}(x_1)-
 \hacalo_{ij}^{\sigma_p}(x_2^-,x_{2_\perp};x_1^-,x_{1_\perp}) 
\hacalo^{ij;\sigma_t}(x_2^+,x_{2_\perp};x_1^+,x_{1_\perp})\Big)
 \nonumber
\end{eqnarray}
where $S_{\matA,\Psi}(A,\psi)$ is a standard QCD action in a background field $\matA$ in the background-Feynman gauge:
\begin{eqnarray}
&&\hspace{-1mm}
S_{\matA,\Psi}(A,\psi)~
\\
&&\hspace{-1mm}
=~S_{\rm cl}(\matA,\Psi)+\!\int\! dz~2{\rm tr} \Big(A^\mu (\matD^2g_{\mu\nu}+2i\matG_{\mu\nu})A^\nu
+ig\matD_\mu A_\nu [A^\mu,A^\nu]+{g^2\over 2} [A^\mu,A^\nu]^2\Big)~+~...
\nonumber
\end{eqnarray}
where dots stand for quark terms which are not relevant for our calculation. 
Hereafter ``Tr'' means color trace in the  adjoint representation.
Note that the term $S_{\rm cl}(\matA,\Psi)$ cancels in the exponent in Eq. (\ref{funtegral1}) so we will ignore it in what follows.
The propagators in the background field $\matA$ can be obtained as an expansion in ``correction field'' $\barC$ since 
it is down by one power of ${m_\perp^2\over s}$ in comparison to $\barA$ and $\barB$. As to propagator in $\barA+\barB$ 
background, it can be in principle obtained as a cluster expansion, but fortunately we will need only a couple of terms 
$\sim U^{-i}$ and  $\sim V^{+i}$ which will be easily identified.

Most frequently we will perform this calculation in the momentum space, so we introduce Fourier transforms 
of projectile and target fields
\begin{eqnarray}
&&\hspace{-1mm}
V^{- i}(z^+,z_\perp)
~=~\!\int\!\dhd{\beta_b}\dhd k_{b_\perp}V^{- i}({\beta_b},k_{b_\perp})e^{-i\varrho{\beta_b}z^++i(k_a,z)_\perp},~~~~~~~
\nonumber\\
&&\hspace{-1mm} 
U^{+ i}(z^-,z_\perp)~=~\!\int\!\dhd{\alpha_a}\dhd {k}_{a_\perp}U^{+ i}({\alpha_a},{k}_{a_\perp})e^{-i\varrho{\alpha_a}z^- +i(k_a,z)_\perp},
\nonumber\\
&&\hspace{-1mm}
V^{-i}({\beta_b},k_{b_\perp})~=~\varrho\!\int\! dz^+ dz_\perp~U^{-i}(z^+,z_\perp)e^{i\varrho{\beta_b}z^+ -i(k_a,z)_\perp},  
\nonumber\\
&&\hspace{-1mm}
U^{+i}({\alpha_a},{k}_{a_\perp})~=~\varrho\!\int\! dz^- dz_\perp~U^{+i}(z^-,z_\perp)
e^{i\varrho{\alpha_a}z^- -i(k_a,z)_\perp}  
\label{normi}
\end{eqnarray}
To avoid cluttering of our formulas, throughout the paper  we use the  
$\hbar$-inspired notation
\begin{equation}
\int\!\dhd^n p~\equiv~\int\!{d^n p\over (2\pi)^n}
\end{equation}
where $n$ is the dimension of corresponding momentum space.

Thus, the object of our calculations is the Fourier transform of Eq. (\ref{master})
\begin{eqnarray}
&&\hspace{-1mm}
\frc_1(x_1,x_2;{\alpha'_a},\alpha_a,{k}_{a_\perp},{k'}_{a_\perp},{\beta'_b},\beta_b,k_{b_\perp},k'_{b_\perp};\sigma_p,\sigma_t)
\nonumber\\
&&\hspace{-1mm}
=~\int\! dz_1^-dz_2^- dw_1^+dw_2^+ dz_{1_\perp} dz_{2_\perp}dw_{1_\perp} dw_{2_\perp}
e^{-i\varrho{\alpha'_a}z_2^- +i(k'_a,z_2)_\perp}e^{-i\varrho\alpha_az_1^- +i(k_a,z_1)_\perp}
\nonumber\\
&&\hspace{5mm}
\times~
e^{-i\varrho{\beta'_b}z_2^++i(k'_b,z_2)_\perp}e^{-i\varrho\beta_bz_1^++i(k_b,z_1)_\perp}
\frc_1(x_1,x_2;z_i^-,z_{i_\perp},w_i^+,w_{i_\perp};\sigma_p,\sigma_t)
\label{ceodin}
\end{eqnarray}
For one-loop calculations in the background field $\matA$ it is convenient to multiply the hadronic tensor (\ref{W}) by
additional factor ${N_c^2-1\over 2g^2N_c}\pi^2$. We define
\begin{eqnarray}
&&\hspace{-1mm}
\pizw(x_1,x_2)~
=~{N_c^2-1\over 2g^2N_c}\pi^2 \langle W^{\rm one-loop}(x_1,x_2)\rangle_{\matA}~
\nonumber\\
&&\hspace{-1mm}=~{N_c^2-1\over N_c}8\pi^2g^2
\langle F^{- i,a}(x_2)F^{+a}_{~i}(x_2)F^{- j,b}(x_1)F^{+b}_{~j}(x_1) \rangle^{\rm one-loop}_{\matA}
\label{pizw}
\ega
The contributions to $\pizw(x_1,x_2)$ will be parametrized as follows
\begin{eqnarray}
&&\hspace{-1mm}
\pizw(x_1,x_2)-\pizw^{\sigma_p,\sigma_t}_{\rm eik}(x_1,x_2)~=~
\!\int\!\dhd{\alpha'_a}\dhd {k'}_{a_\perp}\dhd{\beta'_b}\dhd k'_{b_\perp}\dhd\alpha_a\dhd k_{a_\perp}\dhd\beta_b\dhd k_{b_\perp}
\nonumber\\
&&\hspace{-1mm}
\times~e^{-i{\alpha'_a}\vro x_2^- -i\alpha_a\vro x_1^-}e^{-i{\beta'_b}\vro x_2^+ -i\beta_b\vro x_1^+}
e^{-i(k'_a+k'_a,x_2)_\perp-i(k_a+k_b,x_1)_\perp}
\nonumber\\
&&\hspace{-1mm}
\times~U^{+,b}_{~~i}({\alpha'_a},{k'}_{a_\perp})V^{-i,a}({\beta'_b},k'_{b_\perp})
U^{+,b}_{~j}(\alpha_a,{k}_{a_\perp})V^{-j,a}(\beta_b,k_{b_\perp})
\nonumber\\
&&\hspace{-1mm}
\times~
[I-I^{\sigma_p,\sigma_t}_{\rm eik}]({\alpha'_a}, {k'}_{a_\perp},\alpha_a, k_{a_\perp},{\beta'_b}, k'_{b_\perp},\beta_b, k_{b_\perp},  x_2,x_1)
\label{calcul1}
\ega
where $\pizw^{\sigma_p,\sigma_t}_{\rm eik}(x_1,x_2)$ is the contribution of eikonal TMD operators
$\langle \hacalo^{ij,\sigma_p}\hacalo^{ij;\sigma_t}\rangle_{\matA}$ which has to be subtracted according to
Eq. (\ref{funtegral1}). The coefficient function $\frc_1$ is then
 a Fourier transform of $[I-I^{\sigma_p,\sigma_t}_{\rm eik}]$.

Recall that the kinematical region for hadronic tensor (\ref{region}) translates to Eq. (\ref{kinregion})
in the non-forward case so in our approximation longitudinal distances are smaller than the transverse ones. 
\footnote{It is worth noting that for tree-level calculations, the parameter 
${q_\perp^2\over Q^2}$ is sufficient.  
A more restrictive parameter (\ref{region}) 
 is necessary for our calculation of logarithmical corrections. If ${q_\perp^2\over Q^2}\ll 1$  but Eq. (\ref{region})  
 is not satisfied, the calculation of power corrections 
 in Refs. \cite{Balitsky:2017flc,Balitsky:2017gis} is still valid but the logarithmical corrections will probably 
 be more complicated than our result (\ref{masteresult}).} 

In order to have parameters of the Fourier transformations with ''natural'' scales resembling those of forward
case, in the formulas  (\ref{normi}) we should take the origin somewhere between $x_2$ and $x_1$, so the kinematical region where we calculate Eq. (\ref{ceodin}) is
\begin{equation}
{\alpha'_a}\sim \alpha_a, ~{\beta'_b}\sim \beta_b,~k_{a_\perp}\!\sim k'_{a_\perp}\!\sim k_{b_\perp}\sim\! k'_{b_\perp}\!\sim Q_\perp,
~~~~~\alpha_a\beta_b,\alpha'_a\beta_b,\alpha\beta'_b,\alpha'_a\beta'_b\sim {Q^2\over s}\gg {Q_\perp^2\over s}
\label{kinemregion}
\end{equation}
which corresponds to
\begin{equation}
x_2^+\sim x_1^+,~x_2^-\sim x_1^-,~x_{2_\perp}\sim x_{1_\perp},~~~~~x_2^+x_2^-\sim x_1^+ x_1^-\sim O\big({1\over \zeta}\big)\ll x_{2_\perp}^2\sim x_{1_\perp}^2\sim O(\zeta^0)
\label{coordregion}
\end{equation}
in the coordinate space because $x_i^+\sim{1\over {\beta_b}\vro}$ and $x_i^-\sim{1\over {\alpha_a}\vro}$ and 
$\rho\sim\big({1\over \sqrt\zeta}\big)$.

\section{Virtual contributions \label{sect:virtual}}
It is convenient to start calculation of functional integral (\ref{funtegral1}) from the so-called ``virtual'' contribution to the first term
given by diagrams in Fig. \ref{fig:nlovirt} a-c. (The reason for  ``production'' diagram (d) appearing on this Figure is explained in the end of this Section). 
Let us start with the diagram in Fig.  \ref{fig:nlovirt}a. 
\begin{figure}[htb]
\begin{center}
\includegraphics[width=141mm]{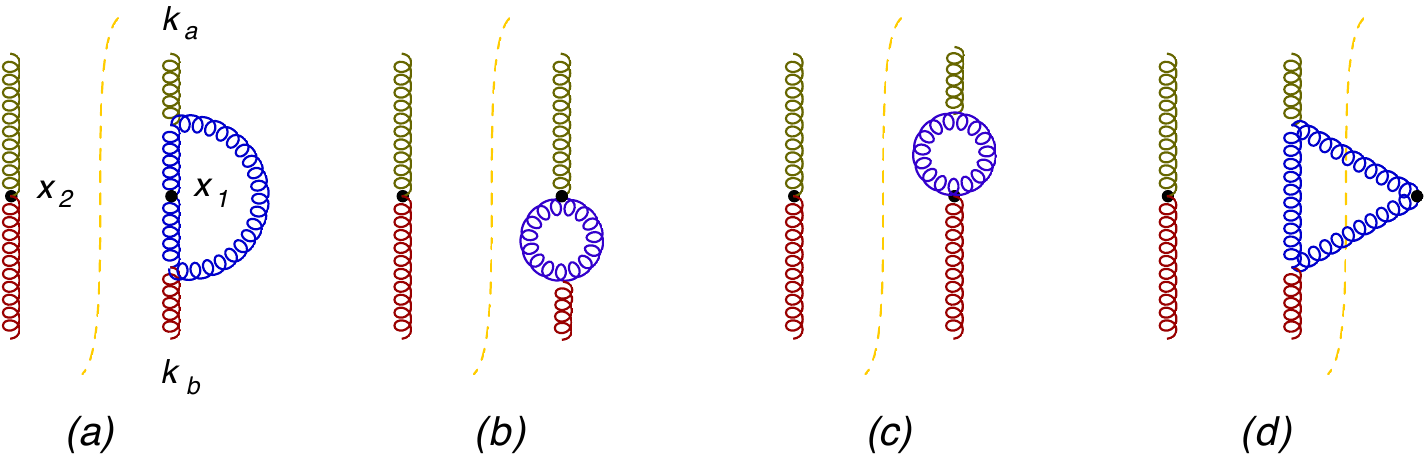}
\end{center}
\caption{Diagrams (a)-(c): virtual diagrams in the right sector.  Diagram (d): related diagram with two-gluon production.\label{fig:nlovirt}}
\end{figure}

In an arbitrary background field $\cala$ we get
$$
F_{\mu\nu}(A+\cala)~=~\calf_{\mu\nu}(\cala)+(\cald_\mu A_\nu-\cald_\nu A_\mu) -ig[A_\mu,A_\nu]
$$
The diagram in Fig.  \ref{fig:nlovirt}a corresponds to 
\begin{equation}
\langle (\cald_\mu A_\nu-\cald_\nu A_\mu)(x_1) (\cald_\mu A_\nu-\cald_\nu A_\mu)(x_1)\rangle
\end{equation}
expanded up to two $\calf_{\mu\nu}(\cala)$. Using the background-field propagator in Schwinger's notations
\begin{equation}
i\langle T\{A_\mu(x) A_\nu(y)\}\rangle_\cala~=~(x|{1\over \calp^2g_{\mu\nu}+2i\calf_{\mu\nu}+\ie}|y)
\end{equation}
and identities
\begin{eqnarray}
&&\hspace{-1mm}
\calp^\mu{1\over \calp^2}\calp_\mu~=~1-g^2{1\over P^2}\calf_{\mu\nu}\calf^{\mu\nu}{1\over P^2}
\\
&&\hspace{22mm}
-~g^2{1\over \calp^2}\calp^\eta \cald^\xi \calf_{\xi\eta}{1\over \calp^2}
+g^2{1\over \calp^2}\{\calp_\alpha, \calf^{\alpha\xi}\}{1\over \calp^2}\{\calp^\beta,\calf_{\beta\xi}\}{1\over \calp^2}
\nonumber\\
&&\hspace{-1mm}
\calp^\mu{1\over \calp^2}\calf_{\mu\nu}{1\over \calp^2}\calp^\nu
~=~{ig^2\over 2}{1\over \calp^2}\calf_{\mu\nu}\calf^{\mu\nu}{1\over \calp^2}
+ig^2{1\over \calp^2}\calp^\nu \cald^\mu \calf_{\mu\nu}{1\over \calp^2}
\nonumber\\
&&\hspace{22mm}
-ig^2{1\over \calp^2}\{\calp_\alpha, \calf^{\alpha\xi}\}{1\over \calp^2}\calf_{\beta\xi}\calp^\beta{1\over \calp^2}
-ig^2{1\over \calp^2}\calp_\alpha \calf^{\alpha\xi}{1\over \calp^2}\{\calf_{\beta\xi},\calp^\beta\}{1\over \calp^2}~+~O(\calf_{\xi\eta}^3)
\nonumber
\end{eqnarray}
one obtains after some algebra
\begin{eqnarray}
&&\hspace{-1mm}
(\calp_\mu \delta^\xi_\nu-\calp_\nu \delta^\xi_\mu){1\over \calp^2 g_{\xi\eta}+2i\calf_{\xi\eta}+\ie}(\calp_\mu \delta^\eta_\nu-\calp_\nu \delta^\eta_\mu)
~=~6-4{g^2\over \calp^2}\calf_{\mu\nu}\calf^{\mu\nu}{1\over \calp^2}
\label{virtu1}\\
&&\hspace{-1mm}
+~4\calf_{\mu\nu}{g^2\over \calp^2}\calf^{\mu\nu}{1\over \calp^2}+4{g^2\over \calp^2}\calf^{\mu\nu}{1\over \calp^2}\calf_{\mu\nu}
-2{g^2\over \calp^2}\calp^\eta \cald^\xi \calf_{\xi\eta}{1\over \calp^2}
+8{g^2\over \calp^2}\cald_\mu \calf_{\lambda\rho}{1\over \calp^2}\cald^\mu\calf^{\lambda\rho}{1\over \calp^2}
\nonumber\\
&&\hspace{-1mm}
-~2i{g^2\over \calp^2}\cald_\alpha \calf_{\beta\xi}{1\over \calp^2}\{\calp^\alpha, \calf^{\beta\xi}\}{1\over \calp^2}
+2i{g^2\over \calp^2}\{\calf_{\alpha\xi},\calp_\beta\}{1\over \calp^2}\cald^\beta \calf^{\alpha\xi}{1\over \calp^2}
~+~O\big(\calf_{\xi\eta}^3,(\cald^\xi \calf_{\xi\eta})^2\big)
\nonumber\\
&&\hspace{-1mm}
=~6
+2\calf_{\mu\nu}{g^2\over \calp^2}\calf^{\mu\nu}{1\over \calp^2}+2{g^2\over \calp^2}\calf^{\mu\nu}{1\over \calp^2}\calf_{\mu\nu}
-2{g^2\over \calp^2}\calp^\eta \cald^\xi \calf_{\xi\eta}{1\over \calp^2}
+4{g^2\over \calp^2}\cald_\mu \calf_{\lambda\rho}{1\over \calp^2}\cald^\mu\calf^{\lambda\rho}{1\over \calp^2}
\nonumber\\
&&\hspace{-1mm}
~+~O\big(\calf_{\xi\eta}^3,(\cald^\xi \calf_{\xi\eta})^2\big)
\nonumber
\end{eqnarray}
(here all $\calp^2$ are  $\calp^2+\ie$).

Next, it is convenient to add to Eq. (\ref{virtu1}) the contribution of diagrams in Fig.  \ref{fig:nlovirt}b,c which has the form
\begin{eqnarray}
&&\hspace{-2mm}
2\langle igf^{abc}A^b_\mu A^c_\nu(x)\calf^{c\mu\nu}(x)\rangle
~=~-4ig^2f^{abc}(x|{1\over \calp^2}G_{\mu\nu}{1\over \calp^2}|x)^{ab}G^{c\mu\nu}(x)~+~O\big(\calf_{\xi\eta}^3,(\cald^\xi \calf_{\xi\eta})^2\big)
\nonumber\\
&&\hspace{43mm}
=~(x|-4{g^2\over \calp^2}\calf_{\mu\nu}{1\over \calp^2}\calf^{\mu\nu}|x)^{aa}
~+~O\big(\calf_{\xi\eta}^3,(\cald^\xi \calf_{\xi\eta})^2\big)
\label{dvaakv}
\end{eqnarray}
We get 
\begin{eqnarray}
&&\hspace{-1mm}
2g^2\langle i(\cald_\mu A^a_\nu-\cald_\nu A^a_\mu)(x_1) \cald^\mu A^{a,\nu}(x_1)+igf^{abc}A^b_\mu A^c_\nu(x_1)\calf^{c\mu\nu}(x_1)\rangle
\nonumber\\
&&\hspace{-1mm}
=~g^4(x_1|-2{1\over \calp^2}\calp^\eta \cald^\xi \calf_{\xi\eta}{1\over \calp^2}
+4{1\over \calp^2}\cald_\mu \calf_{\lambda\rho}{1\over \calp^2}\cald^\mu\calf^{\lambda\rho}{1\over \calp^2}|x_1)^{aa}
~+~O\big(\calf_{\xi\eta}^3,(\cald^\xi \calf_{\xi\eta})^2\big)
\end{eqnarray}
Note that the absence of UV divergent terms follows from one-loop renorm-invariance of local operator $g^2F^a_{\mu\nu}F^{a\mu\nu}$.

Next, the expansion of propagators $(p^2+2\{p,\cala\}+\cala^2)^{-1}$ in powers of external field $\cala$ 
 will bring more powers of $\calf_{\xi\eta}$ in the numerators leading to power 
 corrections to Eq. (\ref{facoord1}) rather than the logarithmical ones. Thus, one can replace all $(\calp^2+\ie)^{-1}$ 
in the r.h.s. of Eq. (\ref{virtu1}) by usual propagators $(p^2+\ie)^{-1}$. Also, for our background field $\matA$ 
\begin{equation}
\hspace{-1mm}
\calf_{\mu\nu}(\matA)~=~\calf_{\mu\nu}(\barA)+\calf_{\mu\nu}(\barB)+\calf_{\mu\nu}(\barC)
-ig([\barA_\mu,\barB_\nu]+[\barA_\mu+\barB_\mu,\barC_\nu]-\mu\leftrightarrow\nu)
\end{equation}
Since the correction field $\barC$ is proportional to commutators of $U^{-i}$ and $V^{+j}$ we can neglect it in the r.h.s. of 
Eq. (\ref{virtu1}) - it will lead to the terms with two $U^{-i}$ and one $V^{+j}$, or vice versa. By the same token,
one can disregard $[\barA_\mu,\barB_\nu]$ since its Lorentz components are either zero or made from $[U^{-i},V^{+j}]$.
Finally, we are interested only in terms with one $U^{-i}$ and one $V^{+j}$ in the r.h.s. of Eq. (\ref{virtu1}) 
corresponding to diagrams in Fig. \ref{fig:nlovirt}
\footnote{The terms with zero $U^{-i}$'s and two $V^{+j}$'s will lead
to zero color trace after integration over projectile fields in the integral (\ref{W2}), and similarly for terms with $U^{+j}$'s 
}

Thus, in our approximation
\begin{eqnarray}
&&\hspace{-1mm}
2g^2\langle (\cald_\mu A^a_\nu-\cald_\nu A^a_\mu)(x_1) \cald^\mu A^{a,\nu}(x_1)+igf^{abc}A^b_\mu A^c_\nu(x_1)\calf^{c\mu\nu}(x_1)\rangle_{\matA}
\label{virt1}\\
&&\hspace{-1mm}
=~-16i(x_1|{1\over p^2+\ie}\barD_\mu U^{+i}{1\over p^2+\ie}\barD^\mu V^-_{~i}{1\over p^2+\ie}|x_1)^{aa}
\nonumber
\end{eqnarray}
This term is proportional  to 
$\barD_\mu U^{+i}\otimes \barD^\mu V^-_{~i}=\partial^+U^{+i}\otimes\partial^-V^-_{~i}+\barD_k U^{-i}\otimes \barD^k V^+_{~i}$.
As we discussed after Eq. (\ref{pc}), $\partial^+\otimes\partial^-$ brings extra factor of $s$  and $\barD_k\otimes\barD^k$ only $q_\perp^2$ 
so we can disregard it. We get
\begin{eqnarray}
&&\hspace{-2mm}
-16i(x_1|{1\over p^2+\ie}\partial^- U^{+i}{1\over p^2+\ie}\partial^+ V^-_{~i}{1\over p^2+\ie}|x_1)^{aa}
\nonumber\\
&&\hspace{-2mm}
=~-8N_c\!\int\!\dhd{\alpha_a}\dhd {k}_{a_\perp}\!\int\!\dhd{\beta_b}\dhd k_{b_\perp}
U^{+ i,a}({\alpha_a},{k}_{a_\perp})V^{-,a}_{~i}({\beta_b},k_{b_\perp})e^{-i\vro{\alpha_a}x_1^- -i\vro{\beta_b}x_1^++i(k_a+k_a,x_1)_\perp}
\nonumber\\
&&\hspace{22mm} 
\times~\int\! {\dhd^4p\over i}~{s{\alpha_a}{\beta_b}\over [(p+k_a)^2+\ie](p^2+\ie)[(p-k_a)^2+\ie]}
\nonumber\\
&&\hspace{-2mm}
=~-{g^2N_c\over 2\pi^2}\!\int\!\dhd{\alpha_a}\dhd {k}_{a_\perp}\!\int\!\dhd{\beta_b}\dhd k_{b_\perp}
U^{+ i,a}({\alpha_a},{k}_{a_\perp})V^{-,a}_{~i}({\beta_b},k_{b_\perp})e^{-i\vro{\alpha_a}x_1^- -i\vro{\beta_b}x_1^++i(k_a+k_a,x_1)_\perp}
\nonumber\\
&&\hspace{22mm} 
\times~\Big(\ln{-{\alpha_a}{\beta_b}s-\ie\over {k}_{a_\perp}^2}\ln{-{\alpha_a}{\beta_b}s-\ie\over k_{b_\perp}^2}+{\pi^2\over 3}\Big)
\label{dudv}
\end{eqnarray}
where we used Eq. (\ref{integrali}) in the last line.
Finally,  we obtain
\begin{eqnarray}
&&\hspace{-1mm}
g^2\langle T\{F_{\mu\nu}^a(x_1)F^{\mu\nu,a}(x_1)\}\rangle_{\rm Fig. \ref{fig:nlovirt}a+b+c}
\nonumber\\
&&\hspace{-1mm}
=~-\!\int\!\dhd{\alpha_a}\dhd {k}_{a_\perp}\!\int\!\dhd{\beta_b}\dhd k_{b_\perp}
U^{+i,a}({\alpha_a},{k}_{a_\perp})V^{-,a}_{~i}({\beta_b},k_{b_\perp})e^{-i\vro{\alpha_a}x_1^- -i\vro{\beta_b}x_1^++i(k_a+k_a,x_1)_\perp}
\nonumber\\
&&\hspace{22mm}
\times~{g^2N_c\over 2\pi^2}
\Big(\ln{-{\alpha_a}{\beta_b}s-\ie\over {k}_{a_\perp}^2}\ln{-{\alpha_a}{\beta_b}s-\ie\over k_{b_\perp}^2}+{\pi^2\over 3}\Big)
\label{figa4abc}
\end{eqnarray}

Next, let us consider the diagram in Fig. \ref{fig:nlovirt}d. This diagram and its left-right permutation  
are the only diagrams with two-gluon cut since  gluon with $k_a$ or $k_b$ cannot solely produce two gluons. As
we will see below, the diagram in Fig. \ref{fig:nlovirt}d  will change Feynman-type singularities in logarithms in Eq. (\ref{figa4abc}) 
to causal-type singularities.

The structure of the diagram in  Fig. \ref{fig:nlovirt}d is the same as in Fig. \ref{fig:nlovirt}a with two Feynman propagators  replaced by cut propagators 
and the left-sector propagator between $U^{-i}$ and $V^{+i}$ being ${i\over p^2-\ie}$ instead of ${-i\over p^2+\ie}$.  Note also that 
the first three terms in the r.h.s. of Eq.  (\ref{virtu1}) do not contribute since our background fields cannot produce particles. We get
\begin{eqnarray}
&&\hspace{-1mm}
g^2\langle T\{F_{\mu\nu}^a(x_1)F^{a,\mu\nu}(x_1)\}\rangle_{\rm Fig. \ref{fig:nlovirt}d}
\label{withcut1}\\
&&\hspace{-1mm}
=~-16i(x_1|
\tilde{\delta}_-(p)\partial^- U^{+i}{1\over p^2-\ie}\partial^+ V^-_{~i}\tilde\delta_+(p)|x_1)^{aa}
\nonumber
\end{eqnarray}
Hereafter we introduce  space-saving notations
\begin{equation}
\tilde\delta_+(p)~\equiv~2\pi\delta(p^2)\theta(p_0),~~~~~~~~\tilde{\delta}_-(p)~\equiv~2\pi\delta(p^2)\theta(-p_0)
\hspace{-1mm}
\end{equation}
Using integral (\ref{imtegral1}) from Appendix \ref{sect:integrals}  one easily obtains
\begin{eqnarray}
&&\hspace{-1mm}
g^2\langle T\{F_{\mu\nu}^a(x_1)F^{\mu\nu,a}(x_1)\}\rangle_{\rm Fig. \ref{fig:nlovirt}d}
\nonumber\\
&&\hspace{-1mm}
=~\!\int\!\dhd{\alpha_a}\dhd {k}_{a_\perp}\!\int\!\dhd{\beta_b}\dhd k_{b_\perp}
U^{+i}({\alpha_a},{k}_{a_\perp})V^-_{~i}({\beta_b},k_{b_\perp})e^{-i\vro{\alpha_a}x_1^- -i\vro{\beta_b}x_1^++i(k_a+k_a,x_1)_\perp}
\nonumber\\
&&\hspace{22mm}
\times~\theta(-{\alpha_a})\theta(-{\beta_b}){g^2N_c\over \pi}(-i)\ln{{\alpha_a}^2{\beta_b}^2s^2\over {k}_{a_\perp}^2k_{b_\perp}^2}
\label{figa4d}
\end{eqnarray}
Next, using the identity
\beq
\ln{-{\alpha_a}{\beta_b}s-\ie\over {k}_{a_\perp}^2}\ln{-{\alpha_a}{\beta_b}s-\ie\over k_{b_\perp}^2}
+2\pi i\theta(-{\alpha_a})\theta(-{\beta_b})\ln{{\alpha_a}^2{\beta_b}^2s^2\over {k}_{a_\perp}^2k_{b_\perp}^2}~=~
\ln{-Q_{ab}^2\over {k}_{a_\perp}^2}\ln{-Q_{ab}^2\over k_{b_\perp}^2}
\eeq
where 
\footnote{Later we will use similar notations $Q_{ab'}^2\equiv(\alpha_a+\ie)({\beta'_b}+\ie)s$, 
 $Q_{a'b}^2\equiv({\alpha'_a}+\ie)(\beta_b+\ie)s$, and $Q_{a'b'}^2\equiv(\alpha'_a+\ie)(\beta'_b+\ie)s$}
\beq
Q_{ab}^2\equiv({\alpha_a}+\ie)({\beta_b}+\ie)s
\label{kuab}
\eeq
we get the contribution of diagrams in Fig. \ref{fig:nlovirt}  in the form
\begin{eqnarray}
&&\hspace{-1mm}
g^2\langle T\{F_{\mu\nu}^a(x_1)F^{\mu\nu,a}(x_1)\}\rangle^{\rm Fig. \ref{fig:nlovirt}}
~=~{g^2N_c\over 2\pi^2}\!\int\!\dhd\alpha_a\dhd {k}_{a_\perp}\!\int\!\dhd\beta_b\dhd k_{b_\perp}
\label{virtfig5}\\
&&\hspace{-1mm} 
\times~U^{+ i,a}(\alpha_a,{k}_{a_\perp})V^{-,a}_{~i}(\beta_b,k_{b_\perp})
e^{-i\vro\alpha_ax_1^- -i\vro\beta_bx_1^++i(k_a+k_b,x_1)_\perp}
I^{\rm virt}_{\rm Fig. \ref{fig:nlovirt}}(\alpha_a,{k}_{a_\perp},\beta_b,k_{b_\perp})
\nonumber
\ega
where
\begin{eqnarray}
&&\hspace{-1mm}
I^{\rm virt}_{\rm Fig. \ref{fig:nlovirt}}(\alpha_a,{k}_{a_\perp},\beta_b,k_{b_\perp})~=~-16\pi^2\int\! {\dhd^4p\over i}~\Big[{s{\alpha_a}{\beta_b}\over [(p+k_a)^2+\ie](p^2+\ie)[(p-k_b)^2+\ie]}
\nonumber\\
&&\hspace{-1mm} 
+~\tilde{\delta}_-(p+k_a){s{\alpha_a}{\beta_b}\over p^2-\ie}\tilde{\delta}_+(p-k_b)\Big]
~=~-\ln{-Q_{ab}^2\over {k}_{a_\perp}^2}\ln{-Q_{ab}^2\over {k}_{b_\perp}^2}-{\pi^2\over 3}~+~O(\lambda)
\label{ivirt5}
\ega
and $\lambda$ is defined in Eq. (\ref{lambdas}). 

In coordinate space, the singularity (\ref{kuab}) means that
\begin{eqnarray}
&&\hspace{1mm}
\int\! \dhd{\alpha'_a} ~e^{-i\varrho{\alpha'_a} x_1^-}f(-{\alpha'_a}-\ie)U({\alpha'_a})~=~\int_{-\infty}^{x_1^-}\! dz_1^-~\tilde{f}(x_1^- -z_1^-)U(z_1^-)
\nonumber\\
&&\hspace{1mm}
\int\! \dhd{\beta'_b} ~e^{-i\varrho{\beta'_b} x_1^+}f(-{\beta'_b}-\ie)V({\beta'_b})
~=~\int_{-\infty}^{x_1^+}\! dz_1^+~\tilde{f}(x_1^+ -z_1^+)V(z_1^-)
\label{causality}
\end{eqnarray}
Thus, after summation of the diagrams  Fig. \ref{fig:nlovirt}a,b,c and  Fig. \ref{fig:nlovirt}d we get the result that the emission of the background-field gluon always preceeds the original point $x_1$.

Actually, it can be seen before the calculation of integrals. To this end, consider the identity
\bega
&&\hspace{-1mm}
(x|{1\over p^2+\ie}\cala{1\over p^2+\ie}\calb{1\over p^2+\ie}
+~\tilde{\delta}_-(p)\cala{1\over p^2-\ie}\calb\tilde\delta_+(p)|y)
+\tilde{\delta}_-(p)\cala\tilde\delta_+(p)\calb{1\over p^2+\ie}
\nonumber\\
&&\hspace{11mm}
+~{1\over p^2+\ie}\cala\tilde{\delta}_-(p)\calb\tilde\delta_+(p)
~=~(x|
{1\over p^2+\ie p_0}\cala{1\over p^2+\ie}\calb{1\over p^2-\ie p_0}
\nonumber\\
&&\hspace{22mm}
-i{1\over p^2+\ie p_0}\cala{1\over p^2+\ie p_0}\calb\tilde\delta_+(p)
-i\tilde{\delta}_-(p)\cala{1\over p^2-\ie p_0}\calb{1\over p^2-\ie p_0}|y)
\ega
valid for any operators $\cala$ and $\calb$. Using this formula,
 the sum of $\partial^- U^{+i}\otimes \partial^+ V^-_{~i}$ terms in Eqs. (\ref{virt1}) and (\ref{withcut1}) can be rewritten as
\begin{eqnarray}
&&\hspace{-1mm}
{1\over p^2+\ie}\partial^- U^{+i}{1\over p^2+\ie}\partial^+ V^-_{~i}{1\over p^2+\ie}
+\tilde{\delta}_-(p)\partial^- U^{+i}{1\over p^2-\ie}\partial^+ V^-_{~i}\tilde\delta_+(p)
\nonumber\\
&&\hspace{-1mm}
=~{1\over p^2+\ie p_0}\partial^- U^{+i}{1\over p^2+\ie}\partial^+ V^-_{~i}{1\over p^2-\ie p_0}
\label{virtsum}\\
&&\hspace{-1mm}
-i\tilde{\delta}_-(p)\partial^- U^{+i}
{1\over p^2-\ie p_0}\partial^+ V^-_{~i}{1\over p^2-\ie p_0}
-i{1\over p^2+\ie p_0}\partial^- U^{+i}{1\over p^2+\ie p_0}\partial^+ V^-_{~i}\tilde\delta_+(p)
\nonumber
\end{eqnarray}
from where the causal structure of the result of summation is evident. Note that we used 
$\tilde{\delta}_-(p)\partial^- U^{+i}\tilde\delta_+(p)=\tilde{\delta}_-(p)\partial^- V^{-i}\tilde\delta_+(p)=0$ following from the fact 
that background fields $U^{+i}(x^-,x_\perp)$ and $V^{-i}(x^-,x_\perp)$ cannot produce two real particles.
This is similar to the case of  tree-level diagrams where 
the summation of the emissions from both sides of the cut leads to the diagrams with retarded propagators, 
see Ref.  \cite{Balitsky:2017flc}.

One can calculate diagrams in Fig.  \ref{fig:nlovirtleft}
\begin{figure}[htb]
\begin{center}
\includegraphics[width=141mm]{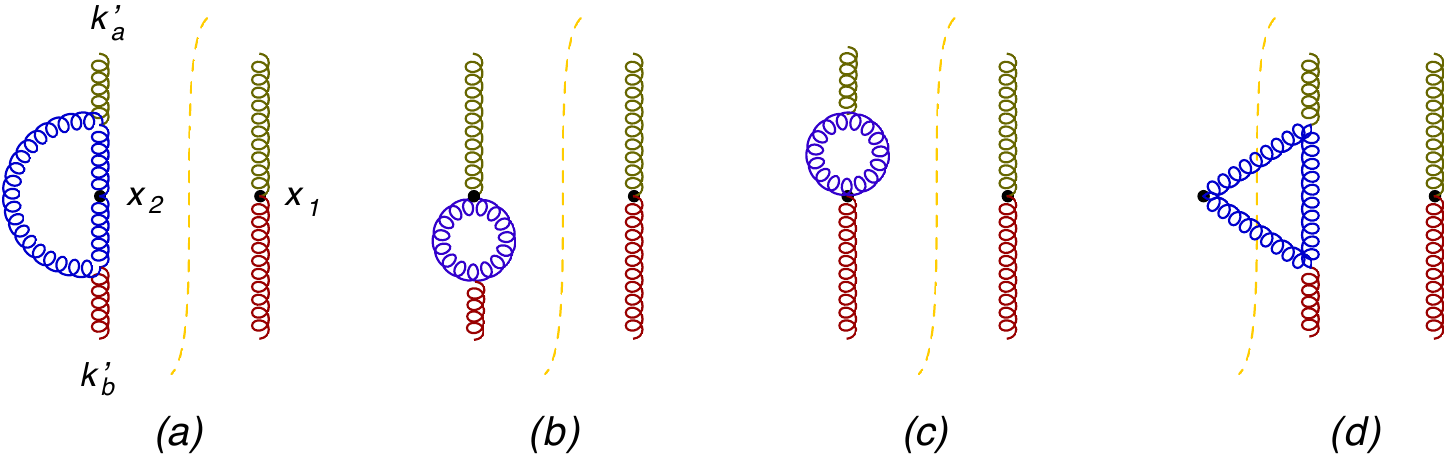}
\end{center}
\caption{Diagrams (a)-(c): virtual diagrams in the left sector.  Diagram (d): related diagram with two-gulon production\label{fig:nlovirtleft}}
\end{figure}
in a similar way. The result for the diagrams in Fig. \ref{fig:nlovirtleft} a,b,c is obtained by complex conjugation of Eq. (\ref{figa4abc})
\begin{eqnarray}
&&\hspace{-1mm}
g^2\langle \tilT\{F_{\mu\nu}^a(x_2)F^{\mu\nu,a}(x_2)\}\rangle_{\rm Fig. \ref{fig:nlovirtleft}a+b+c}
\nonumber\\
&&\hspace{-1mm}
=~-\!\int\!\dhd{\alpha'_a}\dhd {k'}_{a_\perp}\!\int\!\dhd{\beta'_b}\dhd k'_{b_\perp}
U^{+i,a}({\alpha'_a},{k'}_{a_\perp})V^{-,a}_{~i}({\beta'_b},k'_{b_\perp})e^{-i\vro{\alpha'_a}x_2^- -i\vro{\beta'_b}x_2^+ +i(k'_a+k'_b,x_2)_\perp}
\nonumber\\
&&\hspace{22mm}
\times~{g^2N_c\over 2\pi^2}\ln{-{\alpha'_a}{\beta'_b}s+\ie\over {k'}_{a_\perp}^2}\ln{-{\alpha'_a}{\beta'_b}s+\ie\over k_{b_\perp}^2}
\label{figa5abc}
\end{eqnarray}
Next, the diagram in Fig. \ref{fig:nlovirtleft}d
\begin{equation}
\hspace{-0mm}
g^2\langle \tilT\{F_{\mu\nu}^a(x_2)F^{a,\mu\nu}(x_2)\}\rangle_{\rm Fig. \ref{fig:nlovirtleft}d}
~
=~-16i(x_2|\tilde\delta_+(p)\partial^- U^{+i}{1\over p^2+\ie}\partial^+ V^-_{~i}\tilde{\delta}_-(p)|x_2)
\label{withcut2}
\end{equation}
can be obtained from the integrals in Eq.  (\ref{integrali})  by the 
replacements $k_a\rightarrow -k'_a,~k_b\rightarrow -k'_b$.
The result is
\begin{eqnarray}
&&\hspace{-1mm}
g^2\langle \tilT\{F_{\mu\nu}^a(x_2)F^{\mu\nu,a}(x_2)\}\rangle_{\rm Fig. \ref{fig:nlovirtleft}d}
\nonumber\\
&&\hspace{-1mm}
=~\!\int\!\dhd{\alpha'_a}\dhd {k'}_{a_\perp}\!\int\!\dhd{\beta'_b}\dhd k'_{b_\perp}
U^{+i}({\alpha'_a},{k'}_{a_\perp})V^-_{~i}({\beta'_b},k'_{b_\perp})e^{-i\vro{\alpha'_a}x_2^- -i\vro{\beta'_b}x_2^++i(k'_a+k'_b,x_2)_\perp}
\nonumber\\
&&\hspace{22mm}
\times~\theta({\alpha'_a})\theta({\beta'_b}){g^2N_c\over \pi}i\ln{{\alpha'_a}^2{\beta'_b}^2s^2\over {k'}_{a_\perp}^2k_{b_\perp}^2}
\label{figa5d}
\end{eqnarray}
Using now
\beq
\ln{-{\alpha'_a}{\beta'_b}s+\ie\over {k'}_{a_\perp}^2}\ln{-{\alpha'_a}{\beta'_b}s+\ie\over {k'}_{b_\perp}^2}
-2\pi i\theta({\alpha'_a})\theta({\beta'_b})\ln{{\alpha'_a}^2{\beta'_b}^2s^2\over {k'}_{a_\perp}^2{k'}_{b_\perp}^2}~=~
\ln{-Q_{a'b'}^2\over {k'}_{a_\perp}^2}\ln{-Q_{a'b'}^2\over {k'}_{b_\perp}^2}
\eeq
we get
\begin{eqnarray}
&&\hspace{-1mm}
g^2\langle \tilT\{F_{\mu\nu}^a(x_2)F^{\mu\nu,a}(x_2)\}\rangle^{\rm Fig. \ref{fig:nlovirtleft}}
~=~{g^2N_c\over 2\pi^2}\!\int\!\dhd{\alpha'_a}\dhd {k'}_{a_\perp}\!\int\!\dhd{\beta'_b}\dhd k'_{b_\perp}
\nonumber\\
&&\hspace{-1mm} 
U^{+ i,a}({\alpha'_a},{k'}_{a_\perp})V^{-,a}_{~i}({\beta'_b},k'_{b_\perp})e^{-i\vro{\alpha'_a}x_2^- -i\vro{\beta'_b}x_2^++i(k'_a+k'_b,x_2)_\perp}
I^{\rm virt}_{\rm Fig. \ref{fig:nlovirtleft}}
\ega
where
\begin{eqnarray}
&&\hspace{-1mm} 
I^{\rm virt}_{\rm Fig. \ref{fig:nlovirtleft}}
~=~16\pi^2\int\! {\dhd^4p\over i}~\Big[{s{\alpha'_a}{\beta'_b}\over [(p+k'_a)^2-\ie](p^2-\ie)[(p-k'_b)^2-\ie]}
\nonumber\\
&&\hspace{11mm} 
+~\tilde{\delta}_+(p+k'_a){s{\alpha'_a}{\beta'_b}\over p^2+\ie}\tilde{\delta}_-(p-k'_b)\Big]
~=~-\ln{-Q_{a'b'}^2\over {k'}_{a_\perp}^2}\ln{-Q_{a'b'}^2\over {k'}_{b_\perp}^2}-{\pi^2\over 3}
\label{ivirt6}
\ega
and $Q_{ab}^2$ is defined in Eq. (\ref{kuab}).
Again,  the sum of all diagrams in Fig. \ref{fig:nlovirtleft} reveals causal structure in the coordinate space.

 The final result for the virtual contributions can be presented as follows
\bega
&&\hspace{-1mm}
\pizw^{\rm virt}(x_1,x_2)~=~~{N_c^2-1\over N_c}8\pi^2\Big(
V^{- i,a}(x_2)U^{+a}_{~i}(x_2)\langle F^{- j,b}(x_1)F^{+b}_{~j}(x_1) \rangle_{\matA}^{\rm Fig. \ref{fig:nlovirt}}
\nonumber\\
&&\hspace{22mm}
+~\langle F^{- i,b}(x_2)F^{+b}_{~i}(x_2) \rangle_{\matA}^{\rm Fig. \ref{fig:nlovirtleft}}V^{- j,a}(x_1)U^{+a}_{~j}(x_1)\Big)
\nonumber\\
&&\hspace{-1mm}
=~
\!\int\!\dhd{\alpha'_a}\dhd {k'}_{a_\perp}\dhd{\beta'_b}\dhd k'_{b_\perp}\dhd\alpha_a\dhd k_{a_\perp}\dhd\beta_b\dhd k_{b_\perp}
e^{-i{\alpha'_a}\vro x_2^- -i\alpha_a\vro x_1^-}e^{-i{\beta'_b}\vro x_2^+ -i\beta_b\vro x_1^+}
\nonumber\\
&&\hspace{-1mm}
\times~
e^{-i(k_a+k_a,x_1)_\perp-i(k'_a+k'_b,x_2)_\perp}~
U^{+,b}_{~~i}({\alpha'_a},{k'}_{a_\perp})V^{-i,a}({\beta'_b},k'_{b_\perp})
U^{+,b}_{~j}(\alpha_a,k_{a_\perp})V^{-j,a}(\beta_b,k_{b_\perp})
\nonumber\\
&&\hspace{11mm}
\times~I^{\rm virt}({\alpha'_a}, {k'}_{a_\perp},{\beta'_b}, k'_{b_\perp},\alpha_a, k_{a_\perp},\beta_b, k_{b_\perp})
~+~O(\lambda)
\label{virtotvet}
\ega
where
\bega
&&\hspace{-1mm}
I^{\rm virt}({\alpha'_a}, k'_{a_\perp},{\beta'_b}, {k}_{b_\perp},\alpha_a, k'_{a_\perp},{\beta'_b}, {k}'_{b_\perp})
~=~I^{\rm virt}_{\rm Fig. \ref{fig:nlovirt}}(\alpha_a, k_{a_\perp},\beta_b, k_{b_\perp})
\label{ivirt}\\
&&\hspace{-1mm}
+~I^{\rm virt}_{\rm Fig. \ref{fig:nlovirtleft}}({\alpha'_a}, k'_{a_\perp},{\beta'_b}, k'_{b_\perp})
~=~-I^{\rm d.log}({\alpha'_a}, k'_{a_\perp},{\beta'_b}, k'_{b_\perp})
-I^{\rm d.log}(\alpha_a, k'_{a_\perp},\beta_b, k'_{b_\perp})
\nonumber
\ega
For future use,  we introduced the notation
\beq
I^{\rm d.log}({\alpha_a}, k_{a_\perp},{\beta_b}, k_{b_\perp})~=~\ln{-Q_{ab}^2\over k_{a_\perp}^2}\ln{-Q_{ab}^2\over k_{b_\perp}^2}+{\pi^2\over 3}
\label{dlog}
\eeq
for the double-log contributions. The expression for $I^{\rm d.log}(\alpha'_a, k'_{a_\perp},\beta'_b, k'_{b_\perp})$ is similar.

\section{``Production'' diagrams \label{sect:real}}
\subsection{Power counting for production terms}

From power counting (\ref{pc}) it is easy to see that the leading contribution to hadronic tensor (\ref{pizw}) 
with one-gluon production 
comes from the following terms 
\begin{eqnarray}
&&\hspace{-1mm}
\pizw(x_1,x_2)~\equiv~{N_c^2-1\over 2N_c}\pi^2g^2
\matF^a_{\mu\nu}(x_2)\langle F^{\mu\nu,a}(x_2)F^{\alpha\beta,b}(x_1)\rangle_\matA \matF^b_{\alpha\beta}(x_1)
~=~{N_c^2-1\over N_c}8\pi^2
\label{realeadingw}\\
&&\hspace{-1mm}
\times~\big[
V^{-i,a}(x_2)\langle F^{+,a}_{~i}(x_2)F^{-,b}_{~j}(x_1)\rangle_\matA U^{+j,b}(x_1)+
U^{+i,a}(x_2)\langle F^{-,a}_{~i}(x_2)F^{+,b}_{~j}(x_1)\rangle_\matA V^{-j,b}(x_1)
\nonumber\\
&&\hspace{-1mm}
+~
U^{+i,a}(x_2)\langle F^{-,a}_{~i}(x_2)F^{-,b}_{~j}(x_1)\rangle_\matA U^{+j,b}(x_1)
+V^{-i,a}(x_2)\langle F^{+,a}_{~i}(x_2)F^{+,b}_{~j}(x_1)\rangle_\matA V^{-j,b}(x_1)\big]
\nonumber
\end{eqnarray}
In this Section we calculate the first term in this equation. The second term is obtained by trivial replacements
while the third and the fourth terms correspond to ``handbag'' diagrams considered in next Section.

The gluon propagator in the background field $\matA$ is given by Eq. (\ref{gluprop2}) from Appendix \ref{sect:gluprop}.
Also, in Appendix \ref{sect:corfildz} it was proved that the contributions due to background ``correction field'' $\barC$ can 
be neglected so we need to compute
\begin{eqnarray}
&&\hspace{-1mm}
V^{a,-i}(x_2)\langle (\cald^+ A_i-\cald_i A^+)(x_2)(\cald^- A_j-\cald_j A^-)(x_1)\rangle^{ab} U^{b,+j}(x_1)~=~
\nonumber\\
&&\hspace{-1mm}
=~-V^{a,-i}i(x_2|(\calp^+\delta_i^\alpha-\calp_ig^{+\alpha})
{1\over \calp^2g_{\alpha\xi}+2i\calf_{\alpha\xi}-\ie}p^2
\nonumber\\
&&\hspace{11mm}
\times~\tilde\delta_+(p)p^2{1\over \calp^2\delta^\xi_{\beta}+2i\calf^\xi_{~\beta}+\ie}
(\calp^-\delta^\beta_j-\calp_jg^{\beta -})|x_1)^{ab}U^{b,+j}
\nonumber\\
&&\hspace{-1mm}
=~-V^{a,-i}(x_2|(p^+\delta_i^\alpha-\calp_ig^{+\alpha})\Big\{g_{\alpha\beta}{1\over \calp^2-\ie}p^2\tilde\delta_+(p)p^2{1\over \calp^2+\ie}
\nonumber\\
&&\hspace{-1mm}
-~2i{1\over \calp^2-\ie}\Big[p^2\tilde\delta_+(p)p^2{1\over \calp^2+\ie}\calf_{\alpha\beta}
+\calf_{\alpha\beta}{1\over \calp^2-\ie}p^2\tilde\delta_+(p)p^2\Big]{1\over \calp^2+\ie}|y)
\nonumber\\
&&\hspace{-1mm}
+~4{1\over \calp^2-\ie}
\Big[\calf_{\alpha\xi}{1\over \calp^2-\ie}\calf_\beta^{~\xi}{1\over \calp^2-\ie}p^2\tilde\delta_+(p)p^2
+\calf_{\alpha\xi}{1\over \calp^2-\ie}p^2\tilde\delta_+(p)p^2{1\over \calp^2+\ie}\calf_\beta^{~\xi}
\nonumber\\
&&\hspace{-1mm}
+~p^2\tilde\delta_+(p)p^2{1\over \calp^2+\ie}\calf_{\alpha\xi}{1\over \calp^2+\ie}\calf_\beta^{~\xi}\Big]{1\over \calp^2+\ie}
\Big\}(p^-\delta^\beta_j-\calp_jg^{\beta -})|x_1)^{ab}U^{b,+j}(x_1)
\label{fla73}
\end{eqnarray}

The leading contribution, shown in Fig. \ref{fig:real1},
\begin{eqnarray}
&&\hspace{-1mm}
-4V^{-a,i}(x_2)(x_2|{p^+\over p^2-\ie}U^{+i}{1\over p^2-\ie}V^{-j}p^-\tilde\delta_+(p)
+{p^+\over p^2-\ie}U^{+i}\tilde\delta_+(p)V^{-j}{p^-\over p^2-\ie}
\nonumber\\
&&\hspace{-1mm}
+~p^+\tilde\delta_+(p)U^{+i}{1\over p^2+\ie}V^{-j}{p^-\over p^2-\ie}|x_1)^{ab}U^{b,+j}(x_1)
\label{fla74}
\end{eqnarray}
comes from the last term in the r.h.s. of Eq. (\ref{fla73}).
\begin{figure}[htb]
\begin{center}
\includegraphics[width=133mm]{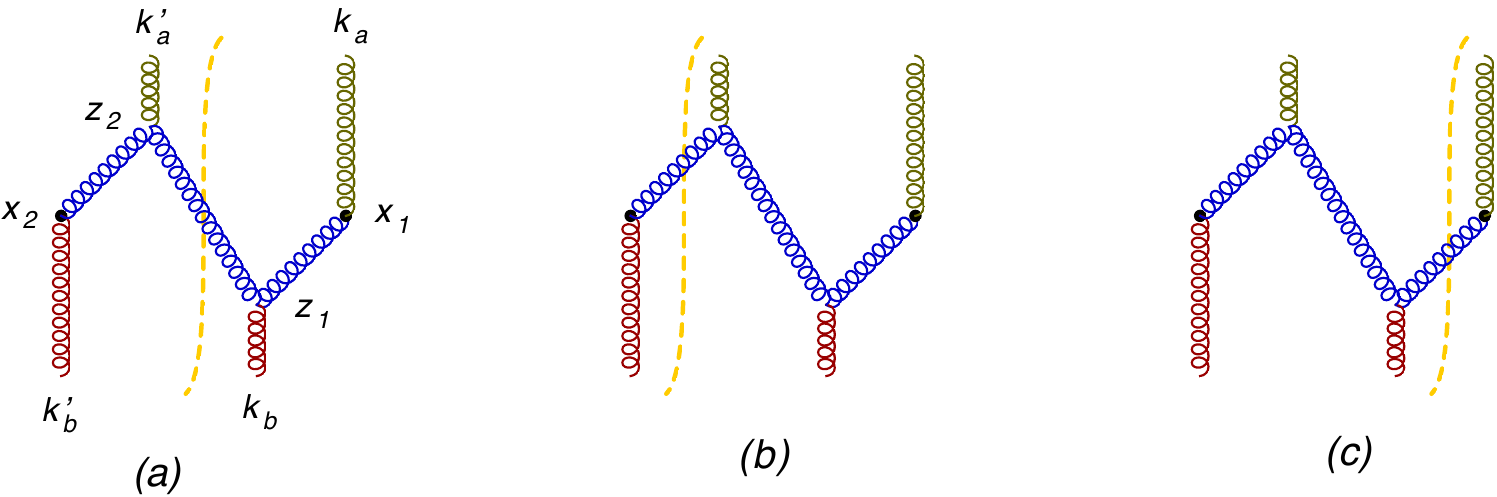}
\end{center}
\caption{First set of leading diagrams with gluon production.  Projectile fields $U^{+i}(z_2)$ are denoted by green
tails while target fields $V^{-i}(z_1)$ by red tails. \label{fig:real1}}
\end{figure}
As we will see in the next Section, this contribution is logarithmic
similarly to Eq. (\ref{dudv}) for virtual diagram.

Next, using power counting (\ref{pc}), we  demonstrate that all other contributions to the r.h.s. of 
Eq. (\ref{fla73}) are power corrections with respect to Eq. (\ref{fla74}). 
To this end, we note that  $p^+p^-\sim {\alpha'_a}\beta_bs=Q_{a'b}^2$ and $U^{+i}V^{-j}\sim Q_\perp^2s$ so the numerator 
$p^+p^-U^{+i}V^{-j}$ in the integral of Eq. (\ref{fla74}) is of order  $Q_{ab}^2Q_\perp^2s$. 
Now, if we take $\sim P_iP_jU^{+k}V^{-}_{~k}$ contribution to the last term in Eq. (\ref{fla73}), it is $\sim Q_\perp^4s$ 
which is down by ${Q_\perp^2\over Q_{ab}^2}$ factor in comparison to $p^+p^-U^{+i}V^{-j}$.  As to 
the term $\sim p^+P_j\calf_{ik}\calf^{-k}~\sim~{\alpha'_a}sQ_\perp^4$, it is $O\big({\alpha'_a}^2{Q_\perp^2/Q_{ab}^2}\big)$ 
in comparison to Eq. (\ref{fla74}).
\footnote{Recall that ${\alpha'_a}$ (and $\beta_b$) are either $\sim 1$ or $\ll 1$ depending on moderate-$x$ or small-$x$ 
kinematics}

Let us now consider the terms in the second line in the r.h.s.  of Eq. (\ref{fla73}). To get the contribution having four gluon tails, we expand ${1\over P^2}$ once and get terms like 
\beq
\hspace{-0mm}
V^{a,-i}(x_2|(p^+\delta_i^\alpha-\calp_ig^{+\alpha}){1\over p^2-\ie}\{p^k,\cala_k\}\tilde\delta_+(p)\calf_{\alpha\beta}
{1\over p^2+\ie}(p^-\delta^\beta_j-\calp_jg^{\beta -})|x_1)^{ab}U^{b,+j}(x_1)
\eeq
where $\cala_k$ is either $U_k$ or $V_k$. The numerator in the integral of this equation is 
again $\sim Q_\perp^4s$, so, as we mentioned above, it is a power correction
in comparison to the leading term (\ref{fla74}). Finally, the term in the first line in the r.h.s.  of Eq. (\ref{fla73})
should be expanded twice to get four gluons, so we have
\beq
\hspace{-0mm}
V^{a,-i}(x_2|(p^+\delta_i^\alpha-\calp_ig^{+\alpha}){1\over p^2-\ie}\{p^k,\cala_k\}\tilde\delta_+(p)\{p^l,\cala_l\}
{1\over p^2+\ie}(p^-\delta^\beta_j-\calp_jg^{\beta -})|y)^{ab}U^{b,+j}(x_1)
\eeq
The numerator in this equation is at best $\sim  Q_\perp^4s$ which is $O(Q_\perp^2/Q_{ab}^2)$ in comparison to the leading term.

Thus, all terms in the r.h.s. of Eq. (\ref{fla73}) are power corrections $\sim O\big({m_\perp^2\over s}\sim\zeta^{-1}\big)$ to the logarithmical leading term  (\ref{fla74}) which we calculate in the next Section. 

\subsection{Calculation of leading production terms \label{maincalc}}
In this Section we will calculate the first term in the r.h.s. of  term in Eq. (\ref{realeadingw}) given by Eq. (\ref{fla74}), see Fig. \ref{fig:real1}. First, it is convenient to use the identity
\begin{eqnarray}
&&\hspace{-1mm}
(x_2|{1\over p^2-\ie}\cala\tilde\delta_+(p)\calb{1\over p^2+\ie}
+\tilde\delta_+(p)\cala{1\over p^2+\ie}\calb{1\over p^2+\ie}
+{1\over p^2-\ie}\cala{1\over p^2-\ie}\calb\tilde\delta_+(p)
\nonumber\\
&&\hspace{5mm}
+~\tilde\delta_+(p)\cala\tilde{\delta}_-(p)\calb\tilde\delta_+(p)|x_1)
~
=~
(x_2|{1\over p^2+\ie p_0}\cala\tilde\delta_+(p)\calb{1\over p^2-\ie p_0}
\nonumber\\
&&\hspace{11mm}
+~\tilde\delta_+(p)\cala{1\over p^2-\ie p_0}\calb{1\over p^2-\ie p_0}
+~{1\over p^2+\ie p_0}\cala{1\over p^2+\ie p_0}\calb\tilde\delta_+(p)|x_1)
\label{kozalfla}
\end{eqnarray}
and rewrite Eq. (\ref{fla74}) 
changing singularities of propagators accordingly
\begin{eqnarray}
&&\hspace{-1mm}
-4V^{-a,i}(x_2)(x_2|{p^+\over p^2+\ie p_0}U^{+i}{1\over p^2+\ie p_0}V^{-j}p^-\tilde\delta_+(p)
+{p^+\over p^2+\ie p_0}U^{+i}\tilde\delta_+(p)V^{-j}{p^-\over p^2-\ie p_0}
\nonumber\\
&&\hspace{-1mm}
+~p^+\tilde\delta_+(p)U^{+i}{1\over p^2-\ie p_0}V^{-j}{p^-\over p^2-\ie p_0}|x_1)^{ab}U^{b,+j}(x_1)
\nonumber\\
&&\hspace{-1mm}
=~{N_c\over 8\pi^2(N_c^2-1)}
\!\int\!\dhd{\alpha'_a}\dhd {k'}_{a_\perp}\dhd{\beta'_b}\dhd k'_{b_\perp}\dhd\alpha_a\dhd {k}_{a_\perp}\dhd\beta_b\dhd {k}_{b_\perp}
\nonumber\\
&&\hspace{-1mm}
\times~e^{-i{\alpha'_a}\vro x_2^- -i\alpha_a\vro x_1^-}e^{-i{\beta'_b}\vro x_2^+ -i\beta_b\vro x_1^+}
e^{-i(k'_a+k'_b,x_2)_\perp-i(k_a+k_b,x_1)_\perp}
\label{calcula1}\\
&&\hspace{-1mm}
\times~U^{+,b}_{~~i}({\alpha'_a},k'_{a_\perp})V^{-i,a}({\beta'_b},k'_{b_\perp})
U^{+,b}_{~j}(\alpha_a,k_{a_\perp})V^{-j,a}(\beta_b,k_{b_\perp})
I_1({\alpha'_a}, k'_{a_\perp},\beta_b, k_{b_\perp}, x_1,x_2)
\nonumber
\ega
where
\bega
&&\hspace{-1mm}
I_1({\alpha'_a}, k'_{a_\perp},\beta_b, k_{b_\perp}, x_1,x_2)
~=~8\pi^2s^2\!\int\!\dhd\alpha\dhd\beta\dhd p_\perp~
e^{i\alpha\vro x_{12}^- +i\beta\vro x_{12}^+ -i(p,x_{12})_\perp}
\label{iodin}\\
&&\hspace{-1mm}
\times~\bigg[{{\alpha'_a}+\alpha\over ({\alpha'_a}+\alpha)\beta s-(p+k'_a)_\perp^2+\ie}\tilde\delta(\alpha\beta s-p_\perp^2)\theta(\alpha)
{\beta-\beta_b\over \alpha(\beta-\beta_b)s-(p-k_b)_\perp^2-\ie}
\nonumber\\
&&\hspace{-1mm}
+~\tilde{\delta}[({\alpha'_a}+\alpha)\beta s-(p+k'_a)_\perp^2]\theta(\beta){1\over \alpha\beta s-p_\perp^2-\ie}
{(\alpha+{\alpha'_a})(\beta-\beta_b)\over\alpha(\beta-\beta_b)s-(p-k_b)_\perp^2-\ie\alpha}
\nonumber\\
&&\hspace{-1mm}
+~{({\alpha'_a}+\alpha)(\beta-\beta_b)
\over ({\alpha'_a}+\alpha)\beta s-(p+k'_a)_\perp^2+\ie({\alpha'_a}+\alpha)}{1\over \alpha\beta s-p_\perp^2+\ie}
\tilde\delta[\alpha(\beta-\beta_b)s-(p-k_b)_\perp^2]\theta(\alpha)\bigg]
\nonumber
\end{eqnarray}
Here we used the fact that after taking matrix elements between nucleon states  only the colorless operators survive.

Note that singularities in denominators in there expressions correspond to ${\alpha'_a}+\ie$ and $\beta_b+\ie$ 
so it is sufficient to perform calculations at, say, positive ${\alpha'_a}$ and $\beta_b$. 

To calculate the integral (\ref{iodin}) it is convenient to split it in two parts using identity
\beq
\delta(\alpha\beta s-p_\perp^2)~=~\delta(\alpha\beta s-p_\perp^2)\bigg[{p_\perp^2\over\alpha^2s\xi+p_\perp^2}+ {p_\perp^2\over\beta^2s\xi^{-1}+p_\perp^2}\bigg]
\eeq
where $\xi$ is an arbitrary positive number of order of 1.
 We get
 \bega
 &&\hspace{-1mm}
 I_1~=~ I_{1a}+I_{1b},
 \ega
where 
\bega
&&\hspace{-1mm}
I_{1a}({\alpha'_a}, k'_{a_\perp},\beta_b, k_{b_\perp}, x_1,x_2)~=~
8\pi^2s^2\!\int\!\dhd\alpha\dhd\beta\dhd p_\perp~\theta(\alpha)~
e^{i\alpha\vro x_{12}^- +i\beta\vro x_{12}^+ -i(p,x_{12})_\perp}
\nonumber\\
&&\hspace{-1mm}
\times~\bigg[{\theta(\alpha)({\alpha'_a}+\alpha)(\beta-\beta_b)\tilde{\delta}(\alpha\beta s-p_\perp^2)
\over [({\alpha'_a}+\alpha)\beta s-(p+k'_a)^2+\ie][\alpha(\beta-\beta_b)s-(p-k_b)_\perp^2-\ie]}{p_\perp^2\over\alpha^2s\xi+p_\perp^2}
\label{iodina}\\
&&\hspace{-1mm}
+~{({\alpha'_a}+\alpha)(\beta-\beta_b)\over [({\alpha'_a}+\alpha)\beta s-(p+k'_a)_\perp^2+\ie({\alpha'_a}+\alpha)](\alpha\beta s-p_\perp^2+\ie)}\tilde{\delta}[\alpha(\beta-\beta_b)s-(p-k_b)_\perp^2]\bigg]
\nonumber
\end{eqnarray}
and 
\bega
&&\hspace{-1mm}
I_{1b}({\alpha'_a}, k'_{a_\perp},\beta_b, k_{b_\perp}, x_1,x_2)
~=~8\pi^2s^2\!\int\!\dhd\alpha\dhd\beta\dhd p_\perp~\theta(\beta)
e^{i\alpha\vro x_{12}^- +i\beta\vro x_{12}^+ -i(p,x_{12})_\perp}
\nonumber\\
&&\hspace{-1mm}
\times~\bigg[{({\alpha'_a}+\alpha)(\beta-\beta_b)\tilde{\delta}(\alpha\beta s-p_\perp^2)\over [({\alpha'_a}+\alpha)\beta s-(p+k'_a)^2+\ie]
[\alpha(\beta-\beta_b)s-(p-k_b)_\perp^2-\ie]}{p_\perp^2\over\beta^2s\xi^{-1}+p_\perp^2}
\nonumber\\
&&\hspace{-1mm}
+~\tilde{\delta}[({\alpha'_a}+\alpha)\beta s-(p+k'_a)^2]{1\over \alpha\beta s-p_\perp^2-\ie}
{(\alpha+{\alpha'_a})(\beta-\beta_b)\over\alpha(\beta-\beta_b)-(p-k_b)_\perp^2-\ie\alpha}
\bigg]
\label{iodinb}
\end{eqnarray}

As we discussed above, the Eq. (\ref{calcula1}) is not yet the contribution to the coefficient function. 
According to Eq. (\ref{facoord2}), one needs to subtract relevant matrix elements of gluon TMDs from the result of calculation 
in the background fields $\cala$ and $\calb$.  The ``projectile'' matrix elements of operator $\hacalo^{ij,\sigma_p}(x_2^-,x_{2_\perp};x_1^-,x_{1_\perp})$ 
are given by the diagrams shown in  Fig. \ref{fig:eik1} and ``target'' matrix elements of operator
$\hacalo^{ij;\sigma_t}(x_2^+,x_{2_\perp};x_1^+,x_{1_\perp})$ are given by the diagrams shown in  Fig. \ref{fig:eik2}. 
Consequently, to get the contribution of Eq. (\ref{iodin}) to the coefficient function one should subtract from the integral (\ref{iodin})
two eikonal-type contributions of TMD matrix elements coming from  diagrams shown in Fig. \ref{fig:eik2}a,b and Fig. \ref{fig:eik1}d,e. 

The first contribution, coming from the ``projectile'' eikonals in Fig.  \ref{fig:eik2}a,b,  corresponds to the 
$\alpha\ll{\alpha'_a}$ asymptotics in Eq. (\ref{iodin}) cut from above according to ``smooth'' cutoff $e^{-i{\alpha\over\sigma_t}}$ 
discussed in Ref. \cite{Balitsky:2019ayf} (see also Appendix \ref{app:eikonals})
\bega
&&\hspace{-2mm}
I_{\rm Fig.  \ref{fig:eik2}a,b}^{\rm eik}(\beta_b, {k}_{b_\perp}, x_1^+,x_{1_\perp},x_2^+,x_{2_\perp})
\label{iodinaeik}\\
&&\hspace{-2mm}
=~8\pi^2s\!\int_0^\infty\!\dhd\alpha~e^{-i{\alpha\over\sigma_t}}
{\dhd\beta\over\beta+\ie}\dhd p_\perp~
e^{i\beta\vro x_{12}^+ -i(p,x_{12})_\perp}
\nonumber\\
&&\hspace{-2mm}
\times~\Big[\tilde{\delta}(\alpha\beta s-p_\perp^2)
{\beta-\beta_b\over \alpha(\beta-\beta_b)s-(p-k_b)_\perp^2-\ie}
+{(\beta-\beta_b)\over \alpha\beta s-p_\perp^2+\ie}
\tilde{\delta}[\alpha(\beta-\beta_b)s-(p-k_b)_\perp^2]\Big]
\nonumber
\end{eqnarray}
The second contribution coming from Fig. \ref{fig:eik1}e,f corresponds to $\beta\ll\beta_b$ 
asymptotics of the integrand in Eq. (\ref{iodin}) integrated
with the upper ``smooth cutoff'' factor $e^{i{\beta\over\sigma_p}}$
\bega
&&\hspace{-2mm}
I_{\rm Fig. \ref{fig:eik1}e,f}^{\rm eik}({\alpha'_a}, k'_{a_\perp},  x_1^-,x_{1_\perp},x_2^-,x_{2_\perp})
\label{iodinbeik}\\
&&\hspace{-2mm}
=~8\pi^2s\!\int_0^\infty\!\dhd\beta~e^{i{\beta\over\sigma_p}}
{\dhd\alpha\over\alpha-\ie}\dhd p_\perp~
e^{i\alpha\vro x_{12}^- -i(p,x_{12})_\perp}
\nonumber\\
&&\hspace{-2mm}
\times~\Big[{{\alpha'_a}+\alpha\over ({\alpha'_a}+\alpha)\beta s-(p+k'_a)^2+\ie}\tilde{\delta}(\alpha\beta s-p_\perp^2)
+~\tilde{\delta}[({\alpha'_a}+\alpha)\beta s-(p+k'_a)^2]{\alpha+{\alpha'_a}\over \alpha\beta s-p_\perp^2-\ie}
\Big]
\nonumber
\end{eqnarray}

To get the contribution to the coefficient function we need to calculate 
\begin{eqnarray}
&&\hspace{-1mm}
J_1({\alpha'_a}, {k'}_{a_\perp},\beta_b, {k}_{b_\perp}, x_2,x_1)
\label{jotodin}
\\
&&\hspace{1mm}
=~I_1({\alpha'_a}, {k'}_{a_\perp},\beta_b, {k}_{b_\perp}, z_2,z_1)-~I_{\rm Fig.  \ref{fig:eik2}a,b}^{\rm eik}(\beta_b, {k}_{b_\perp}, x_2,x_1)
-I_{\rm Fig.  \ref{fig:eik1}e,f}^{\rm eik}({\alpha'_a}, k'_{a_\perp}, x_2x_1)
\nonumber
\end{eqnarray}
As demonstrated in the Appendix \ref{sect:noparallel}, one can neglect $x_{12}^+$ and $x_{12}^-$ in 
the difference
\bega
&&\hspace{-1mm}
I_{1a}({\alpha'_a}, {k'}_{a_\perp},\beta_b, {k}_{b_\perp}, x_1,x_2)
-I_{\rm Fig. \ref{fig:eik2}a,b}^{\rm eik}(\beta_b, {k}_{b_\perp}, x_1,x_2)
\label{differenca}
\end{eqnarray}
and similarly in
\beq
I_{1b}({\alpha'_a}, {k'}_{a_\perp},\beta_b, {k}_{b_\perp}, x_1,x_2)-I_{\rm Fig. \ref{fig:eik1}e,f}^{\rm eik}({\alpha'_a}, k'_{a_\perp}, x_1,x_2)
\label{differenceb}
\eeq
Qualitatively, the argument is as follows. Consider Eq. (\ref{iodinb}). In order for $e^{i\alpha\vro x_{12}^-}$ to be essential, $\alpha$ should be of order of ${\alpha'_a}$ due to Eq. (\ref{coordregion}). Due to $\delta$-functions in  Eq. (\ref{iodinb}), this means that $\beta$ should
be small, of order of ${p_\perp^2\over {\alpha'_a}s}\ll 1$ since $p_\perp\sim {1\over x_{12_\perp}}\sim Q_\perp$. However, the contribution
of small $\beta$ to Eq. (\ref{iodin}) is subtracted by small-$\beta$ eikonal (\ref{iodinbeik}) so the resulting difference 
$I_{1b}-I_{\rm Fig. \ref{fig:eik2}e,f}^{\rm eik}$ is small and the factor $e^{i\alpha\vro x_{12}^-}$ can be neglected.  
Similarly, the factor  $e^{i\beta\vro x_{12}^+}$ in the difference
 $I_1-I_{\rm Fig. \ref{fig:eik1}a,b}^{\rm eik}$ at small $\alpha$ can be replaced by 1. In Appendix \ref{sect:noparallel} it is demonstrated that the corrections due to these 
 approximations are $\sim {Q_\perp^2\over \sigma_t\beta_bs}\sim\lambda_t$  and $\sim {Q_\perp^2\over \sigma_p{\alpha'_a}s}\sim\lambda_p$, respectively. 
 As we discussed in Sect. \ref{sec:funt}, we  neglect such power corrections. 
 
 We get
\bega
&&\hspace{-1mm}
I_{1a}({\alpha'_a}, {k'}_{a_\perp},\beta_b, {k}_{b_\perp}, x_{1_\perp}, x_{2_\perp})~=~
4\pi\!\int_0^\infty\!{d\alpha\over\alpha} \!\int\!\dhd p_\perp~
{e^{-i(p,x_{12})_\perp}\over \alpha\beta_bs+(p-k_b)_\perp^2-p_\perp^2+\ie}
\nonumber\\
&&\hspace{5mm}
\times~\bigg[
{p_\perp^2\over\alpha^2s\xi+p_\perp^2}
{(\alpha+{\alpha'_a})(\alpha\beta_bs-p_\perp^2)
\over \big[(\alpha+{\alpha'_a})p_\perp^2-\alpha(p+k'_a)^2+\ie\big]}
\nonumber\\
&&\hspace{11mm}
+~{(p-k_b)_\perp^2(\alpha+{\alpha'_a})
\over 
\alpha({\alpha'_a}+\alpha)\beta_bs+({\alpha'_a}+\alpha)(p-k_b)_\perp^2-\alpha(p+k'_a)_\perp^2+\ie({\alpha'_a}+\alpha)}
\bigg]
\label{iodinae}
\end{eqnarray}
and 
\bega
&&\hspace{-1mm}
I_{1b}({\alpha'_a}, {k'}_{a_\perp},\beta_b, {k'}_{b_\perp}, x_{1_\perp}, x_{2_\perp})
=~4\pi\!\int_0^\infty\!{d\beta\over\beta} \!\int\!\dhd p_\perp~
{e^{-i(p,x_{12})_\perp}\over {\alpha'_a}\beta s-(p+k'_a)^2+p_\perp^2+\ie}
\nonumber\\
&&\hspace{5mm}
\times~\bigg[
\!\int_0^\infty\!{\dhd\beta\over\beta} {p_\perp^2\xi\over\beta^2s+p_\perp^2\xi}
{(\beta_b-\beta)({\alpha'_a}\beta s+p_\perp^2)
\over \big((\beta_b-\beta)p_\perp^2+\beta(p-k_b)^2+\ie\big)}
\nonumber\\
&&\hspace{11mm}
+~{(\beta_b-\beta)(p+k'_a)_\perp^2
\over
({\alpha'_a}+\ie)(\beta_b-\beta)\beta s-(\beta_b-\beta)(p+k'_a)_\perp^2-\beta(p-k_b)_\perp^2}
\bigg]
\label{iodinbe}
\end{eqnarray}
To calculate the sum of Eqs. (\ref{iodinae}) and (\ref{iodinbe}), it is convenient to perform change of variables 
$\alpha={p_\perp^2\over\beta s}$ in the first term in square brackets in Eq. (\ref{iodinae}) and change
$\alpha={(p-k'_b)_\perp^2\over(\beta-\beta_b)s}$ in the second term. Since this change affects cancellation of logarithmic divergences 
at $\alpha\rightarrow 0$, before the change we replace $\!\int_0^\infty\!{d\alpha\over\alpha}$ by $\!\int_\ve^\infty\!{d\alpha\over\alpha}$.
After some algebra one obtains
\bega
&&\hspace{-1mm}
I_{1}({\alpha'_a}, {k'}_{a_\perp},\beta_b, {k}_{b_\perp}, x_{1_\perp}, x_{2_\perp})~=~
4\pi\!\int\!\dhd p_\perp~e^{-i(p,x_{12})_\perp}\!\int_0^\infty d\beta~\bigg[\theta(\beta-\beta_b)-\theta(\beta)\big]
\nonumber\\
&&\hspace{5mm}
\times~
{(\beta-\beta_b)[(p-k_b)_\perp^2-{\alpha'_a}(\beta_b-\beta)s]
\big[\beta(p-k_b)_\perp^2+p_\perp^2(\beta_b-\beta)s+\ie\big]^{-1}\over \big(-({\alpha'_a}+\ie)\beta(\beta_b-\beta)s+(p-k_b)_\perp^2\beta+(p+k'_a)_\perp^2(\beta_b-\beta)\big)}
\nonumber\\
&&\hspace{11mm}
+~\lim_{\ve\rightarrow 0}{1\over \beta[(p-k_b)_\perp^2-p_\perp^2+\ie]}
\Big[\theta\Big(\beta-{p_\perp^2\over\ve s}\Big)-\theta\Big(\beta-\beta_b-{(p-k_b)_\perp^2\over \ve s}\Big)\Big]\bigg]
\ega
Let us take $\beta_b>0$, then
\bega
&&\hspace{-1mm}
I_{1}({\alpha'_a}, {k'}_{a_\perp},\beta_b, {k}_{b_\perp}, x_{1_\perp}, x_{2_\perp})~=~
4\pi\!\int\!\dhd p_\perp~e^{-i(p,x_{12})_\perp}
\label{fla721}\\
&&\hspace{-1mm}
\times~\bigg[\!\int_0^1\! du~
{\baru[(p-k_b)_\perp^2-Q_{a'b}^2\baru]\over [\baru(p+k'_a)_\perp^2+ u(p-k_b)_\perp^2-Q_{a'b}^2\baru u][p_\perp^2\baru+u(p-k_b)_\perp^2]}+{\ln(p-k_b)_\perp^2/p_\perp^2\over (p-k_b)_\perp^2-p_\perp^2}\bigg]
\nonumber\\
&&\hspace{-1mm}
=~-4\pi\!\int\!\dhd p_\perp~\!\int_0^1\! du~
{e^{-i(p,x_{12})_\perp}\baru Q_{ab'}^2
\over [\baru(p+k'_a)_\perp^2+ u(p-k_b)_\perp^2-Q_{a'b}^2\baru u][p_\perp^2\baru+u(p-k_b)_\perp^2]}
+O(\lambda)
\nonumber\\
&&\hspace{-1mm}
=~-4\pi\!\int\!\dhd p_\perp~\!\int_0^1\! du~
{e^{-i(p,x_{12})_\perp}Q_{ab'}^2\over [\baru(p+k'_a)_\perp^2-Q_{a'b}^2u][p_\perp^2\baru+u(p-k_b)_\perp^2]}
+O(\lambda)
\nonumber\\
&&\hspace{-1mm}
=~4\pi\!\int\!\dhd p_\perp~
e^{-i(p,x_{12})_\perp}{Q_{a'b}^2\over Q_{a'b}^2p_\perp^2+(p+k'_a)_\perp^2(p-k_b)_\perp^2}
\ln{-Q_{a'b}^2p_\perp^2\over (p+k'_a)_\perp^2(p-k_b)_\perp^2}+O(\lambda)
\nonumber
\ega
where $\baru\equiv 1-u$ and $Q_{a'b}^2\equiv({\alpha'_a}+\ie)(\beta_b+\ie)s$ (recall that the analytical properties of integrals 
over ${\alpha'_a}$ and $\beta_b$ are determined by the integral (\ref{iodin}) to be ${\alpha'_a}+\ie$ and $\beta_b+\ie$).

This integral is calculated in the Appendix \ref{app:rints}, see Eq. (\ref{mintegral1}):
\bega
&&\hspace{-1mm}
4\pi\!\int\!\dhd p_\perp~
{e^{-i(p,x)_\perp}Q_{ab}^2\over Q_{ab}^2p_\perp^2+(p+k_a)_\perp^2(p-k_b)_\perp^2}
\ln{-Q_{ab}^2p_\perp^2\over (p+k_a)_\perp^2(p-k_b)_\perp^2}
\label{mintegral}\\
&&\hspace{-1mm}
=~\ln{-Q_{ab}^2\over {k}_{a_\perp}^2}\ln{-Q_{ab}^2\over k_{b_\perp}^2}-\half\Big(\ln{-Q_{ab}^2x_\perp^2\over 4}+2\gamma\Big)^2
+{\pi^2\over 3}
\nonumber\\
&&\hspace{-1mm}
+~\!\int_0^1\! {du\over u} \Big[\ln{{k}_{a_\perp}^2x_\perp^2\baru u\over 4}+2\gamma
+2e^{iu(k,x)_\perp}K_0(\sqrt{{k}_{a_\perp}^2x_\perp^2\baru u})~+~{k}_{a_\perp}\leftrightarrow -k_{b_\perp}\Big]
~+~O(\lambda)
\nonumber
\ega
It is convenient to represent it as a sum of the double-log contribution similar to virtual term, and the remainder.
We get
\bega
&&\hspace{-1mm}
I_{1}~=~
I^{\rm d.log}({\alpha'_a}, {k'}_{a_\perp},\beta_b, {k}_{b_\perp})
+I_{1}^{\rm rem}({\alpha'_a}, k'_{a_\perp},\beta_b, k_{b_\perp}, x_{12_\perp})
\ega
where $I^{\rm d.log}$ was defined in Eq. (\ref{dlog})
\bega
&&\hspace{-1mm}
I^{\rm d.log}({\alpha'_a}, {k'}_{a_\perp},\beta_b, {k}_{b_\perp})~
=~\ln{-Q_{a'b}^2\over {k'}_{a_\perp}^2}\ln{-Q_{a'b}^2\over {k}_{b_\perp}^2}+{\pi^2\over 3}~+~O(\lambda),
\label{iodinvirt}
\ega
and
\bega
&&\hspace{-1mm}
I_{1}^{\rm rem}({\alpha'_a}, k'_{a_\perp},\beta_b, k_{b_\perp}, x_{12_\perp})
\label{iodinremainder}
\\
&&\hspace{-1mm}
=~-\half\Big(\ln{-Q_{a'b}^2x_{12_\perp}^2\over 4}+2\gamma\Big)^2~+~\!\int_0^1\! {du\over u} 
\Big[\Big(\ln{{k'}_{a_\perp}^2x_{12_\perp}^2\baru u\over 4}+2\gamma\big)
\nonumber\\
&&\hspace{5mm}
+~2e^{-iu(k'_a,x_{12})_\perp}K_0(\sqrt{{k'}_{a_\perp}^2x_{12_\perp}^2\baru u})+k'_{a_\perp}\leftrightarrow -{k}_{b_\perp}\Big]
~+~O(\lambda)
\nonumber\\
&&\hspace{-1mm}
=~-\half\Big(\ln{-Q_{a'b}^2x_{12_\perp}^2\over 4}+2\gamma\Big)^2+I_K(k'_{a_\perp},x_{12_\perp})
+I_K(-k_{b_\perp},x_{12_\perp})
~+~O(\lambda)
\nonumber
\ega
where
\beq
I_K(k_\perp,x_\perp)~\equiv~\!\int_0^1\! {du\over u} 
\Big[\ln{k_\perp^2x_\perp^2\baru u\over 4}+2\gamma
+2e^{iu(k,x)_\perp}K_0(\sqrt{k_\perp^2x_\perp^2\baru u})\Big]
\label{ika}
\eeq
and $K_0$ is the Macdonald function.

\begin{figure}[htb]
\begin{center}
\includegraphics[width=133mm]{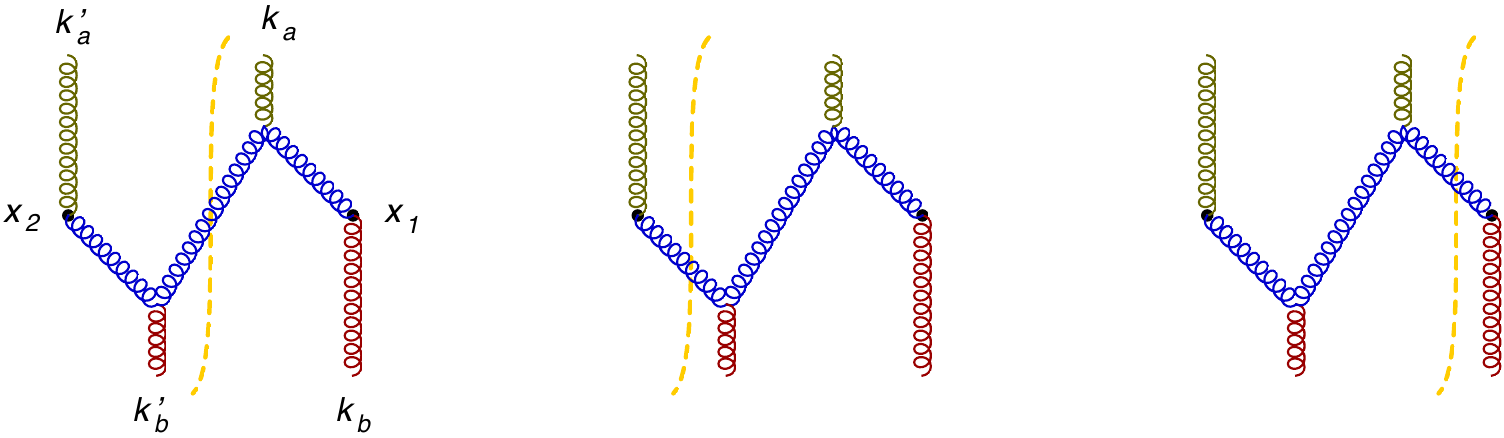}
\end{center}
\caption{Second set of leading diagrams with gluon production.  Projectile fields $U^{-i}(z_2)$ are denoted by green
tails while target fields $V^{+i}(z_1)$ by red tails. \label{fig:real2} }
\end{figure}

The second leading contribution to hadronic tensor (\ref{pizw}) comes from the second term 
in the r.h.s. of Eq. (\ref{realeadingw})
\begin{eqnarray}
&&\hspace{-2mm}
U^{+i,a}(x_2)\langle g^2F^{-,a}_{~i}(x_2)F^{+,b}_{~j}(x_1)\rangle_\matA V^{-j,b}(x_1)
\nonumber\\
&&\hspace{-2mm}
=~-4g^2U^{+i,a}(x_2)(x_2|{p^-\over p^2+\ie p_0}V^{-i}{1\over p^2+\ie p_0}U^{+j}p^+\tilde\delta_+(p)
\nonumber\\
&&\hspace{-2mm}
+~{p^-\over p^2+\ie p_0}V^{-i}\tilde\delta_+(p)U^{+j}{p^+\over p^2-\ie p_0}+p^-\tilde\delta_+(p)V^{-i}{1\over p^2-\ie p_0}U^{+j}{p^+\over p^2-\ie p_0}|x_1)^{ab}V^{-j,b}(x_1)
\nonumber\\
&&\hspace{-2mm}
=~{g^2N_c\over 8\pi^2(N_c^2-1)}
\!\int\!\dhd\alpha_a\dhd k_{a_\perp}\dhd\beta_b\dhd k_{b_\perp}\dhd\alpha'_a\dhd k'_{a_\perp}\dhd\beta'_b\dhd k'_{b_\perp}
\nonumber\\
&&\hspace{-2mm}
\times~e^{-i\alpha_a\vro x_1^- -i\alpha'_a\vro x_2^-}e^{-i\beta_b\vro x_1^+ -i\beta'_b\vro x_2^+}
e^{-i(k_a+k_b,x_1)_\perp-i(k'_a+k'_b,x_2)_\perp}
\label{vnizvverx}\\
&&\hspace{-2mm}
\times~U^{+,b}_{~~i}(\alpha_a,k_{a_\perp})V^{-i,a}(\beta_b,k_{b_\perp})
U^{+,b}_{~j}(\alpha'_a,k'_{a_\perp})V^{-j,a}(\beta'_b,k'_{b_\perp})
I_2(\alpha_a, k_{a_\perp},\beta'_b, k'_{b_\perp}, x_1,x_2)
\nonumber
\end{eqnarray}
The corresponding diagrams are shown in Fig. \ref{fig:real2}.
It is clear that they differ from the diagrams in Fig. \ref{fig:real1} by 
trivial projectile$\leftrightarrow$target replacements
\beq
x^+\leftrightarrow x^-,~~~~
\alpha_a\leftrightarrow\beta_b,~~~~\alpha'_a\leftrightarrow\beta'_b,~~~~ 
k_{a_\perp}\leftrightarrow k_{b_\perp},~~~~ k'_{a_\perp}\leftrightarrow k'_{b_\perp}
\label{projtarepl}
\eeq
so we get
\bega
&&\hspace{-1mm}
I_2({\alpha_a}, k_{a_\perp},\beta'_b, k'_{b_\perp}, x_1,x_2)
~=~8\pi^2s^2\!\int\!\dhd\alpha\dhd\beta\dhd p_\perp~
e^{i\alpha\vro x_{12}^- +i\beta\vro x_{12}^+ -i(p,x_{12})_\perp}
\label{idva}\\
&&\hspace{-1mm}
\times~\bigg[{{\beta'_b}+\beta\over (\beta'_b+\beta)\beta s-(p+k'_b)^2+\ie}\tilde{\delta}(\alpha\beta s-p_\perp^2)\theta(\alpha)
{\alpha-\alpha_a\over (\alpha-\alpha_a)\beta s-(p-k_a)_\perp^2-\ie}
\nonumber\\
&&\hspace{-1mm}
+~\tilde{\delta}[\alpha({\beta'_b}+\beta) s-(p+k'_b)^2]\theta(\alpha){1\over \alpha\beta s-p_\perp^2-\ie}
{(\alpha-\alpha_a)(\beta+\beta'_b)\over(\alpha-\alpha_a)\beta s-(p-k_a)_\perp^2-\ie\beta}
\nonumber\\
&&\hspace{-1mm}
+~{(\alpha-\alpha_a)(\beta'_b+\beta)
\over \alpha(\beta'_b+\beta)s-(p+k'_b)_\perp^2+\ie(\beta'_b+\beta)}{1\over \alpha\beta s-p_\perp^2+\ie}
\tilde{\delta}[\beta(\alpha-\alpha_a)s-(p-k_a)_\perp^2]\theta(\alpha)\bigg]
\nonumber
\end{eqnarray}
Similarly to the previous case, after subtraction of the corresponding ``projectile'' eikonals in Fig.  \ref{fig:eik1}a,b and ``projectile'' eikonals in Fig.  \ref{fig:eik2}e,f one can  set $x_{12}^\parallel=0$ and get 
\bega
&&\hspace{-1mm}
I_2(\alpha_a, k_{a_\perp},{\beta'_b}, k'_{b_\perp}, x_1,x_2)~\stackrel{x_{12}^\parallel=0}=~
I^{\rm d.log}(\alpha_a, k_{a_\perp},{\beta'_b}, k'_{b_\perp})+
I^{\rm rem}_2(\alpha_a, k_{a_\perp},{\beta'_b}, k'_{b_\perp}, x_{12_\perp})
\nonumber\\
\ega
from Eqs. (\ref{iodinvirt}) and (\ref{iodinremainder}) projectile$\leftrightarrow$target replacements (\ref{projtarepl}) so that
\bega
&&\hspace{-1mm}
I^{\rm d.log}(\alpha_a, k_{a_\perp},{\beta'_b}, k'_{b_\perp})
~=~\ln{-Q_{ab'}^2\over k_{a_\perp}^2}\ln{-Q_{ab'}^2\over {k'}_{b_\perp}^2}+{\pi^2\over 3}~+~O(\lambda)
\label{idvavirt}
\ega
(cf. Eq. (\ref{iodinvirt})) and
\bega
&&\hspace{-1mm}
I_2^{\rm rem}(\alpha_a, k_{a_\perp},{\beta'_b}, k'_{b_\perp}, x_{2_\perp}, x_{1_\perp})
\label{idvaremainder}\\
&&\hspace{-1mm}
=~-\half\Big(\ln{-Q_{ab'}^2x_\perp^2\over 4}+2\gamma\Big)^2
+I_K(k'_{b_\perp},x_{21_\perp})+I_K(-k_{a_\perp},x_{21_\perp})
~+~O(\lambda)
\nonumber
\ega
where  $I_K$ is given by Eq. (\ref{ika}).

 The final result for ``production'' contributions (at $x_{12}^+=x_{12}^-=0)$ can be presented as follows
\bega
&&\hspace{-1mm}
\pizw^{\rm prod}(x_1,x_2)~=~~{N_c^2-1\over N_c}8\pi^2\Big(
V^{-i,a}(x_2)\langle F^{+,a}_{~i}(x_2)F^{-,b}_{~j}(x_1)\rangle_\matA^{\rm Fig. \ref{fig:real1}} U^{+j,b}(x_1)
\nonumber\\
&&\hspace{22mm}
+~U^{+i,a}(x_2)\langle F^{-,a}_{~i}(x_2)F^{+,b}_{~j}(x_1)\rangle_\matA^{\rm Fig. \ref{fig:real2}} V^{-j,b}(x_1)\Big)
\nonumber\\
&&\hspace{-1mm}
\stackrel{x_{12}^\parallel=0}=~
\!\int\!\dhd{\alpha_a}\dhd k_{a_\perp}\dhd{\beta_b}\dhd k_{b_\perp}\dhd\alpha'_a\dhd k'_{a_\perp}\dhd\beta_b\dhd k'_{b_\perp}
e^{-i{\alpha'_a}\vro x_2^- -i\alpha_a\vro x_1^-}e^{-i{\beta'_b}\vro x_2^+ -i\beta_b\vro x_1^+}
\nonumber\\
&&\hspace{-1mm}
\times~
e^{-i(k_a+k_a,x_1)_\perp-i(k'_a+k'_b,x_2)_\perp}~
U^{+,b}_{~~i}({\alpha'_a},{k'}_{a_\perp})V^{-i,a}({\beta'_b},k_{b_\perp})
U^{+,b}_{~j}(\alpha_a,k_{a_\perp})V^{-j,a}(\beta_b,k_{b_\perp})
\nonumber\\
&&\hspace{11mm}
\times~I^{\rm prod}({\alpha_a}, k_{a_\perp},{\beta_b}, k_{b_\perp},\alpha'_a, k'_{a_\perp},{\beta'_b}, k'_{b_\perp})
~+~O(\lambda)
\label{prodotvet}
\ega
where
\bega
&&\hspace{-1mm}
I^{\rm prod}(\alpha_a, k_{a_\perp},{\beta_b}, k_{b_\perp},{\alpha'_a}, k'_{a_\perp},{\beta'_b}, k'_{b_\perp})
\label{iprod}\\
&&\hspace{-1mm}
=~
I^{\rm d.log}({\alpha'_a}, k'_{a_\perp},\beta_b, k_{b_\perp})
+I_{1}^{\rm rem}({\alpha'_a}, {k'}_{a_\perp},\beta_b, k_{b_\perp}, x_{21_\perp})
\nonumber\\
&&\hspace{-1mm}
+~I^{\rm d.log}(\alpha_a, k_{a_\perp},{\beta'_b}, k'_{b_\perp})
+I_2^{\rm rem}(\alpha_a, k_{a_\perp},{\beta'_b}, k'_{b_\perp}, x_{21_\perp})
\nonumber
\ega
where $I_{1}^{\rm rem}$ and $I_{2}^{\rm rem}$ are given by Eqs. (\ref{iodinremainder}) and (\ref{idvaremainder}), respectively.

\subsection{Handbag diagrams \label{sec:handbag}}
\begin{figure}[htb]
\begin{center}
\includegraphics[width=111mm]{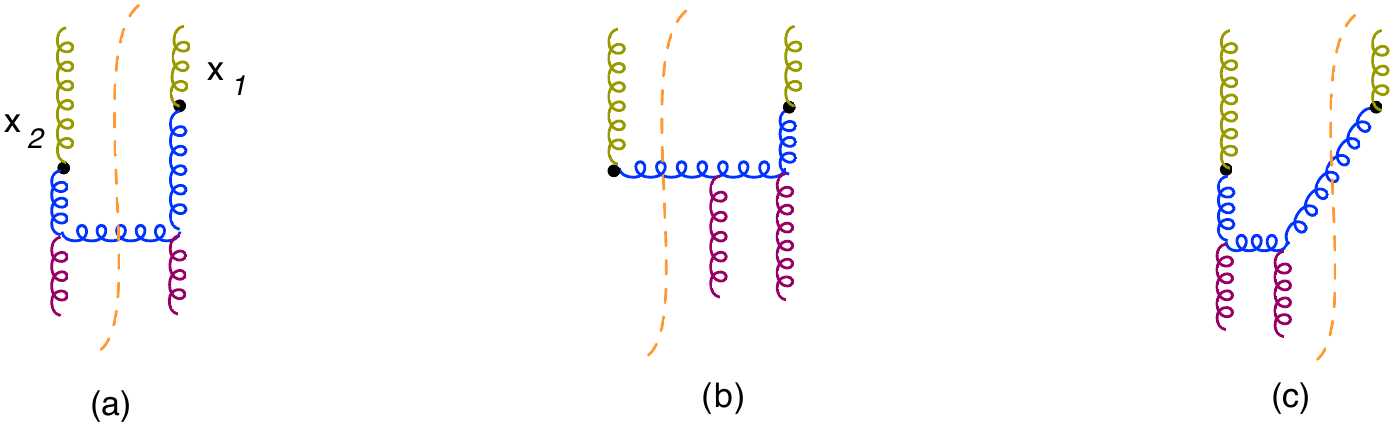}
\end{center}
\caption{``Target'' handbag diagrams\label{fig:hand1}. }
\end{figure}

Let us start with the third term in the r.h.s. of Eq. (\ref{realeadingw}) given by ``target handbag'' diagrams in Fig. 
\ref{fig:hand1}.
We need to subtract from these diagrams the corresponding diagrams coming from ``target'' TMD eikonals in 
Fig. \ref{fig:hand1}.  As was mentioned above (see Appendix \ref{app:eikonals} for details), we use ``point-splitting'' regularization of integrals over $\alpha$ in these contributions. 
The subtracted ``eikonal'' diagrams then
look the same as those in Fig. \ref{fig:hand1} with the only difference in points where gluon fields $V^{-i,a},F^{+,a}_{~i},F^{-,b}_{~j}$, and $U^{+j,b}$ are located. 
\begin{figure}[htb]
\begin{center}
\includegraphics[width=133mm]{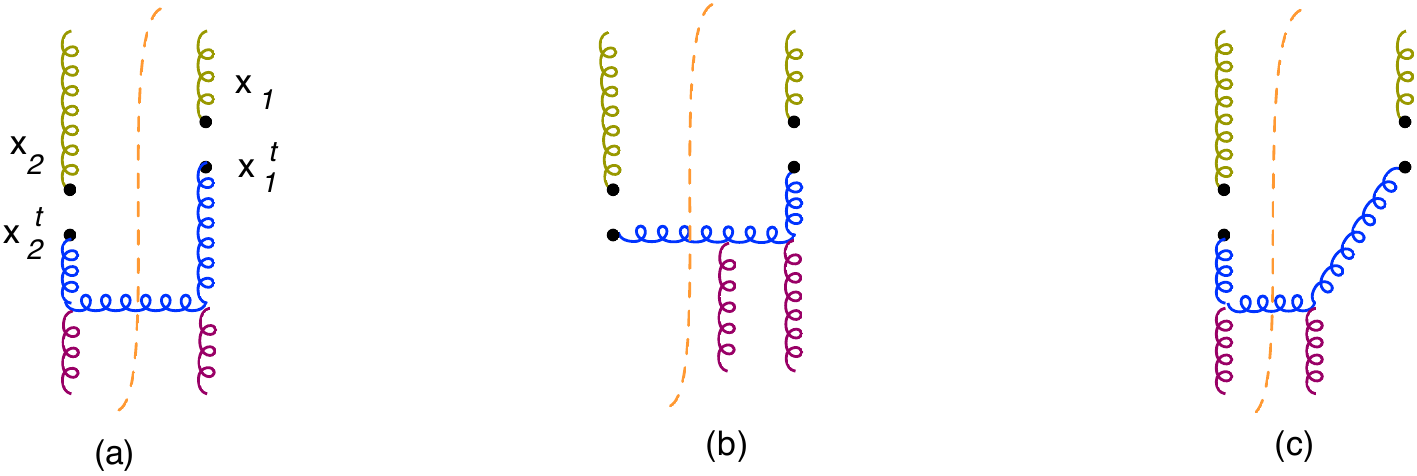}
\end{center}
\caption{``Target eikonal'' handbag diagrams.
Here $x_1^t=x_{1_\perp}+x_1^+$,   $x'_1=x_1^t +\delta^+$ and similarly for $x_2$. 
\label{fig:handeik1}}
\end{figure}

Using Wightman ``cut'' propagator (\ref{gluprop2}) in the background field one easily obtains
\begin{eqnarray}
&&\hspace{-1mm}
\langle g^2\tilF^{- i,a}(x_2)F^{- j,b}(x_1)\rangle~=~
-4(x_2|{p_i\over p^2+\ie p_0}V^-_{~\xi}\tilde\delta_+(p)V^{-\xi}{p_j\over p^2-\ie p_0}
\nonumber\\
&&\hspace{-1mm} 
-~p_i\tilde\delta_+(p)V^-_{~\xi}{1\over  p^2-\ie p_0}V^{-\xi}{p_j\over  p^2-\ie p_0}
-{p_i\over p^2+\ie p_0}V^-_{~\xi}{1\over p^2+\ie p_0}V^{-\xi}p_j\tilde\delta_+(p)|x_1)^{ab}g^2
\nonumber\\
&&\hspace{-1mm}
=~
-2g^2\!\int\!\dhd\beta_b\dhd{\beta'_b}\dhd k_{b_\perp}\dhd k'_{b_\perp}
e^{-ik_ax_1-ik'_bx_2}V^{-,ac}_{~k}({\beta'_b},k'_{b_\perp})V^{-k,cb}(\beta_b,k_{b_\perp})
\!\int\!\dhd p_\perp
\nonumber\\
&&\hspace{-1mm}
\times~\!\int_0^\infty\!{\dhd\alpha\over\alpha}\Big\{
{(p+k'_b)_i(p-k_b)_j\Big(e^{-i{\beta'_b}\vro x_{12}^+ +i{(p+k'_b)_\perp^2\over\alpha s}\vro x_{12}^+}
-
e^{i{p_\perp^2\over\alpha s}\vro x_{12}^+}\Big)e^{i\alpha\vro x_{12}^--i(p,x_{12})_\perp}
\over [\alpha{\beta'_b}s-(p+k'_b)_\perp^2+p_\perp^2+\ie][\alpha({\beta'_b}+\beta_b)s-(p+k'_b)_\perp^2+(p-k_b)_\perp^2+\ie]}
\nonumber\\
&&\hspace{-1mm}
+~{(p+k'_b)_i(p-k_b)_j\Big(e^{i\beta_b\vro x_{12}^+ +i{(p-k_b)_\perp^2\over\alpha s}\vro x_{12}^+}
-e^{i{p_\perp^2\over\alpha s}\vro x_{12}^+}\Big)e^{i\alpha\vro x_{12}^--i(p,x_{12})_\perp}
\over[\alpha({\beta'_b}+\beta_b)s+(p-k_b)_\perp^2-(p+k'_b)_\perp^2+\ie][\alpha\beta_bs+(p-k_b)_\perp^2-p_\perp^2+\ie]}
\Big\}
\label{dvavae}
\end{eqnarray}
so we get (see the definition (\ref{pizw}))
\begin{eqnarray}
&&\hspace{-1mm}
\pizW^{\rm handbag}(x_1,x_2)-\pizW^{\rm handbag}_{\rm eik}(x_1,x_2)~=~
8\pi^2
\!\int\!\dhd{\alpha'_a}\dhd {k'}_{a_\perp}\dhd{\beta'_b}\dhd k_{b_\perp}\dhd\alpha_a\dhd {k'}_{a_\perp}\dhd\beta_b\dhd {k'}_{b_\perp}
\nonumber\\
&&\hspace{-1mm}
\times~e^{-i{\alpha'_a}\vro x_2^- -i\alpha_a\vro x_1^-}e^{-i{\beta'_b}\vro x_2^+ -i\beta_b\vro x_1^+}
e^{-i(k_a+k_a,x_2)_\perp-i(k'_a+k'_b,x_1)_\perp}U^{+,b}_{~~i}({\alpha'_a},{k'}_{a_\perp})
\label{calculat1}\\
&&\hspace{-1mm}
\times~V^{-i,a}({\beta'_b},k_{b_\perp})
U^{+,b}_{~j}(\alpha_a,p'_{A_\perp})V^{-j,a}(\beta_b,k'_{b_\perp})
I^{h1}(\beta_b,{\beta'_b}, k_{b_\perp}, k'_{b_\perp}, x_1,x_2)
\nonumber
\ega
where 
\bega
&&\hspace{-1mm}
I^{h1}({\beta'_b}, k_{b_\perp},\beta_b, k'_{b_\perp}, x_1,x_2)~
=~-2
\!\int\!\dhd p_\perp~e^{-i(p,x_{12})_\perp}\!\int_0^\infty\!{\dhd\alpha\over\alpha}\Big(e^{i\alpha\vro x_{12}^-}-1\Big)
\nonumber\\
&&\hspace{-1mm}
\times~\Big\{
{(p+k'_b)_i(p-k_b)_j\Big(e^{-i{\beta'_b}\vro x_{12}^+ +i{(p+k'_b)_\perp^2\over\alpha s}\vro x_{12}^+}
-
e^{i{p_\perp^2\over\alpha s}\vro x_{12}^+}\Big)
\over [\alpha{\beta'_b}s-(p+k'_b)_\perp^2+p_\perp^2+\ie][\alpha({\beta'_b}+\beta_b)s-(p+k'_b)_\perp^2+(p-k_b)_\perp^2+\ie]}
\nonumber\\
&&\hspace{-1mm}
+~{(p+k'_b)_i(p-k_b)_j\Big(e^{i\beta_b\vro x_{12}^+ +i{(p-k_b)_\perp^2\over\alpha s}\vro x_{12}^+}
-e^{i{p_\perp^2\over\alpha s}\vro x_{12}^+}\Big)
\over[\alpha({\beta'_b}+\beta_b)s+(p-k_b)_\perp^2-(p+k'_b)_\perp^2+\ie][\alpha\beta_bs+(p-k_b)_\perp^2-p_\perp^2+\ie]}
\Big\}
\label{ih1}
\ega
The term ``-1'' in the parentheses in the first line comes from subtraction of ``eikonal'' diagrams in Fig. \ref{fig:handeik1}. 
Note that there is no need for the  additional cutoff for $\alpha$ integrals in those diagrams since the integral (\ref{ih1}) is convergent.

Now, in is easy to see that the integral over $p_\perp$ is convergent so the characteristic $p_\perp\sim Q_\perp\sim x_{12_\perp}^{-1}$. Moreover, the integral
over $\alpha$ is convergent at $\alpha\sim{1\over\vro x_{12}^-}\sim{\alpha'_a}$ which means that even at ${\beta'_b}+\beta_b=0$ the 
integral in the  Eq. (\ref{ih1}) is $\sim {Q_\perp^2\over Q^2}$ which is a power correction. Similarly, the fourth term
in Eq. (\ref{realeadingw}) is given by the same set of diagrams with projectile$\leftrightarrow$target reflection so after subtractions
of ``projectile eikonal'' handbag diagrams of Fig. \ref{fig:eik1} g-i it becomes a power correction.

\section{Result for the sum of diagrams in 
Figs. \ref{fig:nlovirt},\ref{fig:nlovirtleft},\ref{fig:real1},\ref{fig:real2} minus TMD matrix elements in Figs. \ref{fig:eik2},\ref{fig:eik1}
\label{sect:sumdiagrams}}

Assembling Eqs.  (\ref{virtotvet}), (\ref{ivirt}),  (\ref{prodotvet}), (\ref{iprod}) 
and subtracting ``eikonal'' TMD matrix elements given by Eq. (\ref{pizweikotvet}),  we get
\bega
&&\hspace{-3mm}
\pizw(x_1,x_2)~-~\pizw^{\rm eik}(x_1,x_2)
~=~~{N_c^2-1\over N_c}8\pi^2g^2
\nonumber\\
&&\hspace{-3mm}
\times~\Big\{
V^{- i,a}(x_2)\langle F^{+a}_{~i}(x_2)F^{- j,b}(x_1) \rangle_{\matA}^{\rm Fig. \ref{fig:real1}}U^{+b}_{~j}(x_1)
+U^{+i,a}(x_2)\langle F^{-,a}_{~i}(x_2)F^{+,b}_{~j}(x_1)\rangle_\matA^{\rm Fig. \ref{fig:real2}} V^{-j,b}(x_1)
\nonumber\\
&&\hspace{-0mm}
+~V^{- i,a}(x_2)U^{+a}_{~i}(x_2)\langle F^{- j,b}(x_1)F^{+b}_{~j}(x_1) \rangle_{\matA}^{\rm Fig. \ref{fig:nlovirt}}
+~\langle F^{- i,a}(x_2)F^{+a}_{~i}(x_2) \rangle_{\matA}^{\rm Fig. \ref{fig:nlovirtleft}}V^{- j,b}(x_1)U^{+b}_{~j}(x_1)
\nonumber\\
&&\hspace{-0mm}
-~U^{+a}_{~i}(x_2)V^{- i,n}(x_2)\langle [x_2^+,-\infty]_{x_{2_\perp}+\delta^-}^{na}[-\infty,x_1^+]_{x_{1_\perp}+\delta^-}^{bc}F^{-j,c}(x_1^+,x_{1_\perp})\rangle_\cala^{\rm Fig.~\ref{fig:eik2}a-c}
U^{+ j,b}(x_1)
\nonumber\\
&&\hspace{-0mm}
-~U^{+a}_{~i}(x_2)\langle F^{-i,n}(x_2^+,x_{2_\perp})[x_2^+,-\infty]_{x_{2_\perp}+\delta^-}^{na}[-\infty,x_1^+]_{x_{1_\perp}+\delta^-}^{bc}\rangle_\cala^{\rm Fig.~\ref{fig:eik2}d-f}V^{-j,c}(x_1)U^{+ j,b}(x_1)
\nonumber\\
&&\hspace{-0mm}
-~U^{+n}_{~i}(x_2)V^{- i,a}(x_2)\langle [x_2^-,-\infty]_{x_{2_\perp}+\delta^+}^{na}[-\infty,x_1^-]_{x_{1_\perp}+\delta^+}^{bc}F^{+j,c}(x_1^-,x_{1_\perp})\rangle_\cala^{\rm Fig.~\ref{fig:eik1}a-c}V^{-j,c}(x_1)
\nonumber\\
&&\hspace{-0mm}
-~V^{- i,a}(x_2)\langle F^{+i,n}(x_2^-,x_{2_\perp})[x_2^-,-\infty]_{x_{2_\perp}+\delta^+}^{na}[-\infty,x_1^-]_{x_{1_\perp}+\delta^+}^{bc}\rangle_\cala^{\rm Fig.~\ref{fig:eik1}d-f}V^{-j,c}(x_1)U^{+b}_{~j}(x_1)\Big\}
\nonumber\\
&&\hspace{-3mm}
=~
\!\int\!\dhd{\alpha'_a}\dhd {k'}_{a_\perp}\dhd{\beta'_b}\dhd k_{b_\perp}\dhd\alpha_a\dhd {k'}_{a_\perp}\dhd\beta_b\dhd {k'}_{b_\perp}
e^{-i{\alpha'_a}\vro x_2^- -i\alpha_a\vro x_1^-}e^{-i{\beta'_b}\vro x_2^+ -i\beta_b\vro x_1^+}
\nonumber\\
&&\hspace{2mm}
\times~e^{-i(k_a+k_a,x_2)_\perp-i(k'_a+k'_b,x_1)_\perp}
U^{+,b}_{~~i}({\alpha'_a},{k'}_{a_\perp})V^{-i,a}({\beta'_b},k'_{b_\perp})
U^{+,b}_{~j}(\alpha_a,k_{a_\perp})V^{-j,a}(\beta_b,k_{b_\perp})
\nonumber\\
&&\hspace{2mm}
\times~g^2[I-I^{\sigma_p,\sigma_t}_{\rm eik}]
({\alpha_a},\alpha'_a,{\beta_b},\beta'_b,k_{a_\perp}, k'_{a_\perp},k_{b_\perp}, k'_{b_\perp},x_1,x_2)
\label{pizwotvet}
\ega
with
\bega
&&\hspace{-1mm}
[I-I^{\sigma_p,\sigma_t}_{\rm eik}]
({\alpha_a},\alpha'_a,{\beta_b},\beta'_b,k_{a_\perp}, k'_{a_\perp},k_{b_\perp}, k'_{b_\perp},x_1,x_2)
\nonumber\\
&&\hspace{-1mm}
=~-I^{\rm d.log}(\alpha_a,\beta_b,k_{a_\perp},k_{b_\perp})-I^{\rm d.log}(\alpha'_a,\beta'_b,k'_{a_\perp},k'_{b_\perp})
+I^{\rm d.log}(\alpha'_a,\beta_b,k'_{a_\perp},k_{b_\perp})
\nonumber\\
&&\hspace{5mm}
+~I^{\rm d.log}({\alpha_a},{\beta'_b},k_{a_\perp},k'_{b_\perp})
+I_1^{\rm rem}({\alpha'_a}, k'_{a_\perp},\beta_b, k_{b_\perp}, x_{12_\perp})
+I_2^{\rm rem}(\alpha_a, k_{a_\perp},{\beta'_b}, k'_{b_\perp}, x_{12_\perp})
\nonumber\\
&&\hspace{5mm}
-~I^{\sigma_p,\sigma_t}_{\rm eik}(\alpha_a,\alpha'_a,\beta_b,{\beta'_b},k_{a_\perp}, k'_{a_\perp},k_{b_\perp}, k'_{b_\perp},x_{12_\perp})
\nonumber\\
&&\hspace{-1mm}
=~-\ln{-Q_{ab}^2\over {k}_{a_\perp}^2}\ln{-Q_{ab}^2\over {k}_{b_\perp}^2}
-\ln{-Q_{a'b'}^2\over {k'}_{a_\perp}^2}\ln{-Q_{a'b'}^2\over {k'}_{b_\perp}^2}
+\ln{-Q_{a'b}^2\over {k'}_{a_\perp}^2}\ln{-Q_{a'b}^2\over {k}_{b_\perp}^2}
\nonumber\\
&&\hspace{5mm}
+~\ln{-Q_{ab'}^2\over {k}_{a_\perp}^2}\ln{-Q_{ab'}^2\over {k'}_{b_\perp}^2}
-\half\Big(\ln{-Q_{a'b}^2x_\perp^2\over 4}+2\gamma\Big)^2-\half\Big(\ln{-Q_{ab'}^2x_\perp^2\over 4}+2\gamma\Big)^2
\nonumber\\
&&\hspace{5mm}
+~\half\ln^2\Big(-{i\over 4}({\alpha'_a}+\ie)\sigma_p s x_\perp^2e^\gamma\Big)
+\half\ln^2\Big(-{i\over 4}(\alpha_a+\ie)\sigma_p s x_\perp^2e^\gamma\Big)
\nonumber\\
&&\hspace{5mm}
+~
\half\ln^2\Big(-{i\over 4}(\beta_b+\ie)\sigma_t s x_\perp^2e^\gamma\Big)
+\half\ln^2\Big(-{i\over 4}(\beta_b+\ie)\sigma_t s x_\perp^2e^\gamma\Big)+\pi^2
\ega
Note that the contribution proportional to integral (\ref{ika}) canceled.
After some algebra this result can be represented as
\bega
&&\hspace{-1mm}
[I-I^{\sigma_p,\sigma_t}_{\rm eik}]({\alpha'_a},\alpha_a,{\beta'_b},\beta_b,{k'}_{a_\perp}, {k'}_{a_\perp},k_{b_\perp}, k'_{b_\perp},x_2,x_1)
\nonumber\\
&&\hspace{11mm}
=~
-\ln{(-i{\alpha'_a}){k'}_{a_\perp}^2\over(-i\alpha_a){k'}_{a_\perp}^2}\ln{(-i{\beta'_b}){k'}_{b_\perp}^2\over(-i\beta_b)k_{b_\perp}^2}
+~\ln^2{x_{12_\perp}^2s\sigma_p\sigma_t\over 4}
\nonumber\\
&&\hspace{18mm}
-~\ln{(-i{\alpha'_a})e^{\gamma}\over\sigma_t}\ln{(-i{\beta'_b})e^{\gamma}\over\sigma_p}
-\ln{(-i\alpha_a)e^{\gamma}\over\sigma_t}\ln{(-i\beta_b)e^{\gamma}\over\sigma_p}+\pi^2
\label{resbezsub}
\ega
However, this formula is not the final result for the coefficient function (\ref{facoord2}) since the integrals 
$I^{\rm virt}$ get contributions form soft/Glauber gluons (sG-gluons) which need to be subtracted. Indeed, the coefficient function
 (\ref{facoord2}) was defined as a result of integration over $C$-fields with $\alpha>\sigma_t$ and $\beta>\sigma_p$. 
 Since we did not impose these restrictions while calculating the loop integrals like Eq. (\ref{dudv}) and Eq. (\ref{iodin}),
 we need to subtract $\alpha<\sigma_t,\beta<\sigma_p$ contributions to these integrals. This will be done in the next Section.

\section{Subtraction of soft/Glauber contributions \label{sec:sgsubtract}} 
As we mentioned above, the coefficient function $\frc_1$ in Eq. (\ref{facoord1}) was defined as  an integral 
over large $|\alpha|>\sigma_t$ and $|\beta|>\sigma_p$ so
the contributions to our background-field diagrams with $|\alpha|<\sigma_t$ and/or $|\beta|<\sigma_p$ should be subtracted
from the result (\ref{resbezsub}). We have already subtracted the TMD matrix elements:  ``target eikonals'' with $|\alpha|<\sigma_t$ and ``projectile eikonals'' with $|\beta|<\sigma_p$. I. Sect. \ref{sec:sgeikonals} below we prove that 
sG-contributions to TMD matrix elements are power corrections so there is no double counting. Still, we need to subtract 
sG-contributions from $\pizw(x_1,x_2)$ itself.

As for the case of  subtractions of eikonal TMD matrix elements, we use smooth $\alpha$ and $\beta$ cutoffs 
which do not change the analytical properties of the diagrams. Let us again start with virtual contributions.

\subsection{sG-contributions to virtual diagrams}
The virtual contribution of diagrams in Fig. \ref{fig:nlovirt}  is given by Eqs. (\ref{virtfig5}) and (\ref{ivirt5}).
Let us now calculate contribution of sG-gluons to virtual diagram. Integral (\ref{ivirt5}) with ``smooth'' $\alpha$ and $\beta$ restrictions has the form
\begin{eqnarray}
&&\hspace{-1mm}
I^{\rm virt~sG}_{\rm Fig. \ref{fig:nlovirt}}~=~-16\pi^2\int\! {\dhd^4p\over i}~\Big[{s{\alpha_a}{\beta_b}\over [(p+k_a)^2+\ie](p^2+\ie)[(p-k_b)^2+\ie]}
\nonumber\\
&&\hspace{-1mm} 
+~\tilde{\delta}_-(p+k_a){s{\alpha_a}{\beta_b}\over p^2-\ie}\tilde{\delta}_+(p-k_b)\Big]
e^{-i{\alpha\over\sigma_t}+i{\beta\over\sigma_p}}
\label{virtsG}
\ega
Note that the choice of signs signs exponential cutoffs in 
sG contributions should be correlated with the choice in TMD matrix elements in order to
have the same sG subtractions in Eq. (4.2).
In addition,
as we will see below, the above choice of signs of the exponential cutoffs agrees with analytical properties of the original uncut diagram, namely that the  background fields 
are emitted before the point $x_1$, see Eq. (\ref{causality}). 

Let us perform the calculation for the most complicated case $\alpha_a,\beta_b<0$ where we need both terms in the r.h.s. of Eq. (\ref{virtsG}).
\begin{eqnarray}
&&\hspace{-1mm}
I^{\rm virt~sG}_{\rm Fig. \ref{fig:nlovirt}}~=~-16\pi^2\!\int{\dhd^4p\over i}e^{-i{\alpha\over\sigma_t}+i{\beta\over\sigma_p}}
\Big\{{\alpha_a\beta_bs
[\alpha(\beta-\beta_b)s-(p-k_b)_\perp^2+\ie]^{-1}\over [(\alpha+\alpha_a)\beta s-(p+k_a)_\perp^2+\ie](\alpha\beta s-p_\perp^2+\ie)
}
\nonumber\\
&&\hspace{-1mm}
+~\tilde{\delta}[(\alpha_a+\alpha)\beta s-(p+k_a)^2]\theta(-\beta){\alpha_a\beta_bs\over \alpha\beta s-p_\perp^2-\ie}
\tilde{\delta}[\alpha(\beta-\beta_b)s-(p-k_b)_\perp^2]\theta(\alpha)\Big\}
\nonumber\\
&&\hspace{-1mm}
\simeq~16\pi^2\int\!{\dhd p\over i}\Big[{\alpha_a\over \alpha_a\beta s-(p+k_a)_\perp^2+\ie}
{s\over \alpha\beta s -p_\perp^2+\ie}{\beta_b\over \alpha\beta_bs+(p-k_b)_\perp^2-\ie}
\nonumber\\
&&\hspace{-1mm}
-~\tilde{\delta}[\alpha_a\beta s-(p+k_a)_\perp^2]\theta(-\beta){\alpha_a\beta_bs\over \alpha\beta s-p_\perp^2-\ie}
\tilde{\delta}[\alpha\beta_bs+(p-k_b)_\perp^2]\theta(\alpha)\Big]e^{-i{\alpha\over\sigma_t}+i{\beta\over\sigma_p}}
\label{virtsG1}
\ega
Here we neglected $\alpha\sim\sigma_t$ in comparison to $\alpha_a$ and $\beta\sim\sigma_p$ in comparison to $\beta_b$. 

Next, we take residue over $\alpha$ and obtain
\bega
&&\hspace{-1mm}
{\rm Eq. ~(\ref{virtsG1})}~\stackrel{\beta_b<0}=~
-8\pi^2\!\int\!\dhd^2p\!\int\!\dhd\beta\bigg[{\alpha_a\beta_bs\theta(\beta)\over  p_\perp^2\beta_b+(p-k_b)_\perp^2\beta-\ie}
{e^{-i{p_\perp^2\over\beta \sigma_ts}+i{\beta\over\sigma_p}}\over \alpha_a\beta s-(p+k_a)_\perp^2+\ie}
\nonumber\\
&&\hspace{-1mm}
-~{\alpha_a\beta_bs\over  p_\perp^2\beta_b+(p-k_b)_\perp^2\beta-\ie}{\theta(\alpha_a)\over \alpha_a\beta s-(p+k_a)_\perp^2+\ie}
e^{i{(p-k_b)_\perp^2\over\beta_b \sigma_ts}+i{\beta\over\sigma_p}}
\nonumber\\
&&\hspace{-1mm}
-2\pi i\theta(-\alpha_a)|\alpha_a\beta_b|s~\tilde{\delta}\big[(p-k_b)_\perp^2\beta-p_\perp^2|\beta_b|\big]
{e^{-i{p_\perp^2\over\beta \sigma_ts}+i{\beta\over\sigma_p}}\over \alpha_a\beta s-(p+k_a)_\perp^2+\ie}
\bigg]
\nonumber\\
&&\hspace{-1mm}
=~-8\pi^2\!\int\!\dhd^2p\!\int\!\dhd\beta{-\alpha_a|\beta_b|s\theta(\beta)\over  -p_\perp^2|\beta_b|+(p-k_b)_\perp^2\beta+\ie}
{1\over \alpha_a\beta s-(p+k_a)_\perp^2+\ie}e^{-i{p_\perp^2\over\beta \sigma_ts}+i{\beta\over\sigma_p}}
\nonumber\\
&&\hspace{-1mm}
\stackrel{\beta=v^2|\beta_b|}=~4\pi\!\int\!\dhd^2p\!\int_0^\infty\!\!dv^2{\alpha_a|\beta_b|s\over  p_\perp^2-(p-k_b)_\perp^2v^2-\ie}
{e^{-i{p_\perp^2\over v^2|\beta_b|\sigma_ts}+iv^2{|\beta_b|\over\sigma_p}}\over (p+k_a)_\perp^2-\alpha_a|\beta_b|sv^2-\ie}
\nonumber\\
&&\hspace{-1mm}
\stackrel{p_\perp=k_\perp v}=~4\pi\!\int\!\dhd^2k\!\int_0^\infty\!\!dv^2{\alpha_a|\beta_b|s\over  k_\perp^2-(k_b-kv)_\perp^2-\ie}
{e^{-i{k_\perp^2\over |\beta_b|\sigma_ts}+iv^2{|\beta_b|\over\sigma_p}}\over (k_a+kv)_\perp^2-\alpha_a|\beta_b|sv^2-\ie}
\label{virtegralsG}
\ega
This  integral can be rescaled by change $l_\perp^2={k_\perp^2\over |\beta_b|\sigma_ts}$ and 
$t^2=v^2{|\beta_b|\over\sigma_p}$ as follows
\bega
-4\pi\!\int\!\dhd^2l_\perp\!\int_0^\infty\!\!dt^2
{e^{-il^2}\over  l_\perp^2-{(k_b-lt\mu_\sigma)_\perp^2\over |\beta_b|\sigma_ts}
-\ie}
{e^{it^2}\over t^2-{(k_a+lt\mu_\sigma)_\perp^2\over \alpha_a\sigma_ps}+\ie}
\label{sgintegral}
\ega
where $\mu_\sigma\equiv \sqrt{\sigma_p\sigma_ts}$. As we assumed, $\mu_\sigma\ll q_\perp$ 
(see Eq. (\ref{musigma})) so one can neglect
$lt\mu_\sigma$ in the denominators and get
\bega
&&\hspace{-1mm}
{\rm Eq. ~(\ref{virtsG1})}~\stackrel{\beta_b<0}=~
-\Big(\ln{-i(\alpha_a+\ie)\sigma_ps\over k_{a_\perp}^2}-\gamma\Big)\Big(\ln{i |\beta_b|\sigma_ts\over k_{b_\perp}^2}-\gamma\Big)
\label{isgotvetodin}
\ega
where we used integral 
\beq
\int_0^\infty\!dx ~{e^{-x}\over x+a}~=~\ln {1\over a}-\gamma~+~O(a)
\label{int7}
\eeq

Performing similar calculation at $\beta_b>0$  we get sG-contribution to the virtual diagram in the form
\bega
&&\hspace{-1mm}
I^{\rm virt~sG}_{\rm Fig. \ref{fig:nlovirt}}~=~
-\Big(\ln{-i(\alpha_a+\ie)\sigma_ps\over k_{a_\perp}^2}-\gamma\Big)\Big(\ln{-i(\beta_b+\ie)\sigma_ts\over k_{b_\perp}^2}-\gamma\Big)
\label{isgotvet1}
\ega
Note that this double-log contribution comes from the region $1\gg l^2\gg{k_{b_\perp}^2\over \sigma_t|{\beta'_b}|s}$ and
$1\gg t^2\gg{{k'}_{a_\perp}^2\over \sigma_p|{\alpha'_a}|s}$ in the integral (\ref{sgintegral}) which corresponds to the region
\beq
\sigma_p\gg\beta\gg {{k'}_{a_\perp}^2\over |{\alpha'_a}|s},~~~~~\sigma_t\gg\alpha\gg {k_{b_\perp}^2\over |{\beta'_b}|s},~~~~~
\sigma_p\sigma_ts\gg p_\perp^2\gg{{k'}_{a_\perp}^2k_{b_\perp}^2\over |{\alpha'_a}{\beta'_b}|s}
\label{charalbevirt}
\eeq
in the original integral (\ref{virtegralsG}).

The result for  Fig. \ref{fig:nlovirtleft} integral (\ref{ivirt6}) is obtained from Fig. \ref{fig:nlovirt} result (\ref{ivirt5}) by complex conjugation
and $k_a\leftrightarrow-k'_a,k_b\leftrightarrow-k'_b$ replacement so the sG-contribution can be obtained in a similar way
\begin{eqnarray}
&&\hspace{-1mm}
I^{\rm virt~sG}_{\rm Fig. \ref{fig:nlovirtleft}}~=~16\pi^2\int\! {\dhd^4p\over i}~
\Big[{s{\alpha'_a}{\beta'_b}\over [(p+k'_a)^2-\ie](p^2-\ie)[(p-k'_b)^2-\ie]}
\nonumber\\
&&\hspace{12mm} 
+~\tilde{\delta}_+(p+k'_a){s{\alpha'_a}{\beta'_b}\over p^2+\ie}\tilde{\delta}_-(p-k'_b)\Big]
e^{-i{\alpha'\over\sigma_t}+i{\beta'\over\sigma_p}}
\nonumber\\
&&\hspace{-1mm}
=~-16\pi^2\int\!{\dhd p\over i}\Big[{\alpha'_a\over \alpha'_a\beta s-(p+k'_a)_\perp^2-\ie}
{s\over \alpha\beta s -p_\perp^2-\ie}{\beta'_b\over \alpha\beta'_bs+(p-k'_b)_\perp^2+\ie}
\nonumber\\
&&\hspace{-1mm}
-~\tilde{\delta}[\alpha'_a\beta s-(p+k'_a)_\perp^2]\theta(\beta){\alpha'_a\beta'_bs\over \alpha\beta s-p_\perp^2+\ie}
\tilde{\delta}[\alpha\beta'_bs+(p-k'_b)_\perp^2]\theta(-\alpha)\Big]e^{-i{\alpha\over\sigma_t}+i{\beta\over\sigma_p}}
\nonumber\\
&&\hspace{12mm} 
=~-\Big(\ln{-i({\alpha'_a}+\ie)\sigma_ps\over {k'}_{a_\perp}^2}-\gamma\Big)\Big(\ln{-i({\beta'_b}+\ie)\sigma_ts\over k_{b_\perp}^2}-\gamma\Big)
\label{virtsG6}
\ega
To get the last line, we performed complex conjugation of Eq. (\ref{virtsG1}), replaced $k_a\rightarrow -k'_a, k_b\rightarrow -k'_b$, and changed the sign of $p$.

\subsection{sG-contributions to production diagrams}
The sG-contribution to the integral (\ref{iodin}) has the form
\bega
&&\hspace{-1mm}
I_1^{\rm sG}({\alpha'_a}, k'_{a_\perp},\beta_b,k_{b_\perp}, x_{12})
~=~8\pi^2s^2\!\int\!\dhd\alpha\dhd\beta\dhd p_\perp~
e^{-i{\alpha\over\sigma_t} +i{\beta\over\sigma_p} -i(p,x_{12})_\perp}
\label{iodinsg}\\
&&\hspace{9mm}
\times~\bigg[{{\alpha'_a}\over {\alpha'_a}\beta s-(p+k'_a)^2+\ie}\tilde{\delta}(\alpha\beta s-p_\perp^2)\theta(\alpha)
{\beta_b\over \alpha\beta_bs+(p-k_b)_\perp^2+\ie}
\nonumber\\
&&\hspace{13mm}
+~\tilde{\delta}[{\alpha'_a}\beta s-(p+k'_a)^2]\theta(\beta){1\over \alpha\beta s-p_\perp^2-\ie}
{\alpha'_a\beta_b\over \alpha\beta_bs+(p-k_b)_\perp^2-\ie\beta_b}
\nonumber\\
&&\hspace{13mm}
+~{-{\alpha'_a}\beta_b
\over {\alpha'_a}\beta s-(p+k'_a)_\perp^2+\ie{\alpha'_a}}{1\over \alpha\beta s-p_\perp^2+\ie}
\tilde{\delta}[\alpha\beta_bs+(p-k_b)_\perp^2]\theta(\alpha)\bigg]
\nonumber
\end{eqnarray}
To get the above equation, we neglected
 $\alpha\sim\sigma_t$ in comparison to ${\alpha'_a}$ and $\beta\sim\sigma_p$ in comparison to ${\beta_b}$ in the denominators in Eq. (\ref{iodin}). Also, in the exponent in Eq. (\ref{iodin}) we neglected 
$\alpha\vro x_{12}^-$ in comparison to ${\alpha\over \sigma_t}=\alpha\vro\delta^-$ and $\beta\vro x_{12}^+$ in comparison to ${\beta\over \sigma_p}=\beta\vro\delta^+$.
It is easy to see that the integral over $\alpha$ in the second term and over $\beta$ in the last term  in the above equation vanish so we get
\bega
&&\hspace{-1mm}
I_1^{\rm sG}({\alpha'_a}, {k'}_{a_\perp},\beta_b, k_{b_\perp}, x_2,x_1)
~=~8\pi^2 s\!\int_0^\infty\!\dhd\beta\dhd p_\perp~
e^{-i{p_\perp^2\over\beta\sigma_ts} +i{\beta\over\sigma_p} -i(p,x_{12})_\perp}
\label{iodinsG}\\
&&\hspace{47mm}
\times~{{\alpha'_a}\beta_bs\over [p_\perp^2\beta_b+(p-k_b)\beta+\ie][{\alpha'_a}\beta s-(p+k'_a)^2+\ie]}
\nonumber
\end{eqnarray}
Let us again consider $\beta_b<0$, then change of variables $\beta=v^2|\beta_b|$ yields 
\bega
&&\hspace{-1mm}
{\rm Eq. ~(\ref{iodinsG})}~\stackrel{\beta_b<0}=~
-4\pi\!\int_0^\infty 
\! dv^2\!\int\!\dhd p_\perp~
e^{-i{p_\perp^2\over v^2|{\beta_b}|\sigma_ts} +iv^2{|{\beta_b}|\over\sigma_p} -i(p,x_{12})_\perp}
\nonumber\\
&&\hspace{20mm}
\times~{{\alpha'_a}|\beta_b|s\over [p_\perp^2-(p-k_b)v^2-\ie][(p+k'_a)^2-{\alpha'_a}|\beta_b|sv^2-\ie]}
\end{eqnarray}
This integral differs from the fifth line in Eq. (\ref{virtegralsG}) by extra factor $e^{-i(p,x_{12})_\perp}$. Doing the same rescaling, we obtain
\bega
&&\hspace{-1mm}
{\rm Eq. ~(\ref{iodinsG})}~\stackrel{\beta_b<0}=~
4\pi\!\int\!\dhd^2l_\perp\!\int_0^\infty\!dt^2
{e^{-it(l,x_{12_\perp})\mu_\sigma}\over  l_\perp^2-{(k'_a-lt\mu_\sigma)_\perp^2\over |{\beta_b}|\sigma_ts}-\ie}
{e^{-il^2+it^2}\over t^2-{(k'_a+lt\mu_\sigma)_\perp^2\over {\alpha'_a}\sigma_ps}+\ie}
\ega
Again, since $\mu_\sigma\ll q_\perp$, we can neglect all $lt\mu_\sigma$ terms and get
\begin{eqnarray}
&&\hspace{-1mm}
I^{\rm sG}_{\rm Fig.\ref{fig:real1}}~=~\Big(\ln{-i({\alpha'_a}+\ie)\sigma_ps\over {k'}_{a_\perp}^2}-\gamma\Big)\Big(\ln{-i(\beta_b+\ie)\sigma_ts\over k_{b_\perp}^2}-\gamma\Big)
\label{iodinsgotvet}
\ega

Similarly to the integral  (\ref{idva}) itself, the sG-contribution to the integral (\ref{idva}) can be obtained from the result for the integral
(\ref{iodinsgotvet}) by trivial projectile$\leftrightarrow$target replacements 
\begin{eqnarray}
&&\hspace{-1mm}
I^{\rm sG}_{\rm Fig.\ref{fig:real2}}~=~\Big(\ln{-i(\alpha_a+\ie)\sigma_ps\over {k'}_{a_\perp}^2}-\gamma\Big)\Big(\ln{-i({\beta'_b}+\ie)\sigma_ts\over k_{b_\perp}^2}-\gamma\Big)
\label{idvasgotvet}
\ega
\subsection{The sum of sG-terms}
Assembling Eqs. (\ref{isgotvet1}), (\ref{virtsG}), (\ref{iodinsgotvet}), and (\ref{idvasgotvet}) we get
\bega
&&\hspace{-1mm}
\pizw(x_1,x_2)^{\rm sG}~
=~~{N_c^2-1\over N_c}16\pi^2g^2\Big\{
V^{- i,a}(x_2)\langle F^{+a}_{~i}(x_2)F^{- j,b}(x_1) \rangle_{\matA}^{\rm Fig. \ref{fig:real1}}U^{+b}_{~j}(x_1)
\nonumber\\
&&\hspace{21mm}
+~U^{+i,a}(x_2)\langle F^{-,a}_{~i}(x_2)F^{+,b}_{~j}(x_1)\rangle_\matA^{\rm Fig. \ref{fig:real2}} V^{-j,b}(x_1)
\nonumber\\
&&\hspace{21mm}
+~V^{- i,a}(x_2)U^{+a}_{~i}(x_2)\langle F^{- j,b}(x_1)F^{+b}_{~j}(x_1) \rangle_{\matA}^{\rm Fig. \ref{fig:nlovirt}}
\nonumber\\
&&\hspace{21mm}
+~\langle F^{- i,a}(x_2)F^{+a}_{~i}(x_2) \rangle_{\matA}^{\rm Fig. \ref{fig:nlovirtleft}}V^{- j,b}(x_1)U^{+b}_{~j}(x_1)\Big\}^{\rm sG}
\nonumber\\
&&\hspace{-1mm}
=~
\!\int\!\dhd{\alpha_a}\dhd k_{a_\perp}\dhd{\beta_b}\dhd k_{b_\perp}\dhd\alpha'_a\dhd k_{a_\perp}\dhd\beta'_b\dhd k'_{b_\perp}
e^{-i{\alpha'_a}\vro x_2^- -i\alpha_a\vro x_1^- -i{\beta'_b}\vro x_2^+ -i\beta_b\vro x_1^+}
\nonumber\\
&&\hspace{-1mm}
\times~
e^{-i(k_a+k_a,x_1)_\perp-i(k'_a+k'_b,x_2)_\perp}
U^{+,b}_{~~i}({\alpha'_a},{k'}_{a_\perp})V^{-i,a}({\beta'_b},k'_{b_\perp})
U^{+,b}_{~j}(\alpha_a,k_{a_\perp})V^{-j,a}(\beta_b,k_{b_\perp})
\nonumber\\
&&\hspace{21mm}
\times~g^2I^{\rm sG}(\alpha_a,\alpha'_a,\beta_b,{\beta'_b},k_{a_\perp}, k'_{a_\perp},k_{b_\perp}, k'_{b_\perp})
\label{pizwotvetsg}
\ega
where 
\bega
&&\hspace{-1mm}
I^{\sigma_p,\sigma_t}_{\rm sG}(\alpha_a,\alpha'_a,\beta_b,{\beta'_b},k_{a_\perp}, k'_{a_\perp},k_{b_\perp}, k'_{b_\perp})~=~
I^{\rm virt~sG}_{\rm Fig. \ref{fig:nlovirt}}+I^{\rm virt~sG}_{\rm Fig. \ref{fig:nlovirtleft}}+I^{\rm sG}_{\rm Fig.\ref{fig:real1}}
+I^{\rm sG}_{\rm Fig.\ref{fig:real2}}
\nonumber\\
&&\hspace{-1mm}
=~-\ln{(-i{\alpha'_a}){k'}_{a_\perp}^2\over (-i\alpha_a){k}_{a_\perp}^2}\ln{(-i{\beta'_b}){k'}_{b_\perp}^2\over (-i\beta_b){k}_{b_\perp}^2}
~+~O\big({\mu_\sigma^2\over Q_\perp^2}\big)
\label{sGotvet}
\ega
It is worth noting that the soft-Glauber contribution (\ref{sGotvet}) is actually a soft contribution with characteristic 
transverse momenta $p_\perp^2\sim {k_{a_\perp}^2k_{b_\perp}^2\over |{\alpha_a}{\beta_b}|s}$. Indeed, while the momenta in the individual 
 integrals for $I^{\rm sG}$ are given by Eq. (\ref{charalbevirt}), the characteristic transverse momenta in their sum (\ref{sGotvet}) are
 of the order of low limit in Eq. (\ref{charalbevirt}). To see that, we rewrite the sum (\ref{sGotvet}) as follows
 \bega
 &&\hspace{-1mm}
 I^{\rm virt~sG}_{\rm Fig. \ref{fig:nlovirt}}+I^{\rm virt~sG}_{\rm Fig. \ref{fig:nlovirtleft}}+I^{\rm sG}_{\rm Fig.\ref{fig:real1}}
+I^{\rm sG}_{\rm Fig.\ref{fig:real2}}
=~-\!\int_0^\infty\!dl^2~ e^{-il^2}
\bigg[{1\over  l_\perp^2-{k_{b_\perp}^2\over |{\beta_b}|\sigma_ts}-\ie}
-{1\over  l_\perp^2-{{k'}_{b_\perp}^2\over |\beta'_b|\sigma_ts}-\ie}\bigg]
\nonumber\\
&&\hspace{22mm}
\times~\!\int_0^\infty\!dt^2~e^{it^2}
\bigg[{1\over t^2-{k_{a_\perp}^2\over {\alpha_a}\sigma_ps}+\ie}
-{1\over t^2-{{k'}_{a_\perp}^2\over \alpha'_a\sigma_ps}+\ie}\bigg]
~+~O\big({\mu_\sigma^2\over Q_\perp^2}\big)
\ega
It is easy to see that the integral over $l^2$ is determined by $l^2\sim {k_{b_\perp}^2\over |{\beta_b}|\sigma_ts}$
and  the integral over $t^2$ by $t^2\sim {k_{a_\perp}^2\over |{\alpha_a}|\sigma_ps}$ which translates to
\beq
\beta\sim {k_{a_\perp}^2\over |{\alpha_a}|s},~~~~~\alpha\sim {k_{b_\perp}^2\over |{\beta_b}|s},~~~~~
p_\perp^2\sim{k_{a_\perp}^2k_{b_\perp}^2\over |{\alpha_a}{\beta_b}|s}
\label{charalbe}
\eeq
which is a soft contribution since Eq. (\ref{charalbe}) means that
\beq
p^+\sim p^- \sim p_\perp\sim O\big({1\over\lambda}\Big)
\label{softcon}
\eeq
in terms of rescaling (\ref{rescale}). Thus, the soft-Glauber contribution (\ref{sGotvet}) is actually a soft contribution 
in accordance with general statement that contributions from Glauber gluons cancel.

Actually, the statement that the soft-Glauber contribution (\ref{sGotvet}) is a soft contribution can  be checked independently. Let us calculate the contribution of small $p_\perp\ll k_{i_\perp}$ to a non-restricted 
integrals of Fig. \ref{fig:nlovirt}. Neglecting $p_\perp$ in comparison to   $k_{i_\perp}$ and using dimensional regularization for UV integrals
obtained as a result of this approximation, we get instead of Eq. (\ref{virtsG1})
\begin{eqnarray}
&&\hspace{-1mm}
I^{\rm virt~soft}_{\rm Fig. \ref{fig:nlovirt}}~=~16\pi^2\int\!{\dhd^{2+2\ve} p\over i}\Big[{\alpha_a\over \alpha_a\beta s-k_{a_\perp}^2+\ie}
{s\over \alpha\beta s -p_\perp^2+\ie}{\beta_b\over \alpha\beta_bs+k_{b_\perp}^2-\ie}
\nonumber\\
&&\hspace{14mm}
-~\tilde{\delta}[\alpha_a\beta s-k_{a_\perp}^2]\theta(-\beta){\alpha_a\beta_bs\over \alpha\beta s-p_\perp^2-\ie}
\tilde{\delta}[\alpha\beta_bs+k_{b_\perp}^2]\theta(\alpha)\Big]
\label{virtsG1e}
\ega
Repeating all the steps in derivation of Eq. (\ref{virtegralsG}) we get
\bega
&&\hspace{-1mm}
{\rm Eq. ~(\ref{virtsG1e})}~\stackrel{\beta_b<0}=~-4\pi\!\int\!\dhd^{2+2\ve}k{1\over  k_\perp^2-k_{b_\perp}^2+\ie}\!\int_0^\infty\!\!dv^2
{v^{2\ve}\over v^2- {k_{a_\perp}^2\over(\alpha_a+\ie)|\beta_b|s}}
\nonumber\\
&&\hspace{-1mm}
=~-{\Gamma^2(-\ve)\Gamma(1+\ve)\over (4\pi)^\ve}(-k_{b_\perp}^2+\ie)^\ve\big(- {k_{a_\perp}^2\over(\alpha_a+\ie)|\beta_b|s}\big)^\ve
\label{virtegralsGe}\\
&&\hspace{-1mm}
=~-{1\over\ve^2}-{\pi^2\over 4}-\gamma^2-\Big({1\over\ve}+\gamma\Big)\big[\ln k_{a_\perp}^2\ln k_{b_\perp}^2-\ln(\alpha_a+\ie)|\beta_b|s\big]-\half\ln^2{(\alpha_a+\ie)|\beta_bs\over k_{a_\perp}^2k_{b_\perp}^2}
\nonumber
\ega
The  similar contribution of diagrams in Fig. \ref{fig:nlovirtleft} is obtained by replacement  ${\alpha_a}\leftrightarrow \alpha'_a$, 
${\beta_b}\leftrightarrow \beta'_b$ and  $k_{a,b_\perp}\leftrightarrow k'_{a,b_\perp}$. Moreover, for soft contributions one can neglect 
$e^{-ipx_{12}}$ in the ``production'' terms and obtain 
\begin{eqnarray}
&&\hspace{-1mm}
I^{\rm soft}~=~{\rm Eq. (\ref{virtegralsGe})}-\big({\alpha_a}\rightarrow \alpha'_a,k_{a_\perp}\rightarrow k'_{a_\perp}\big)
-\big({\beta_b}\rightarrow \beta'_b,k_{b_\perp}\rightarrow k'_{b_\perp}\big)
\\
&&\hspace{14mm}
+~\big({\alpha_a}\rightarrow \alpha'_a,{\beta_b}\rightarrow \beta'_b, k_{a_\perp}\rightarrow k'_{a_\perp}, k_{b_\perp}\rightarrow k'_{b_\perp}\big)~=~-\ln{(i{\alpha'_a}) k^2_{a_\perp}\over (i\alpha_a) {k'}^2_{a_\perp}}
\ln{(i{\beta'_b}) {k'}^2_{b_\perp}\over (i\beta_b) {k'}^2_{b_\perp}}
\nonumber
\ega
which coincides with Eq. (\ref{sGotvet}).

 As we mentioned above, in Appendix \ref{app:softs} it is demonstrated that in the first perturbative order such soft contributions cancel in the sum of all diagrams.  Non-perturbatively,
soft contributions form wave functions of hadrons and also presumably lead to non-perturbative power corrections 
to scattering amplitude $\sim x_{12_\perp}^2\Lambda_{\rm QCD}^2$. 

\subsection{sG-contributions to TMD matrix elements \label{sec:sgeikonals}}

In this section we demonstrate that sG-contributions to TMD matrix elements are power corrections.
Let us start with the ``target eikonals'' of Fig. \ref{fig:eik2} 
given by Eq. (\ref{ieik11ade}). A ``smooth'' cutoff $|\beta|<\sigma_p$ is obtained by inserting an extra 
$e^{i{\beta\over\sigma_p}}=e^{i\beta\vro\delta^+}$ in the integrand
\bega
&&\hspace{-1mm}
I^{{\rm eik},\beta<\sigma_p}_{\rm Fig.  \ref{fig:eik2}a-d}(\beta_b, {k}_{b_\perp}, x_1^+,x_{1_\perp},x_2^+,x_{2_\perp})
~
=~8\pi^2s\!\int_0^\infty\!\dhd\alpha~e^{-i{\alpha\over\sigma_t}+i{\beta\over\sigma_p}}
{\dhd\beta\over\beta+\ie}\dhd p_\perp~
e^{i\beta\vro x_{12}^+ -i(p,x_{12})_\perp}
\nonumber\\
&&\hspace{-1mm}
\times~\Big[\tilde{\delta}(\alpha\beta s-p_\perp^2)
{\beta-\beta_b\over \alpha(\beta-\beta_b)s-(p-k_b)_\perp^2-\ie}
+{(\beta-\beta_b)\over \alpha\beta s-p_\perp^2+\ie}
\tilde{\delta}[\alpha(\beta-\beta_b)s-(p-k_b)_\perp^2]\Big]
\nonumber\\
&&\hspace{-1mm}
-~8\pi^2i\!\int\!\dhd\alpha\dhd\beta\dhd p_\perp
{e^{-i{\alpha\over\sigma_t}+i{\beta\over\sigma_p}}s(\beta-\beta_b)
\over (\beta+\ie)(\alpha\beta s-p_\perp^2+\ie)[\alpha(\beta-\beta_b)s-(p-k_b)_\perp^2+\ie]}
\nonumber\\
&&\hspace{-1mm}
=~8\pi^2
\!\int_0^\infty\!\dhd\alpha~e^{-i{\alpha\over\sigma_t}}\!\int\!\dhd p_\perp 
\bigg(
{{\beta_b}s\over  p_\perp^2}{e^{i{p_\perp^2\over\alpha s}\vro x_{12}^+ -i(p,x_{12})_\perp}-1
\over  \alpha\beta_bs+(p-k_b)_\perp^2+\ie}e^{i{p_\perp^2\over\alpha s}\vro\delta^+}
\nonumber\\
&&\hspace{33mm}
+~{(p-k_b)_\perp^2e^{i(p,x_{12})_\perp}\big[e^{i({\beta_b}+{(p-k_b)_\perp^2\over\alpha s})\vro (x_{12}^+ +\delta^+)}
-e^{i{p_\perp^2\over\alpha s}\vro (x_{12}^+ +\delta^+)}\big]
\over \alpha[\alpha\beta_b s+(p-k_b)_\perp^2+\ie][\alpha\beta_b s+(p-k_b)_\perp^2-p_\perp^2]}\bigg)
\label{ieik11adee}
\end{eqnarray}
Since  $x_{12}^+\ll\delta^+$ we can neglect $x_{12}^+$ and get
\begin{eqnarray}
&&\hspace{-0mm}
I^{\rm eik}_{\rm Fig.~\ref{fig:eik2}a-d}(\beta_b,k_{b_\perp},x_{12_\perp})
~=~
8\pi^2\!\int_0^\infty\!\dhd\alpha\!\int\!{\dhd p_\perp\over  p_\perp^2}
{\beta_bse^{-i{\alpha\over\sigma_t}+i{p_\perp^2\over\alpha\sigma_ps}}\big(e^{-i(p,x_{12})_\perp}-1\big)
\over \alpha\beta_bs+(p-k_b)_\perp^2+\ie}
\nonumber\\
&&\hspace{-0mm}
\stackrel{\alpha=v^2\alpha_a}=~4\pi\!\int_0^\infty\!dv^2\!\int\!{\dhd p_\perp\over  p_\perp^2}
{\alpha_a\beta_bse^{-i{\alpha_a\over\sigma_t}v^2+i{p_\perp^2\over v^2\alpha_a\sigma_ps}}\big(e^{-i(p,x_{12})_\perp}-1\big)
\over v^2\alpha_a\beta_bs+(p-k_b)_\perp^2+\ie}
\nonumber\\
&&\hspace{-1mm}
\stackrel{p_\perp=k_\perp v}=~4\pi\!\int_0^\infty\!\!dv^2\!\int\!{\dhd^2k\over k_\perp^2}\big(e^{-iv(k,x_{12})_\perp}-1\big)
~{\alpha_a\beta_bse^{-i{\alpha_a\over\sigma_t}v^2+i{k_\perp^2\over \alpha_a\sigma_ps}}
\over v^2\alpha_a\beta_bs+(k_\perp v-k_b)_\perp^2+\ie}
\label{ieiktargetsg1}
\end{eqnarray}
Performing the same rescaling as in Eq. (\ref{sgintegral}) we get
\begin{eqnarray}
&&\hspace{-1mm}
I^{\rm eik}_{\rm Fig.~\ref{fig:eik2}a-d}(\beta_b,k_{b_\perp},x_{12_\perp})
~=~4\pi\!\int_0^\infty dt^2\!\int\! {d^2l_\perp\over l_\perp^2}\big(e^{-it\mu_\sigma(l,x_{12})_\perp}-1\big)
{e^{-it^2+il^2}\over t^2+{(k_b-\mu_\sigma tl)_\perp^2\over \sigma_t\beta_bs}}
\nonumber\\
&&\hspace{-1mm}
\simeq~4\pi\!\int_0^\infty dt^2{e^{-it^2}\over t^2+{k_{b_\perp}^2\over \sigma_t\beta_bs}}\!\int\! {d^2l_\perp\over l_\perp^2}\big(e^{-it\mu_\sigma(l,x_{12})_\perp}-1\big)e^{il_\perp^2}
\label{ieiktargetsg2}
\end{eqnarray}
where we again used  $\mu_\sigma\ll q_\perp\sim k_{b_\perp}$. The first integral in the r.h.s. is $\ln{-i\beta_b\sigma_ts\over k_{b_\perp}^2}-\gamma$ (see Eq. (\ref{int7}) but the second is obviously $O(\mu_\sigma^2x_{12_\perp}^2)\sim O\big({\mu_\sigma^2\over q_\perp^2}\big)$ so the sG-contribution to ``target'' eikonal TMD matrix elements is a power correction. Similarly, one 
can demonstrate that the sG-contribution to ``proijectile'' eikonal TMD matrix elements of Fig.  \ref{fig:eik2} is a power correction
$O\big({\mu_\sigma^2\over q_\perp^2}\big)$. Thus, with power accuracy there is no double counting and we should subtract
from the amplitude (\ref{resbezsub}) only the sG-contributions (\ref{sGotvet}).

\section{Result for the coefficient function \label{sec:coefresult}} 
According to Eq. (\ref{facoord2}), the coefficient function is given by the functional integral over central fields
with $\alpha>\sigma_t,\beta>\sigma_p$ minus the eikonal contributions. It is determined by 
\bega
&&\hspace{-6mm}
\pizw(x_1,x_2)~-~\pizw^{\rm sG}(x_1,x_2)~-~\pizw^{\rm eik}(x_1,x_2)
\label{result}\\
&&\hspace{-1mm}
=~
\!\int\!\dhd\alpha_a\dhd k_{a_\perp}\dhd\beta_b\dhd k_{b_\perp}\dhd\alpha'_a\dhd {k'}_{a_\perp}\dhd\beta'_b\dhd {k'}_{b_\perp}
\nonumber\\
&&\hspace{5mm}
\times~e^{-i\alpha_a\vro x_1^- -i\alpha'_a\vro x_2^-}e^{-i\beta_b\vro x_1^+ -i\beta'_b\vro x_2^+}
e^{-i(k_a+k_b,x_1)_\perp-i(k'_a+k'_b,x_2)_\perp}U^{+,b}_{~~i}(\alpha_a,k_{a_\perp})
\nonumber\\
&&\hspace{5mm}
\times~
V^{-i,a}(\beta_b,k_{b_\perp})
U^{+,b}_{~j}(\alpha'_a,k'_{a_\perp})V^{-j,a}(\beta'_b,k'_{b_\perp})
\frc_1(\alpha_a,\alpha'_a,\beta_b,\beta'_b,x_{12_\perp};\sigma_p,\sigma_t)
\nonumber\\
&&\hspace{-1mm}
=~\!\int\!\dhd\alpha_a\dhd\beta_b\dhd\alpha'_a\dhd\beta'_b\dhd {k'}_{b_\perp}
~e^{-i\alpha_a\vro x_1^- -i\alpha'_a\vro x_2^-}e^{-i\beta_b\vro x_1^+ -i\beta'_b\vro x_2^+}
U^{+,b}_{~~i}(\alpha_a,x_{1_\perp})
\nonumber\\
&&\hspace{5mm}
\times~
V^{-i,a}(\beta_b,x_{1_\perp})U^{+,b}_{~j}(\alpha'_a,x_{2_\perp})V^{-j,a}(\beta'_b,x_{2_\perp})
\frc_1(\alpha_a,\alpha'_a,\beta_b,\beta'_b,x_{12_\perp};\sigma_p,\sigma_t)
\nonumber
\ega
where the coefficient function in the momentum representation is
\bega
&&\hspace{-2mm}
\frc_1(\alpha_a,\alpha'_a,\beta_b,\beta'_b,x_{12_\perp};\sigma_p,\sigma_t)~=~I-I^{\sigma_p,\sigma_t}_{\rm eik}-I^{\sigma_p,\sigma_t}_{\rm sG}
\label{momresultat}
\nonumber
\ega
The explicit form of $\frc_1$ is easily found from Eq. (\ref{resbezsub}) and Eq. (\ref{sGotvet})
\bega
&&\hspace{-2mm}
\frc_1(\alpha_a,\alpha'_a,\beta_b,\beta'_b,x_{12_\perp};\sigma_p,\sigma_t)~=~\ln^2{x_{12_\perp}^2s\sigma_p\sigma_t\over 4}
\label{masteresult}\\
&&\hspace{-2mm}
-~\ln{(-i\alpha_a+\epsilon)e^{\gamma}\over\sigma_t}\ln{(-i\beta_b+\epsilon)e^{\gamma}\over\sigma_p}
-\ln{(-i{\alpha'_a}+\epsilon)e^{\gamma}\over\sigma_t}\ln{(-i{\beta'_b}+\epsilon)e^{\gamma}\over\sigma_p}
+\pi^2+O(\lambda_p,\lambda_t)
\nonumber
\ega
This formula is the main technical result of the paper. 

The very important property of the coefficient function $\frc_1$ is that the r.h.s. (\ref{masteresult}) does not actually depend on transverse momenta 
so all the dynamics at the one-loop level proceeds in the longitudinal direction. This fact can be used to check the algebra and approximations leading to the result (\ref{masteresult}). In the Appendix \ref{onmashell} the coefficient function is calculated using the on-shell
background fields with zero transverse momenta 
\begin{eqnarray}
&&\hspace{-1mm} 
U^{+ i}(z^-)~=~\!\int\!\dhd\alpha_a~U^{+ i}(\alpha_a)~e^{-i\varrho\alpha_az^- }
~~~~\Leftrightarrow~~~~
U^{+i}(\alpha_a)~=~\varrho\!\int\! dz^- dz_\perp~U^{+i}(z^-)~
e^{i\varrho\alpha_az^-} ,
\nonumber\\
&&\hspace{-1mm}
V^{- i}(z^+)
~=~\!\int\!\dhd\beta_b~V^{- i}(\beta_b)~e^{-i\varrho\beta_bz^+}~~~~\Leftrightarrow~~~~
V^{-i}(\beta_b)~=~\varrho\!\int\! dz^+ dz_\perp~V^{-i}(z^+)~e^{i\varrho\beta_bz^+ }
\nonumber\\
\label{bezmassi}
\end{eqnarray}
and the result (\ref{masteresult}) is confirmed. 

In the coordinate space our result reads
\begin{eqnarray}
&&\hspace{-1mm}
\langle \hatW(x_1,x_2)\rangle_\matA
~=~\langle \hacalo_{ij}^{\sigma_p}(x_2^-,x_{2_\perp};x_1^-,x_{1_\perp})
\hacalo^{ij;\sigma_t}(x_2^+,x_{2_\perp};x_1^+,x_{1_\perp})\rangle_\matA
\nonumber\\
&&\hspace{5mm}
+~\int\! dz_2^-dz_1^-dw_1^+dw_2^+
{\alpha_sN_c\over 2\pi}\frc_1(x_{2_\perp},x_{1_\perp};z_2^-,z_1^-,z_2^+,z_1^+;\sigma_p,\sigma_t)
\nonumber\\
&&\hspace{11mm}
\times~U^{+,b}_{~~i}(z_2^-,x_{2_\perp})V^{-i,a}(z_2^+,x_{2_\perp})
U^{+,b}_{~j}(z_1^-,x_{1_\perp})V^{-j,a}(z_1^+,x_{1_\perp})
~+...
\label{facoordi1}
\end{eqnarray}
where
\begin{eqnarray}
&&\hspace{-6mm}
\frc_1(x_{2_\perp},x_{1_\perp};z_2^-,z_1^-,z_2^+,z_1^+;\sigma_p,\sigma_t)~
\label{cooresult}\\
&&\hspace{-6mm}
=~\Big[\ln^2{2\delta^+\delta^-\over x_{12_\perp}^2}+\pi^2\Big]
\delta(x_2-z_2)^-\delta(x_1-z_1)^-\delta(x_2-z_2)^+\delta(x_1-z_1)^+
\nonumber\\
&&\hspace{-1mm}
-~\delta(x_1-z_1)^-\delta(x_1-z_1)^+
\nonumber\\
&&\hspace{7mm}
\times~\Big[{\theta(x_2-z_2)^-\over (x_2-z_2)^-}-\delta(x_2-z_2)^-\!\int_0^{\delta^-}{dz_2^-\over z_2^-} \Big]
 \Big[{\theta(x_2-z_2)^+\over (x_2-z_2)^+}-\delta(x_2-z_2)^+\!\int_0^{\delta^+}{dz_2^+\over z_2^+} \Big]
 \nonumber\\
&&\hspace{-1mm}
-~\delta(x_2-z_2)^-\delta(x_2-z_2)^+
\nonumber\\
&&\hspace{7mm}
\times~ \Big[{\theta(x_1-z_1)^-\over (x_1-z_1)^-}-\delta(x_1-z_1)^-\!\int_0^{\delta^-}{dz_1^-\over z_1^-} \Big]
 \Big[{\theta(x_1-z_1)^+\over (x_1-w_2)^+}-\delta(x_1-z_1)^+\!\int_0^{\delta^+}{dz_1^+\over z_1^+} \Big]
 \nonumber
\ega

Let us check matching of the cutoffs, namely that the r.h.s. of Eq. (\ref{facoordi1}) does not depend on $\sigma_p$ and $\sigma_t$.  We start with $\sigma_t{d\over d\sigma_t}=-\delta^-{d\over\delta^-}$. Since 
\begin{eqnarray}
&&\hspace{-1mm}
-\delta^-{d\over\delta^-}\frc_1(x_{2_\perp},x_{1_\perp};z_i^-,z_{i_\perp},z_i^+,z_{i_\perp};\sigma_p,\sigma_t)~
\label{dercooresult}\\
&&\hspace{-1mm}
=~-\delta(x_2-z_2)^-\delta(x_1-z_1)^-\Big\{2\ln{2\delta^+\delta^-\over x_{12_\perp}^2}
\delta(x_2-z_2)^+\delta(x_1-z_1)^+
 \nonumber\\
&&\hspace{11mm}
+~
\delta(x_1-z_1)^+
 \Big[{\theta(x_2-z_2)^+\over (x_2-z_2)^+}-\delta(x_2-z_2)^+\!\int_0^{\delta^+}{dz_2^+\over z_2^+} \Big]
 \nonumber\\
&&\hspace{22mm}
+~
\delta(x_2-z_2)^+
 \Big[{\theta(x_1-z_1)^+\over (x_1-z_1)^+}-\delta(x_1-z_1)^+\!\int_0^{\delta^+}{dz_1^+\over z_1^+} \Big]\Big\}
 \nonumber\\
 &&\hspace{-1mm}
=~\delta(x_2-z_2)^-\delta(x_1-z_1)^-\Big\{2\ln{sx_{12_\perp}^2\over 4}
\delta(x_2-z_2)^+\delta(x_1-z_1)^+
 \nonumber\\
&&\hspace{11mm}
-~
\delta(x_1-z_1)^+
 \Big[{\theta(x_2-z_2)^+\over (x_2-z_2)^+}-\delta(x_2-z_2)^+\!\int_0^{2\over s\delta^-}{dz_2^+\over z_2^+} \Big]
 \nonumber\\
&&\hspace{22mm}
-~
\delta(x_2-z_2)^+
 \Big[{\theta(x_1-z_1)^+\over (x_1-z_1)^+}-\delta(x_1-z_1)^+\!\int_0^{2\over s\delta^-}{dz_1^+\over z_1^+} \Big]\Big\}
 \nonumber
\ega
we get 
\begin{eqnarray}
&&\hspace{-1mm}
\sigma_t{d\over d\sigma_t}\big[{\rm r.h.s.~of~Eq.~(\ref{facoordi1})}\big]~=~
U^{+,b}_{~~i}(x_2^-,x_{2_\perp})U^{+,b}_{~j}(x_1^-,x_{1_\perp})
 \label{chekevol}\\
&&\hspace{-1mm}
\times~\bigg[\sigma_t{d\over d\sigma_t}\langle\hacalo^{ij;\sigma_t}(x_2^+,x_{2_\perp};x_1^+,x_{1_\perp})\rangle_B
+{\alpha_sN_c\over 2\pi}\Big\{2\ln{sx_{12_\perp}^2\over 4}V^{-i,a}(x_2^+,x_{2_\perp})V^{-j,a}(x_1^+,x_{1_\perp})
 \nonumber\\
&&\hspace{3mm}
-~
\int\! dz_2^+
 \Big[{\theta(x_2-z_2)^+\over (x_2-z_2)^+}-\delta(x_2-z_2)^+\!\int_0^{\sigma_t/\vro}{dz_2^+\over z_2^+} \Big]
 V^{-,b}_{~~i}(z_2^+,x_{2_\perp})V^{-,b}_{~j}(x_1^+,x_{1_\perp})
 \nonumber\\
&&\hspace{3mm}
-~
\int\! dz_1^+ \Big[{\theta(x_1-z_1)^+\over (x_1-z_1)^+}-\delta(x_1-z_1)^+\!\int_0^{\sigma_t/\vro}{dz_1^+\over z_1^+} \Big]
V^{-,b}_{~~i}(x_2^+,x_{2_\perp})V^{-,b}_{~j}(z_1^+,x_{1_\perp})\Big\}\bigg]~=~0
 \nonumber
\end{eqnarray}
due to Eq. (\ref{evoleqtarcoor}).  
Similarly, the r.h.s. of Eq. (\ref{facoordi1}) does not depend on $\sigma_p$ cutoff.

It is possible to represent our result for the coefficient function as  evolution equations with respect to $\sigma_t$ and 
$\sigma_p$. Since the differentiation over $\sigma_t$ is represented by the integration operator in the coordinate space 
and by simple multiplication in the momentum space, we will use the latter. We define the Fourier transform of the operator
$\hatW(x_1,x_2)$ in Eq. (\ref{operw}) as follows
\bega
&&\hspace{0mm}
\hatW({\alpha'_a},\alpha_a,{\beta'_b},\beta_b,x_{1_\perp},x_{2_\perp})~
\\
&&\hspace{0mm}
=~\rho^4\int\! dx_2^-dx_1^-dx_2^+dx_1^+ ~e^{i{\alpha'_a}\vro x_2^- +i{\alpha'_a}\vro x_2^-+i{\beta'_b}\vro x_2^+ +i{\beta'_b}\vro x_2^+}
\hatW(p_A,p'_A,p_B,p'_B;x_{1_\perp},x_{2_\perp})
\nonumber
\end{eqnarray}
The general TMD factorization formula (\ref{facoord}) for  $\hatW({\alpha'_a},\alpha_a,{\beta'_b},\beta_b,x_{1_\perp},x_{2_\perp})$  can be written as
\begin{eqnarray}
&&\hspace{-0mm}
\hatW({\alpha'_a},\alpha_a,{\beta'_b},\beta_b,x_{1_\perp},x_{2_\perp})
~=~\int\! \dhd{\alpha'_a} \dhd\alpha_a\dhd{\beta'_b}\dhd\beta_b
~\frc(x_{1_\perp},x_{2_\perp};{\alpha'_a},\alpha_a,{\beta'_b},\beta_b;\sigma_p,\sigma_t)
\nonumber\\
&&\hspace{5mm}
\times~\hacalo_{ij}^{\sigma_p}({\alpha'_a},\alpha_a,x_{2_\perp},x_{1_\perp}) 
\hacalo^{ij;\sigma_t}({\beta'_b},\beta_b,x_{2_\perp},x_{1_\perp})~+...
\label{facmomentum}
\end{eqnarray}
where 
\begin{eqnarray}
&&\hspace{-0mm}
\hacalo_{ij}^{\sigma_p}({\alpha'_a},\alpha_a,x_{2_\perp},x_{1_\perp}) 
~\equiv~\rho^2\!\int\! dx_2^-dx_1^-e^{i{\alpha'_a}\vro x_2^- +i\alpha_a\vro x_1^-}
\hacalo_{ij}^{\sigma_p}(x_2^-,x_{2_\perp};x_1^-,x_{1_\perp})
\nonumber\\
&&\hspace{-0mm}
\hacalo_{ij}^{\sigma_t}({\beta'_b},\beta_b,x_{2_\perp},x_{1_\perp}) 
~\equiv~\rho^2\!\int\! dx_2^+dx_1^+e^{i{\beta'_b}\vro x_2^+ +i\beta_b\vro x_1^+}
\hacalo_{ij}^{\sigma_t}(x_2^+,x_{2_\perp};x_1^+,x_{1_\perp})
\end{eqnarray}
Here we took into account the absence of dynamics in the transverse space (and tacitly assumed that such 
property survives in higher orders of perturbation theory).  Since the evolution equations for TMD operators
in the momentum space are given by Eqs. (\ref{evoleqtar}) and (\ref{evoleqproj}), the coefficient function should satisfy matching evolution equations
\bega
&&\hspace{-0mm}
\sigma_t{d\over d\sigma_t}\frc(x_{1_\perp},x_{2_\perp};{\alpha'_a},\alpha_a,{\beta'_b},\beta_b;\sigma_p,\sigma_t)
=~{\alpha_sN_c\over 2\pi}\Big[2\ln{sx_{12_\perp}^2\over 4}~\label{evoleqcoeff}\\
&&\hspace{5mm}
+~\ln(-i{\beta'_b}\sigma_t+\epsilon)+\ln(-i\beta_b\sigma_t+\epsilon)+2\gamma\Big]
\frc(x_1,x_2;{\alpha'_a},\alpha_a,{\beta'_b},\beta_b;\sigma_p,\sigma_t)
\nonumber\\
&&\hspace{-0mm}
\sigma_p{d\over d\sigma_p}
\frc(x_{1_\perp},x_{2_\perp};{\alpha'_a},\alpha_a,{\beta'_b},\beta_b;\sigma_p,\sigma_t)
=~{\alpha_sN_c\over 2\pi}\Big[2\ln{sx_{12_\perp}^2\over 4}
\nonumber\\
&&\hspace{5mm}
+~\ln(-i{\alpha'_a}\sigma_p+\epsilon)+\ln(-i\alpha_a\sigma_p+\epsilon)+2\gamma\Big]
~\frc(x_{1_\perp},x_{2_\perp};{\alpha'_a},\alpha_a,{\beta'_b},\beta_b;\sigma_p,\sigma_t)
 \nonumber
\ega
The solution of this equations compatible with first-order result (\ref{masteresult}) is
\bega
&&\hspace{-0mm}
\frc(x_{1_\perp},x_{2_\perp};{\alpha'_a},\alpha_a,{\beta'_b},\beta_b;\sigma_p,\sigma_t)
~=~e^{{\alpha_sN_c\over 2\pi}
\frc_1(x_{12_\perp},{\alpha'_a},\alpha_a,{\beta'_b},\beta_b;\sigma_p,\sigma_t)}
\nonumber\\
&&\hspace{5mm}
+~O\bigg(\alpha_s^2\times\Big[\ln{{\alpha'_a}\over\sigma_t}\ln{{\beta'_b}\over\sigma_p},
\ln{{\alpha'_a}\over\sigma_t},\ln{{\beta'_b}\over\sigma_p},{\rm const}\Big]
\bigg)
\label{frecodin}
\ega
and the final form of ``double operator expansion'' reads
\begin{eqnarray}
&&\hspace{-0mm}
\hatW({\alpha'_a},\alpha_a,{\beta'_b},\beta_b,x_{1_\perp},x_{2_\perp})
~=~\int\! \dhd{\alpha'_a} \dhd\alpha_a\dhd{\beta'_b}\dhd\beta_b
~e^{{\alpha_sN_c\over 2\pi}
\frc_1(x_{12_\perp},{\alpha'_a},\alpha_a,{\beta'_b},\beta_b;\sigma_p,\sigma_t)}
\nonumber\\
&&\hspace{5mm}
\times~\hacalo_{ij}^{\sigma_p}({\alpha'_a},\alpha_a,x_{2_\perp},x_{1_\perp}) 
\hacalo^{ij;\sigma_t}({\beta'_b},\beta_b,x_{2_\perp},x_{1_\perp})~+...
\label{factoper}
\end{eqnarray}
In the next Section we will consider matrix elements of the operator equation (\ref{factoper}) 
between initial and final protons' states and demonstrate that 
\beq
\langle p'_A,p'_B|\hacalo_{ij}^{\sigma_p}\hacalo^{ij;\sigma_t}|p_A,p_B\rangle
~=~\langle p'_A|\hacalo_{ij}^{\sigma_p}|p_A\rangle\langle p'_B|\hacalo^{ij;\sigma_t}|p_B\rangle
\eeq
To prove the above equation, we need to check that the contribution of sG-gluons cancel up to
power corrections terms. 

\subsection{Factorization of integral over $A\cup B$ fields \label{sec:unravel}}
\subsubsection{Cancellation of soft and Glauber gluons \label{sec:sGcancel}}
The functional integral form of our result for hadronic tensor (\ref{WC}) reads
\begin{eqnarray}
&&\hspace{-2mm}
{1\over 16}(N_c^2-1)
\rho^4\int\! dx_1^-dx_2^-dx_1^+dx_2^+ 
~e^{i\alpha_a\vro x_1^- +i{\alpha'_a}\vro x_2^-  +i\beta_b\vro x_1^+ +i{\beta'_b}\vro x_2^+}
\label{funfacoord1}\\
&&\hspace{-2mm}
\times~
\langle p'_A,p'_B| g^2F^a_{\mu\nu}F^{a\mu\nu}(x_2)  g^2F^b_{\lambda\rho}F^{b\lambda\rho}(x_1)
|p_A,p_B\rangle
~
=~
e^{{\alpha_sN_c\over 2\pi}
\frc_1(x_{12_\perp},{\alpha_a},\alpha'_a,{\beta_b},\beta'_b;\sigma_p,\sigma_t)}
\nonumber\\
&&\hspace{-2mm}
\times~
\!\int\!\scrad\Phi_\scra~\Psi^\ast_{p'_A}(t_i)\Psi_{p_A}(t_i)
\Psi^\ast_{p'_B}(t_i)\Psi_{p_B}(t_i)
\hacalo_{ij}^{\sigma_p}({\alpha'_a},\alpha_a,x_{12_\perp})
\hacalo^{ij;\sigma_t}({\beta'_b},\beta_b,x_{12_\perp})~+~...
\nonumber
\end{eqnarray}
where the TMD operators $\hacalo_{ij}^{\sigma_p}({\alpha'_a},\alpha_a,x_{12_\perp})$ 
and  $\hacalo_{ij}^{\sigma_t}(\beta'_b,\beta_b,x_{12_\perp})$ are made of $A$ and $B$  fields, respectively. 
However, as we mentioned above (see Fig. \ref{fig:albe1}), there are $\scras=A\cap B$ fields with both $\alpha<\sigma_t$ and $\beta<\sigma_p$ 
so to get the desired factorization (\ref{facoord}) we need to discuss  the interactions between  $A$ and  $B$ fields.
The integrals over $A$ and $B$ fields give matrix elements of TMD operators (\ref{kalo}) between projectile and target fields
while the integral over $\scras$ fields cancels due to unitarity with power corrections accuracy. To understand this, let us discuss the 
$\scras$ fields which are defined as gluons (and, in principle, quarks) with  both $|\alpha|<\sigma_t$ and $|\beta|<\sigma_p$, see Fig. \ref{fig:albe1}. 
As we mentioned above, depending on the scale of characteristic
transverse momenta they may be Glauber gluons with $p_\perp\sim q_\perp$ or soft  gluons with $p_\perp\ll q_\perp$).  

The cancellation of Glauber gluons is proved in Ref. \cite{Collins:1984kg} (see also Ref. \cite{Diehl:2015bca} for recent discussion). 
\footnote{The discussion of this cancellation in the functional-integral language used in this paper
is presented in  Refs. \cite{Balitsky:2017flc,Balitsky:2017gis}
}  As to soft gluons, it is demonstrated in Refs. \cite{Collins:2011zzd,Echevarria:2011epo,Manohar:2006nz} that they form the correlation function of 
four Wilson lines going form points $x_1$ and $x_2$ in the light-like directions so the result of the 
integration over soft gluons in Eq.  (\ref{funfacoord1}) is
\begin{eqnarray}
&&\hspace{-1mm}
\!\int\!\scrad\Phi_\scra~\Psi^\ast_{p'_A}(t_i)\Psi_{p_A}(t_i)
\Psi^\ast_{p'_B}(t_i)\Psi_{p_B}(t_i)
\hacalo_{ij}^\scra(x_2^+,x_{2_\perp};x_1^+,x_{1_\perp})\hacalo_{ij}^\scra(x_2^-,x_{2_\perp};x_1^-,x_{1_\perp})
\nonumber\\
&&\hspace{-1mm}
=~S(x_2,x_1;\sigma_p,\sigma_t)
\int\!\scrad\Phi_A~\Psi^\ast_{p'_A}(t_i)\Psi_{p_A}(t_i)
\hacalo_{ij}^A(x_2^+,x_{2_\perp};x_1^+,x_{1_\perp})
\\
&&\hspace{33mm}
\times~
\int\!\scrad\Phi_B \Psi^\ast_{p'_B}(t_i)\Psi_{p_B}(t_i)
\hacalo_{ij}^B(x_2^-,x_{2_\perp};x_1^-,x_{1_\perp})
\nonumber\\
&&\hspace{-1mm}
=~S(x_2,x_1;\sigma_p,\sigma_t)\langle p'_A|\hacalo_{ij}(x_2^+,x_{2_\perp};x_1^+,x_{1_\perp})|p_A\rangle
\langle p'_B|\hacalo_{ij}(x_2^-,x_{2_\perp};x_1^-,x_{1_\perp})|p_B\rangle
\nonumber
\ega
where
\begin{equation}
\hspace{0mm}
S(x_2,x_1;\sigma_p,\sigma_t)~=~{1\over N_c^2-1}{\rm Tr}\!\int\! \scrad\scras ~
 \{-\infty^+,x_2^+\}_{x_2}\{x_2^+,-\infty^+\}_{x_2}[-\infty^+,z_1^+]_{z_1}[z_1^+,-\infty^+]_{z_1}
 \label{sGfactor}
\end{equation}
 is the soft factor. Here
$\scrad\scras$ is defined as in Eq. (\ref{deFi}) with $\scra\rightarrow\scras$ replacement.  As we will see in Sect. \ref{app:softs},
with the rapidity-only cutoff the dependence 
of the soft factor on $\sigma_p$ and $\sigma_t$ gives 
power corrections $\sim {Q_\perp^2\over\sigma_ps}$ and/or ${Q_\perp^2\over\sigma_ts}$ hence, contrary to CSS approach 
based on double UV+rapidity cutoff,  there is no logarithmic dependence on the cutoffs in the soft factor. Of course, there may
be the non-perturbative power corrections $\sim \Lambda_{\rm QCD}^2x_{12_\perp}^2$ which should be studied by some non-perturbative methods,
but the claim is that the soft factor with rapidity-only 
regularization does not have perturbative contributions which can mix with the TMD evolution.

Thus, we can neglect the  integration over $\scras$ fields in Eq. (\ref{funfacoord1}) and get the factorized result
\begin{eqnarray}
&&\hspace{-2mm}
\rho^4\int\! dx_2^-dx_1^-dx_2^+dx_1^+ ~e^{i{\alpha'_a}\vro x_2^- +i\alpha_a\vro x_1^-+i{\beta'_b}\vro x_2^+ +i\beta_b\vro x_1^+}
W(p_A,p_B,p'_A,p'_B; x_1,x_2)
\label{factoriz}\\
&&
\hspace{-2mm}
=~
e^{{\alpha_sN_c\over 2\pi}
\frc_1(x_{12_\perp},{\alpha'_a},\alpha_a,{\beta'_b},\beta_b;\sigma_p,\sigma_t)}
\langle p'_A|\hacalo_{ij}^{\sigma_p}({\alpha'_a},\alpha_a,x_{12_\perp})|p_A\rangle
\langle p'_B|\hacalo^{ij;\sigma_t}({\beta'_b},\beta_b,x_{12_\perp})|p_B\rangle
\nonumber
\end{eqnarray}

\subsubsection{Factorization in terms of generalized TMDs \label{sec:gTMDs}}
Let us rewrite our result in terms of generalized TMDs (gTMDs). They can be defined as
follows \cite{Lorce:2013pza}
\begin{eqnarray}
&&\hspace{-5mm}
\pizg_{ij}^{\sigma_p}(x_A,b_\perp;p'_A,p_A)~
=~-{g^{-2}\over\pi\vro x_A}\!\int\!
dz^-e^{ix_A\vro z^-}\langle p'_A|\hacalo_{ij}^{\sigma_p}(-{z^-\over 2}-{b_\perp\over 2},{z^-\over 2}+{b_\perp\over 2}\big)|p_A\rangle,
\nonumber\\
&&\hspace{-5mm}
\pizg_{ij}^{\sigma_t}(x_B,b_\perp;p'_B,p_B)~
=~-{g^{-2}\over\pi\vro x_A}\!\int\!
dz^-e^{ix_B\vro z^-}\langle p'_B|\hacalo_{ij}^{\sigma_p}(-{z^-\over 2}-{b_\perp\over 2},{z^-\over 2}+{b_\perp\over 2}\big)|p_B\rangle
\label{gtmdnorm}
\end{eqnarray}
The above choice of  normalization  reproduces 
 gluon TMDs for unpolarized hadrons defined in Ref. \cite{Mulders:2000sh} at $p'_A=p_A$.
\begin{eqnarray}
&&\hspace{-1mm}
\langle p_A|\hacalo_{ij}^{\sigma_p}(z^-,0^-,b_\perp) |p_A\rangle~
=~-g^2\vro^2\!\int_0^1\! du~u \pizg_{ij}^{\sigma_p}(u,b_\perp)\cos u\vro z^-,
\nonumber\\
&&\hspace{-1mm}
\langle p_A|\hacalo_{ij}^{\sigma_p}(\alpha_q,b_\perp) |p_A\rangle~
\equiv~\rho\!\int\! dz^-e^{i\alpha_q\vro z^-}\langle p_A|\hacalo_{ij}^{\sigma_p}(z^-,0^-,b_\perp) |p_A\rangle
\nonumber\\
&&\hspace{35mm}
=~-\pi g^2\vro^2|\alpha_q|\pizg_{ij}^{\sigma_p}(|\alpha_q|,b_\perp,p_A),
\nonumber\\
&&\hspace{-1mm}
\pizg_{ij}^{\sigma_p}(u,b_\perp)~=~g_{ij}D_g(u,b_\perp;\sigma_p)
+{1\over 2m_N^2}(2\partial_i\partial_j+g_{ij}\partial_\perp^2)H(u,b_\perp;\sigma_p)
\label{operde}
\end{eqnarray}
and similarly for $\langle p_B|\hacalo_{ij}^{\sigma_t}(\beta_q,b_\perp) |p_B\rangle$. 
Note that
at $b_\perp=0$ the TMD $D_g(x_B,\sigma)$ is the  gluon PDF with the rapidity-only cutoff discussed
in Ref. \cite{Balitsky:2015qba}. At the leading order, this is equivalent to usual UV regularization 
of light-ray operator
$\hacalo_{ij}^{\sigma}(z^\pm)$ and
reproduces LO DGLAP equation \cite{Balitsky:2015qba}. At the 
NLO level,  the two-loop DGLAP equation should be reproduced by the combination of rapidity-only evolution of light-ray operator
$\hacalo_{ij}^{\sigma}(z^\pm)$ and usual $\mu^2$ evolution for self-energy and vertex $Z$-factors.

With the normalization (\ref{gtmdnorm}) we get
\begin{eqnarray}
&&\hspace{-1mm}
\langle p'_A|\hacalo_{ij}^{\sigma_p}(\alpha'_a,\alpha_a,x_{2_\perp},x_{1_\perp})|p_A\rangle
\nonumber\\
&&\hspace{11mm}
=~-2\pi^2\delta(\alpha_a+\alpha'_a)g^2\rho^2|\alpha_a|e^{-{i\over 2}(l,x_1+x_2)_\perp}
\pizg_{ij}^{\sigma_p}(|\alpha_a|,x_{12_\perp};p_A,p'_A)
\nonumber\\
&&\hspace{-1mm}
\langle p'_B|\hacalo_{ij}^{\sigma_t}(\beta'_b,\beta_b,x_{2_\perp},x_{1_\perp})  |p_B\rangle
\nonumber\\
&&\hspace{11mm}
=~-2\pi^2\delta(\beta_b+\beta'_b)g^2\rho^2|\beta_b|e^{{i\over 2}(l,x_1+x_2)_\perp}
\pizg_{ij}^{\sigma_t}(|\beta_b|,x_{12_\perp};p_B,p'_B)
\end{eqnarray}
The $\delta$-functions in the above expressions for $\langle p'_A|\hacalo_{ij}^{\sigma_p} |p_A\rangle$ and
$\langle p'_B|\hacalo_{ij}^{\sigma_tp}|p_B\rangle$ are present also in the l.h.s. of Eq. (\ref{factoriz}) because
$p'_A+p'_B=p_A+p_B$. Canceling them, one obtains
\begin{eqnarray}
&&\hspace{-2mm}
\int\! dx_{12}^-dx_{12}^+ ~e^{i\alpha_a\vro x_{12}^- +i\beta_b\vro x_{12}^+}
{N_c^2-1\over 16}\langle p'_A,p'_B|F^2\big(-{x_{12}\over 2}\big)F^2\big({x_{12}\over 2}\big)|p_A,p_B\rangle
\label{factoriza}\\
&&
\hspace{-2mm}
=~{\pi^2\over 2}Q^2e^{{\alpha_sN_c\over 2\pi}\big[
\ln^2{b_\perp^2s\sigma_p\sigma_t\over 4}
-2\ln{\alpha_ae^\gamma\over\sigma_t}\ln{\beta_be^\gamma\over\sigma_p}+{\pi^2\over 2}\big]}
\pizg_{ij}^{\sigma_p}(\alpha_a,x_{12_\perp};p_A,p'_A)\pizg^{ij;\sigma_t}(\beta_b,x_{12_\perp};p_B,p'_B)
\nonumber
\end{eqnarray}
This is the final formula for rapidity-only TMD factorization of hadronic tensor. 

\section{Conclusions and outlook.}
In conclusion let us present our final formula (\ref{factoriza}) for the practical case of  hadronic tensor (\ref{W}) 
which corresponds to ``forward'' matrix element  with $p'_A=p_A$ and $p'_B=p_B$. It reads
\begin{eqnarray}
&&\hspace{-2mm}
W(p_A,p_B;q)~=~\int\! db_\perp~e^{i(q,b)_\perp}
W(p_A,p_B;\alpha_q,\beta_q,b_\perp),
\nonumber\\
&&\hspace{-1mm}
W(p_A,p_B;\alpha_q,\beta_q,b_\perp)
~=~{\pi^2\over 2}Q^2\pizg_{ij}^{\sigma_p}(\alpha_q,b_\perp;p_A)\pizg^{ij;\sigma_t}(\beta_q,b_\perp;p_B)
\nonumber\\
&&\hspace{36mm}
\times~\exp\Big\{{\alpha_sN_c\over 2\pi}\Big[
\ln^2{b_\perp^2s\sigma_p\sigma_t\over 4}
-2\big(\ln{\alpha_q\over\sigma_t}+\gamma\big)\big(\ln{\beta_q\over\sigma_p}+\gamma\big)+{\pi^2\over 2}\Big]\Big\}
\nonumber\\
&&\hspace{36mm}
+~{\rm NLO~ terms}\sim O\big(\alpha_s^2)~+~{\rm power~corrections}
\label{momentumresult}
\end{eqnarray}
where gluon TMDs $\pizg_{ij}^{\sigma_p}(\alpha_q,b_\perp)$ and $\pizg^{ij}_{\sigma_t}(\beta_q,b_\perp)$ are defined in Eq. 
(\ref{operde}) above. Note that  this formula is actually our goal - TMD factorization (\ref{TMDf}) with
the coefficient function (\ref{aim}) at $\eta_a=\ln\sigma_p$ and $\eta_b=\ln\sigma_t$.  
Also, note that up to ${\pi^2\over 2}$ constant, this formula can be restored from the rapidity-only evolution of gluon 
TMD calculated in Refs. \cite{Balitsky:2019ayf,Balitsky:2022vnb}. Since leading-order evolutions of quark and gluon TMDs 
differ only in replacement of color factors $N_c\rightarrow c_F$,  
  one should expect a similar Sudakov-type formula  for the Drell-Yan process, probably with a different constant in place of ${\pi^2\over 2}$.

Let us discuss now the region of applicability of Eq. (\ref{momentumresult}).
The r.h.s. of the evolution formula (\ref{momentumresult}) does not depend on  cutoffs $\sigma_p$ and $\sigma_t$ as long as  
$\sigma_p\geq \tigma_p={4b_\perp^{-2}\over\alpha_qs}$ and $\sigma_t\geq \tigma_t\equiv{4b_\perp^{-2}\over\beta_qs}$, see Eq. (\ref{condisigmas}).  Thus, the result of double-log Sudakov evolution reads
\begin{eqnarray}
&&\hspace{-1mm}
W(p_A,p_B;\alpha_q,\beta_q,b_\perp)~=~
{\pi^2\over 2}Q^2\pizg_{ij}^{\tigma_p}(\alpha_q,b_\perp;p_A)\pizg^{ij;\tigma_t}(\beta_q,b_\perp;p_B)
\label{masterresult}\\
&&\hspace{-1mm}
\times~\exp\Big\{-{\alpha_sN_c\over 2\pi}\Big[
\big(\ln {Q^2b_\perp^2\over 4}+2\gamma\big)^2-2\gamma^2-{\pi^2\over 2}\Big]\Big\}    
+~ O\big(\alpha_s^2){\rm~ terms}~+~{\rm power~corrections}
\nonumber
\end{eqnarray}
This result is universal for moderate $x$ and small-$x$ hadronic tensor. The difference lies in the continuation of the evolution beyond Sudakov region. This is discussed in Appendix G of Ref. \cite{Balitsky:2022vnb} and here I briefly sum up the 
main points of that discussion.
First, if $x_B\sim 1$ and $q_\perp^2\geq m_N^2$, there is no room for any evolution and one should turn to
 phenomenological models of TMDs like  the replacement of $b$ by $b_\ast$ in Refs. \cite{Collins:1984kg,Collins:2017oxh}.
If  $x_B\sim 1$ and $q_\perp^2\geq m_N^2$, there is a room for DGLAP-type evolution summing logs 
$\big(\alpha_s\ln{q_\perp^2/m_N^2}\big)^n$. Similarly, if $x_B={\beta_b}\ll 1$, then even at ${\beta_b}\sigma s=q_\perp^2$ there can be the BFKL-type evolution 
from $\sigma ={q_\perp^2\over {\beta'_b}s}$ to $\sigma ={q_\perp^2\over s}$ which sums up logs 
$(\alpha_s\ln x_B)^n$. The matching between double-log Sudakov evolution (\ref{momentumresult}) and single-log 
DGLAP or BFKL evolutions can in principle be performed by solving general rapidity evolution equations discussed in  Ref. \cite{Balitsky:2015qba}.

There is another issue that should be addressed before matching to BFKL and especially to DGLAP evolutions. 
As usually for rapidity-only factorization, the argument of coupling constant in Eq. (\ref{momentumresult}) is undetermined
in the leading order and should be obtained from higher orders of perturbative expansion. Typically, 
argument of coupling constant in the small-$x$ evolution equations is fixed using the BLM/renormalon approach \cite{Brodsky:1982gc}, see for example Ref. 
\cite{Brodsky:1998kn} for the BFKL equation and Refs. \cite{Balitsky:2006wa,Kovchegov:2006vj} for the BK equation
\cite{Balitsky:1995ub,Kovchegov:1999yj,Kovchegov:1999ua}.
In recent paper \cite{Balitsky:2022vnb} G.A. Chirilli and the author
used this  BLM optimal scale setting \cite{Brodsky:1982gc} to fix the argument of coupling constant in the rapidity-only TMD evolution 
(\ref{evoleqtar}). The result is that the effective argument of a coupling constant is halfway in the logarithmical scale 
between the transverse momentum and energy of TMD distribution.  One of the future directions of this research is to 
use BLM prescription to fix the argument of coupling constant in the coefficient function 
$\frc(x_{12_\perp};{\alpha_a},{\beta_b};\sigma_p,\sigma_t)$ and obtain the running-coupling generalization of 
Sudakov-type  formula (\ref{momentumresult}).

Another outlook is to connect to usual CSS/SCET-type evolution of TMDs at moderate $x$ where 
the two \cite{Gutierrez-Reyes:2017glx,Li:2016axz,Gehrmann:2014yya,Gutierrez-Reyes:2018iod} and three-loop results are available \cite{Luo:2019szz,Ebert:2020yqt,Luo:2020epw}.  It should be noted 
that the ``double operator expansion'' method recently used in Ref.  \cite{Vladimirov:2021hdn}  is very similar 
to the approach of this paper and also uses calculation of Feynman diagrams in two background fields. However, the UV+rapidity cutoff of TMD operators in Ref.  \cite{Vladimirov:2021hdn} is very
different from the rapidity-only cutoff used here so the hope is to fix  the argument of coupling constant 
in Eq. (\ref{momentumresult}) and compare the final results for the evolution. 

Summarizing, the proposed rapidity-only factorization may serve as a bridge between classical TMD factorization  \cite{Collins:1984kg}
at moderate $x$ and $k_T$-factorization \cite{Catani:1990eg} at small $x$. 
It would be interesting to compare this rapidity-only factorization/evolution with other approaches to unification of TMD evolution 
based on small-$x$ improvement of usual $Q_\perp^2$ evolution (see e.g. \cite{Forte:2015gve,Marzani:2015oyb}) and on 
various saturation - inspired methods, see e.g. \cite{Mueller:2013wwa,Mueller:2012uf} and  `improved TMD'' discussed in Refs. 
 \cite{Kotko:2015ura,Altinoluk:2019wyu,Altinoluk:2021ygv}. The study is in progress.

\begin{acknowledgments}

\label{sec:acknowledgments}
The author is grateful to G.A. Chirilli, A. Radyushkin, T. Rogers, and A. Vladimirov for valuable  discussions. 
This work is supported by DOE contract DE-AC05-06OR23177  and by the grant DE-FG02-97ER41028.

\end{acknowledgments}

\section{Appendix}
\subsection{Gluon ``cut'' propagator in the background field $\matA$ \label{sect:gluprop}}

In general, the ``cut'' gluon propagator from left to right sector  in the background-Feynman gauge is given by 
the double functional integral
\begin{eqnarray}
&&\hspace{-1mm}
\langle A_\alpha(x) A_\beta(y)\rangle_\matA
\label{cutprop}\\
&&\hspace{-1mm}
=~\!\int^{\tilA(t_f)=A(t_f)}\!\!  D\tilA_\mu DA_\mu ~\tilA_\alpha(x) A_\beta(y)~
e^{i\!\int\! dz~\half\big(\tilA^{\mu,a} (\tilde\matD^2g_{\mu\nu}-2i\tilde\matF_{\mu\nu})^{ab}\tilA^{\nu,b}
+A^a_\mu (\matD^2g^{\mu\nu}-2i\matF^{\mu\nu})A^b_\nu\big)}
\nonumber
\end{eqnarray}
(recall that in our case background field  $\matA$ is the same for both left and right sector).
In Schwinger's notations, in can be written down as
\beq
\langle A_\alpha(x) A_\beta(y)\rangle_\matA~=~-(x|\Big({1\over \matP^2g_{\alpha\xi}+2i\matF_{\alpha\xi}-\ie}
p^2\Big)\tilde\delta_+(p)\Big(p^2{1\over \matP^2\delta^\xi_{\beta}+2i\matF^\xi_{~\beta}+\ie}\Big)|y)
\label{gluprop1}
\eeq
where  expressions in parenthesis in the r.h.s. can be understood as a series 
\begin{eqnarray}
&&\hspace{-3mm}
p^2{1\over (p+\matA)^2\delta^\xi_{\beta}+2i\matF^\xi_{~\beta}+\ie}
~=~1-\matO^\xi_{~\beta}{1\over p^2+\ie}
+~\matO^\xi_{~\eta}{1\over p^2+\ie}
\matO^\eta_{~\beta}{1\over p^2+\ie}~+~...
\nonumber\\
&&\hspace{-3mm}
{1\over (p+\matA)^2g_{\alpha\xi}+2i\matF_{\alpha\xi}-\ie}p^2~=~
1-{1\over p^2+\ie}\matO_{\alpha\xi}+{1\over p^2-\ie}\matO_{\alpha\eta}{1\over p^2-\ie}\matO^\eta_{~\beta}
~+~...
\label{propser1}
\end{eqnarray}
with
\begin{equation}
\matO^{\alpha\beta}~\equiv~(\{p^\lambda,\matA_\lambda\}+\matA^2)g^{\alpha\beta}+2i\matF^{\alpha\beta}.
\end{equation}

 By rearranging the series, it is possible to prove that 
\begin{eqnarray}
&&\hspace{-1mm}
\langle A^a_\alpha(x) A^b_\beta(y)\rangle_\matA~=~(x|{1\over p^2+\ie p_0}\matO\tilde{\delta}_+(p^2)
+\tilde{\delta}_+(p^2)\matO{1\over p^2-\ie p_0}
\nonumber\\
&&\hspace{-1mm}
+~{1\over p^2+\ie p_0}\matO\tilde\delta_+(p)\matO{1\over p^2-\ie p_0}
+~\tilde\delta_+(p)\matO{1\over p^2-\ie p_0}\matO{1\over p^2-\ie p_0}
\nonumber\\
&&\hspace{-1mm}
+~{1\over p^2+\ie p_0}\matO{1\over p^2+\ie p_0}\matO\tilde\delta_+(p)
-\tilde\delta_+(p)\matO\tilde{\delta}_-(p)\matO\tilde\delta_+(p)~+~O(\matO^3)|y)_{\alpha\beta}^{ab}
\label{gluprop2}
\end{eqnarray}
Thus, the only term which spoils the ``retardiness'' property is the last term, but it can be proportional only to 
correction field $\barC$ since neither projectile no target field can solely produce a pair of gluons. However,
two fields $\barC$ involve four $\matF_{\xi\eta}$ (two $U^{+i}$ and two $V^{-j}$, see Eq. (\ref{corefs})) which exceeds 
our accuracy. Thus, we use formula (\ref{gluprop2}) without the last term.

\subsection{TMD matrix elements \label{app:eikonals}}
In this Section we list the necessary results for the ``eikonal'' contributions - one-loop TMD matrix elements
\begin{figure}[htb]
\begin{center}
\includegraphics[width=141mm]{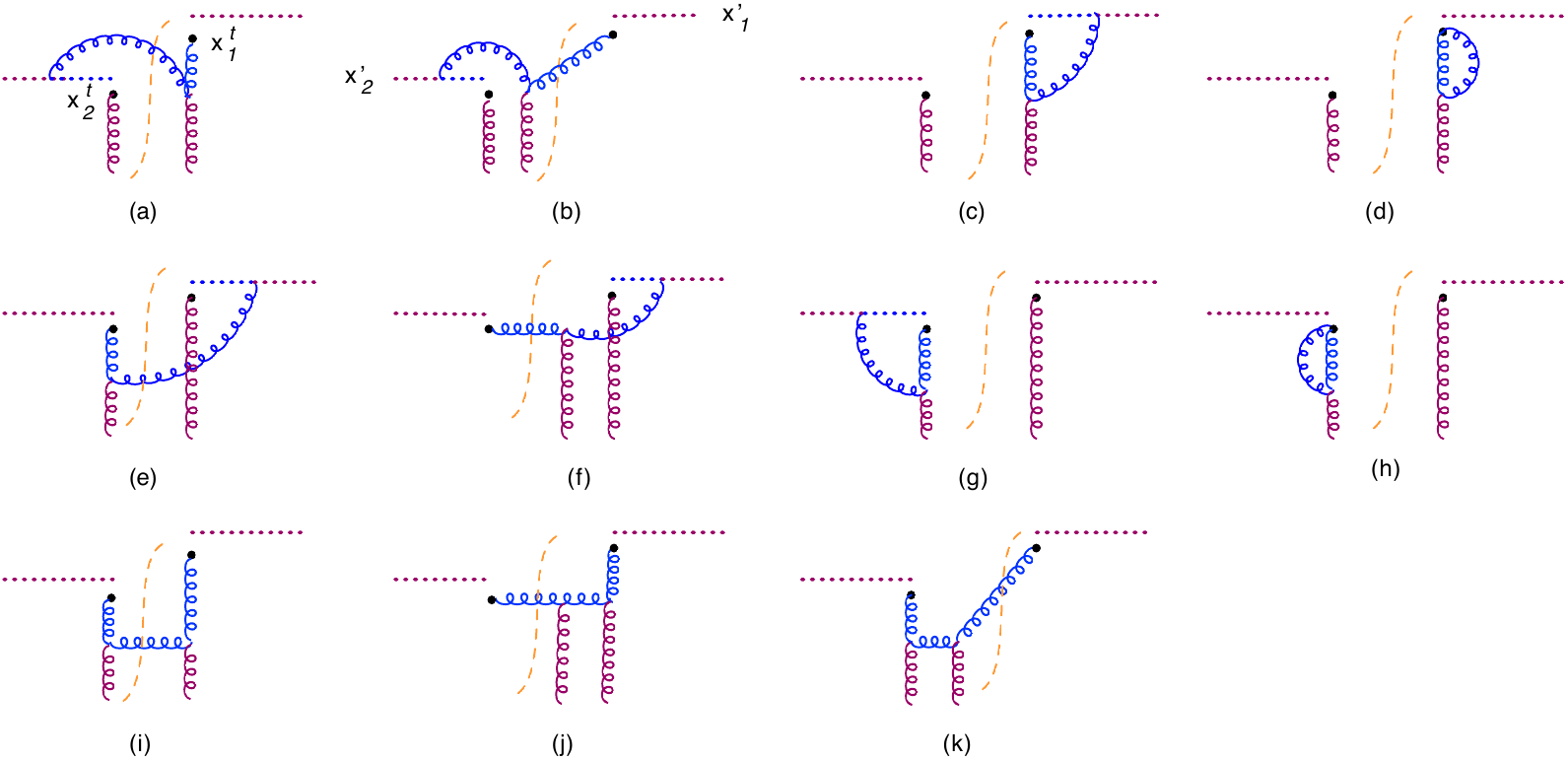}
\end{center}
\caption{``Target''  TMD matrix elements. 
The $e^{-i{\alpha\over\sigma_t}}$ regularization is depicted by point splitting: $F^{-k}$ shown 
by dots stand at $x_1^t=x_{1_\perp}+x_1^+$ 
and $x_2^t=x_{2_\perp}+x_2^+$ while Wilson lines start from  $x'_1=x_1^t+\delta^-$ 
and $x'_2=x_2^t+\delta^-$ where $\delta^-={1\over \vro\sigma_t}$ \label{fig:eik2}.
}
\end{figure}
 calculated in Ref. \cite{Balitsky:2022vnb}. As discussed there, the cutoffs in $\alpha$ and $\beta$ respecting analytical properties of Feynman diagrams (and hence IR real-virtual cancellations) are obtained by ``smooth'' cuts $e^{\pm i{\alpha\over \sigma_t}}$ 
and $e^{\pm i{\beta\over \sigma_t}}$.  These cutoffs are visualized with ``point-splitting'' regularization as shown in Figs.  \ref{fig:eik2} and
\ref{fig:eik1}.
\footnote{As demonstrated in Ref. \cite{Balitsky:2022vnb}, the violations of gauge invariance due to this point-splitting are power corrections $\sim\lambda_p$ or $\lambda_t$.}
As demonstrated in Ref. \cite{Balitsky:2022vnb}, $\delta^+$ and $\delta^-$ should be positive 
which follows  from the requirement that the distances between the ``splitted'' operators should be space-like.

The results of calculation of diagrams in Fig.  \ref{fig:eik2} are:
\begin{eqnarray}
&&\hspace{-0mm} 
\langle [x_2^+,-\infty^+]_{x_{2_\perp}+\delta^-}^{ab}[-\infty^+,x_1^+]_{x_{1_\perp}+\delta^-}^{bc}F^{-j,c}(x_1^+,x_{1_\perp})\rangle_\cala^{\rm Fig.~\ref{fig:eik2}a-d}
\label{glnapravot}\\
&&\hspace{0mm}
=~{g^2N_c\over 8\pi^2}\!\int\!\!\dhd\beta_b\dhd k_{b_\perp}V^{-j,a}({\beta_b},k_{b_\perp}\!) e^{-i\beta_b\vro x_1^++i(k'_b,x_1)_\perp}
I^{\rm eik}_{\rm Fig.~\ref{fig:eik2}a-d}(\beta_b,k_{b_\perp},x_{12})
\nonumber
\ega
where 
\bega
&&\hspace{-1mm}
I^{\rm eik}_{\rm Fig.  \ref{fig:eik2}a-d}(\beta_b, {k}_{b_\perp}, x_1^+,x_{1_\perp},x_2^+,x_{2_\perp})
~
=~8\pi^2s\!\int_0^\infty\!\dhd\alpha~e^{-i{\alpha\over\sigma_t}}
{\dhd\beta\over\beta+\ie}\dhd p_\perp~
e^{i\beta\vro x_{12}^+ -i(p,x_{12})_\perp}
\nonumber\\
&&\hspace{-1mm}
\times~\Big[\tilde{\delta}(\alpha\beta s-p_\perp^2)
{\beta-\beta_b\over \alpha(\beta-\beta_b)s-(p-k_b)_\perp^2-\ie}
+{(\beta-\beta_b)\over \alpha\beta s-p_\perp^2+\ie}
\tilde{\delta}[\alpha(\beta-\beta_b)s-(p-k_b)_\perp^2]\Big]
\nonumber\\
&&\hspace{-1mm}
-~8\pi^2i\!\int\!\dhd\alpha\dhd\beta\dhd p_\perp
{e^{-i{\alpha\over\sigma_t}}s(\beta-\beta_b)
\over (\beta+\ie)(\alpha\beta s-p_\perp^2+\ie)[\alpha(\beta-\beta_b)s-(p-k_b)_\perp^2+\ie]}
\nonumber\\
&&\hspace{-1mm}
=~8\pi^2
\!\int_0^\infty\!\dhd\alpha~e^{-i{\alpha\over\sigma_t}}\!\int\!\dhd p_\perp 
\bigg(
{{\beta_b}s\over  p_\perp^2}{e^{i{p_\perp^2\over\alpha s}\vro x_{12}^+ -i(p,x_{12})_\perp}-1
\over  \alpha\beta_bs+(p-k_b)_\perp^2+\ie}
\nonumber\\
&&\hspace{33mm}
+~{(p-k_b)_\perp^2e^{i(p,x_{12})_\perp}\big[e^{i({\beta_b}+{(p-k_b)_\perp^2\over\alpha s})\vro x_{12}^+}
-e^{i{p_\perp^2\over\alpha s}\vro x_{12}^+}\big]
\over \alpha[\alpha\beta_b s+(p-k_b)_\perp^2+\ie][\alpha\beta_b s+(p-k_b)_\perp^2-p_\perp^2]}\bigg)
\label{ieik11ade}
\end{eqnarray}
At $x_{12}^+=0$ this integral is simplified to
\begin{eqnarray}
&&\hspace{-0mm}
I^{\rm eik}_{\rm Fig.~\ref{fig:eik2}a-d}(\beta_b,k_{b_\perp},x_{12_\perp})
~=~
8\pi^2\!\int_0^\infty\!\dhd\alpha\!\int\!{\dhd p_\perp\over  p_\perp^2}
{\beta_bse^{-i{\alpha\over\sigma_t}}\big(e^{-i(p,x_{12})_\perp}-1\big)
\over \alpha\beta_bs+(p-k_b)_\perp^2+\ie}
\label{napravo2}
\nonumber\\
&&\hspace{-0mm}
=~8\pi^2\!\int_0^\infty\!\dhd\alpha~e^{-i{\alpha\over\sigma_t}}\!\int\!{\dhd p_\perp\over  p_\perp^2}
{{\beta_b}s\big(e^{-i(p,x_{12})_\perp}-1\big)\over  \alpha\beta_bs+p_\perp^2+\ie}
\Big[1-{p_\perp^2-(p-k_b)_\perp^2\over \alpha\beta_bs+(p-k_b)_\perp^2+\ie}\Big]
\nonumber\\
&&\hspace{-0mm}
=~8\pi^2\!\int_0^\infty\!\dhd\alpha~e^{-i{\alpha\over\sigma_t}}\!\int\!{\dhd p_\perp\over  p_\perp^2}
{{\beta_b}s\big(e^{-i(p,x_{12})_\perp}-1\big)\over  \alpha\beta_bs+p_\perp^2+\ie}
+4\pi\!\int\!{\dhd p_\perp\over p_\perp^2}\big(e^{-i(p,x_{12})_\perp}-1\big)\ln{p_\perp^2\over(p-k_b)_\perp^2}
\nonumber\\
&&\hspace{-0mm}
=~-\half\ln^2\Big(-{i\over 4}(\beta_b+\ie)\sigma_t s x_{12}^2e^\gamma\Big)-{\pi^2\over 4}
+I_K(-k_{b_\perp},x_{12_\perp})
\label{ieiktarget1}
\end{eqnarray}
where  $I_K$ is defined in Eq. (\ref{ika}).
Here  we neglected the $e^{-i{\alpha\over\sigma}}$ cutoff in the second integral in the third line since it 
converges at $\alpha\sim{Q_\perp^2\over |{\beta'_b}|s}$ so ${\alpha\over\sigma_t}\sim\lambda_t$.  
The first term in the 
last line is given by Eq. (C4) from Ref. \cite{Balitsky:2022vnb} and the second by Eq. (\ref{eikfla}) from Appendix \ref{app:rints}.  

Similarly, the result for diagrams in Fig. \ref{fig:eik2} e-h reads
\begin{eqnarray}
&&\hspace{-1mm} 
\langle F^{-i,a}(x_2^+,x_{2_\perp})[x_2^+,-\infty]_{x_{2_\perp}+\delta^-}^{ab}[-\infty,x_1^+]_{x_{1_\perp}+\delta^-}^{bc}\rangle_\cala^{\rm Fig.~\ref{fig:eik2}e-h}
\label{glnalevot}\\
&&\hspace{-1mm} 
=~{g^2N_c\over 8\pi^2}\!\int\!\dhd{\beta'_b}\dhd k'_{b_\perp}
e^{-i{\beta'_b} x_{12}^+ +i(k'_b,x_{12_\perp})}V^{-i,c}({\beta'_b},k'_{b_\perp})
I^{\rm eik}_{\rm Fig.~\ref{fig:eik2}e-h}({\beta'_b},k'_{b_\perp}x_{12_\perp}),
\nonumber\\
&&\hspace{-1mm}
I^{\rm eik}_{\rm Fig.~\ref{fig:eik2}e-h}({\beta'_b},k'_{b_\perp},x_{12})~=~8\pi^2
\!\int_0^{\infty}\!\dhd\alpha\!\int\!\dhd p_\perp~e^{i{\alpha\over \sigma_t}}\bigg({{\beta'_b} s\over p_\perp^2}
{\big(e^{-i(p,x_{12})_\perp+i{p_\perp^2\over\alpha s}\vro x_{12}^+}-1\big)\over \alpha{\beta'_b} s-(p+k'_b)_\perp^2+\ie}
\nonumber\\
&&\hspace{11mm}
+~{(p+k'_b)_\perp^2e^{-i(p,x_{12})_\perp}
\big[e^{i{(p-k_a)_\perp^2\over\alpha s}\vro x_{12}^+
-i\beta_b\vro x_{12}^+}-e^{i{p_\perp^2\over\alpha s}\vro x_{12}^+}\big]
\over\alpha[\alpha{\beta'_b} s+p_\perp^2-(p+k'_b)_\perp^2+\ie][\alpha{\beta'_b} s-(p+k'_b)_\perp^2+\ie]}
\bigg)
\nonumber
\end{eqnarray}
which simplifies to 
\begin{eqnarray}
&&\hspace{-0mm} 
I^{\rm eik}_{\rm Fig.~\ref{fig:eik2}e-h}({\beta'_b},k'_{b_\perp},x_{12})~=~8\pi^2
\!\int_0^{\infty}\!\dhd\alpha\!\int\!\dhd p_\perp~e^{i{\alpha\over \sigma_t}}{{\beta'_b} s\over p_\perp^2}
{\big(e^{-i(p,x_{12})_\perp}-1\big)\over \alpha{\beta'_b} s-(p+k'_b)_\perp^2+\ie}
\label{ieiktarget2a}
\end{eqnarray}
at $x_2^+=x_1^+$. This  integral can be obtained from Eq. (\ref{ieiktarget1}) by complex conjugation and
replacement $k_b\rightarrow -k'_b$ so one obtains
\begin{eqnarray}
&&\hspace{-0mm} 
I^{\rm eik}_{\rm Fig.~\ref{fig:eik2}e-h}(\beta'_b,k'_{b_\perp},x_{12_\perp})
~
\nonumber\\
&&\hspace{-0mm} 
=~-\half\ln^2\Big(-{i\over 4}(\beta'_b+\ie)\sigma_t s x_{12_\perp}^2e^\gamma\Big)-{\pi^2\over 4}
+I_K(k'_{b_\perp},x_{12_\perp})~+~O(\lambda_t)
\label{ieiktarget2}
\end{eqnarray}
As to  `handbag'' diagrams in Fig. \ref{fig:eik2}i-k, they were already discussed  in Sect. \ref{sec:handbag}. 

The diagrams in Fig. \ref{fig:eik1}  are obtained by simple target$\leftrightarrow$projectile replacements (\ref{projtarepl}) in Eqs. (\ref{ieiktarget1}) and (\ref{glnalevot}).
\begin{figure}[htb]
\begin{center}
\includegraphics[width=141mm]{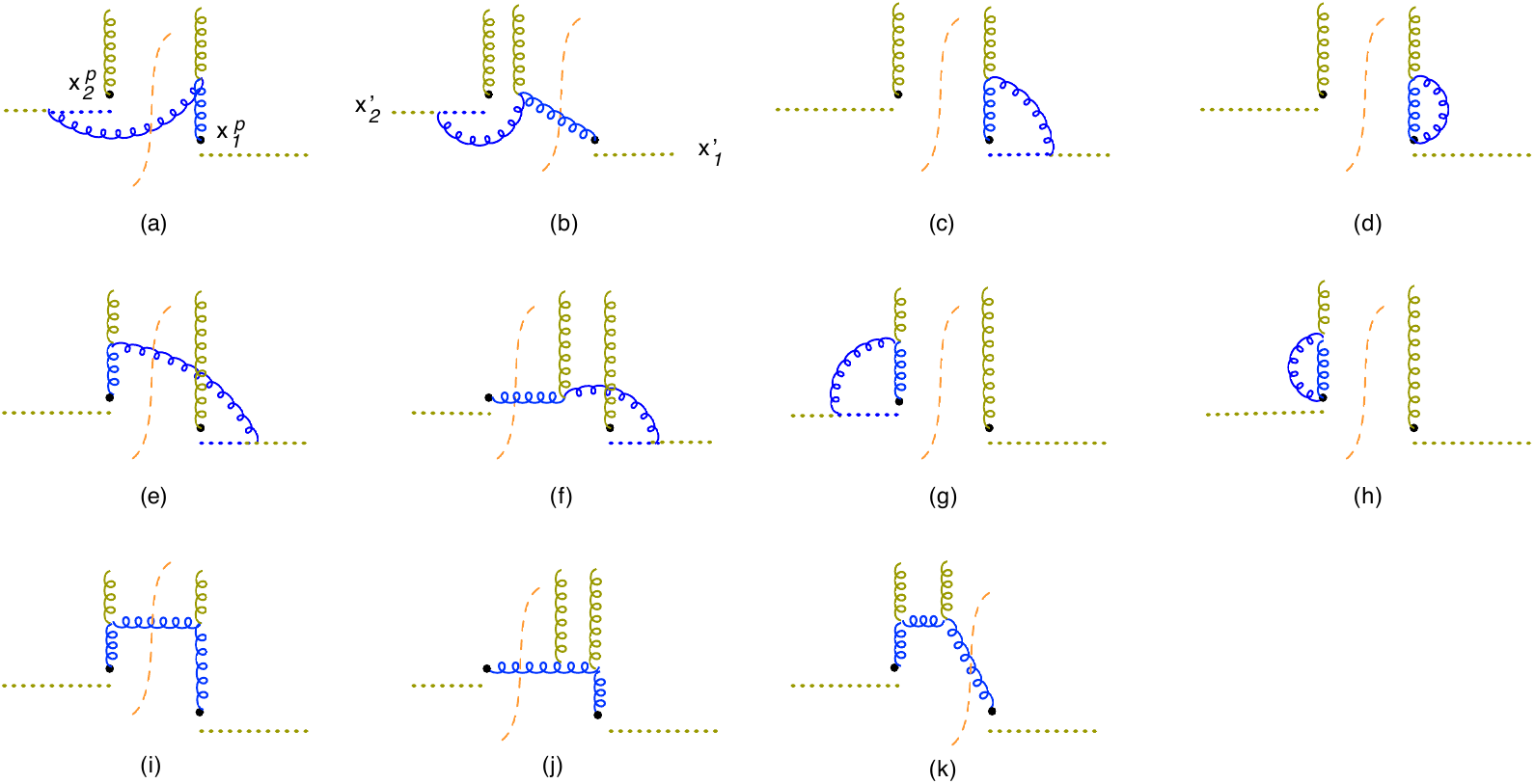}
\end{center}
\caption{``Projectile''  TMD matrix elements. 
The $e^{-i{\beta\over\sigma_p}}$ regularization is depicted by point splitting: $F^{+k}$ shown by dots stand at  $x^p_1=x_{1_\perp}+x_1^-$ and $x^p_2=x_{2_\perp}+x_2^-$ 
while Wilson lines start from  $x'_1=x_2+\delta^+$ 
and $x'_2=x_1+\delta^+$ where $\delta^+={1\over \vro\sigma_p}$ \label{fig:eik1}. 
}
\end{figure}
We get 
\begin{eqnarray}
&&\hspace{-0mm} 
\langle [x_2^-,-\infty]_{x_{2_\perp}+\delta^+}^{ab}[-\infty,x_1^-]_{x_{1_\perp}+\delta^+}^{bc}F^{+j,c}(x_1^-,x_{1_\perp})\rangle_\cala^{\rm Fig.~\ref{fig:eik1}a-d}
\label{glnalevop}\\
&&\hspace{0mm}
=~{g^2N_c\over 8\pi^2}\!\!\int\!\!\dhd{\alpha_a}\dhd k_{a_\perp}U^{+j,b}({\alpha_a},k_{a_\perp}\!) e^{-i{\alpha_a}\vro x_1^- +i(k_a,x_1)_\perp}
I^{\rm eik}_{\rm Fig.~\ref{fig:eik1}a-d}({\alpha_a},k_{a_\perp},x_{12}), 
\nonumber\\
&&\hspace{0mm}
I^{\rm eik}_{\rm Fig.~\ref{fig:eik1}a-d}({\alpha_a},k_{a_\perp},x_{12})~=~8\pi^2
\!\int_0^\infty\!\dhd\beta~e^{-i{\beta\over\sigma_p}}\!\int\!\dhd p_\perp 
\bigg(
{{\alpha_a}s\over  p_\perp^2}
{e^{i{p_\perp^2\over\beta s}\vro x_{12}^- -i(p,x_{12})_\perp}-1
\over  {\alpha_a}\beta s+(p-k_a)_\perp^2+\ie}
\nonumber\\
&&\hspace{0mm}
+~{(p-k_a)_\perp^2e^{-i(p,x_{12})_\perp}\big[e^{i({\alpha_a}+{(p-k_a)_\perp^2\over\beta s})\vro x_{12}^-}
-e^{i{p_\perp^2\over\beta s}\vro x_{12}^-}\big]
\over \alpha[{\alpha_a}\beta s+(p-k_a)_\perp^2+\ie][{\alpha_a}\beta s+(p-k_a)_\perp^2-p_\perp^2]}\bigg)
\nonumber
\end{eqnarray}
simplified to 
\begin{eqnarray}
&&\hspace{-0mm} 
I^{\rm eik}_{\rm Fig.~\ref{fig:eik1}a-d}({\alpha_a},k_{a_\perp},x_{12_\perp})
~=~-\half\ln^2\Big(-{i\over 4}({\alpha_a}+\ie)\sigma_p s x_{12_\perp}^2e^\gamma\Big)-{\pi^2\over 4}
+I_K(-k_{a_\perp},x_{12_\perp})
\nonumber\\
\label{ieikproj1}
\end{eqnarray}
at $x_{12}^-=0$, and
\begin{eqnarray}
&&\hspace{-1mm} 
\langle F^{+i,a}(x_2^-,x_{2_\perp})[x_2^-,-\infty]_{x_{2_\perp}+\delta^+}^{ab}[-\infty,x_1^-]_{x_{1_\perp}+\delta^+}^{bc}\rangle_\cala^{\rm Fig.~\ref{fig:eik1}e-h}
\label{glnalevote}\\
&&\hspace{-1mm} 
=~g^2N_c\!\int\!\dhd\alpha_a\dhd {k'}_{a_\perp}
e^{-i\alpha_a x_{12}^+ +i(k'_a,x_{12_\perp})}U^{+i,c}(\alpha_a,{k'}_{a_\perp})
I^{\rm eik}_{\rm Fig.~\ref{fig:eik1}e-h}(\alpha_a,{k'}_{a_\perp},x_2,x_1),
\nonumber\\
&&\hspace{-1mm}
I^{\rm eik}_{\rm Fig.~\ref{fig:eik1}e-h}(\alpha'_a,k'_{a_\perp},x_{12})
~=~8\pi^2
\!\int_0^{\infty}\!\dhd\beta\!\int\!\dhd p_\perp
~e^{i{\beta\over \sigma_p}}\bigg({\alpha'_a s\over p_\perp^2}
~{\big(e^{-i(p,x_{12})_\perp+i{p_\perp^2\over\beta s}\vro x_{12}^-}-1\big)\over \alpha'_a\beta s-(p+k'_a)_\perp^2+\ie}
\nonumber\\
&&\hspace{40mm}
+~{(p+k'_a)_\perp^2
e^{-i(p,x_{12})_\perp}\big[e^{i{(p+k'_a)_\perp^2\over\alpha s}\vro x_{12}^-
-i\alpha'_a\vro x_{12}^-}-e^{i{p_\perp^2\over\beta s}\vro x_{12}^-}\big]
\over\beta[\alpha'_a\beta s+p_\perp^2-(p+k'_a)_\perp^2+\ie][\alpha'_a\beta s-(p+k'_a)_\perp^2+\ie]}
\bigg)
\nonumber
\end{eqnarray}
which similarly simplifies to 
\begin{eqnarray}
&&\hspace{-0mm}   
I^{\rm eik}_{\rm Fig.~\ref{fig:eik1}e-h}(\alpha_a,{k'}_{a_\perp},x_{12_\perp})~
\nonumber\\
&&\hspace{-0mm} 
=~-\half\ln^2\Big(-{i\over 4}(\alpha'_a+\ie)\sigma_p s x_{12}^2e^\gamma\Big)
-{\pi^2\over 4}+I_K(k'_{a_\perp},x_{12_\perp})~+~O(\lambda_p)
\label{ieikproj2}
\end{eqnarray}
at $x_{12}^+=0$.  Again, the  ``handbag'' diagrams in Fig. \ref{fig:eik1}i-k were already accounted for in Sect. \ref{sec:handbag}.

Assembling Eqs. (\ref{ieiktarget1}), (\ref{ieiktarget2}),   (\ref{ieikproj1}), and (\ref{ieikproj2}), we get the matrix elements of eikonal TMD operators which have to be subtracted from $\pizw$ according 
to Eq. (\ref{facoord2}):
\bega
&&\hspace{-1mm}
\pizw(x_1,x_2)^{\rm eik}=~~{N_c^2-1\over N_c}8\pi^2
\nonumber\\
&&\hspace{-1mm}
\times~\Big\{U^{+a}_{~i}(x_2)V^{- i,n}(x_2)\langle [x_2^+,-\infty]_{x_{2_\perp}+\delta^-}^{na}[-\infty,x^+]_{x_{1_\perp}+\delta^-}^{bc}F^{-j,c}(x^+,x_{1_\perp})\rangle_\cala^{\rm Fig.~\ref{fig:eik2}a-c}
U^{+ j,b}(x_1)
\nonumber\\
&&\hspace{-1mm}
+~U^{+a}_{~i}(x_2)\langle F^{-i,n}(x_2^+,x_{2_\perp})[x_2^+,-\infty]_{x_{2_\perp}+\delta^-}^{na}[-\infty,x_1^+]_{x_{1_\perp}+\delta^-}^{bc}\rangle_\cala^{\rm Fig.~\ref{fig:eik2}d-f}V^{-j,c}(x_1)U^{+ j,b}(x_1)
\nonumber\\
&&\hspace{-1mm}
+~U^{+n}_{~i}(x_2)V^{- i,a}(x_2)\langle [x^-,-\infty]_{x_{2_\perp}+\delta^+}^{na}[-\infty,x_1^-]_{x_{1_\perp}+\delta^+}^{bc}F^{+j,c}(x^-,x_{1_\perp})\rangle_\cala^{\rm Fig.~\ref{fig:eik1}a-c}V^{-j,c}(x_1)
\nonumber\\
&&\hspace{-1mm}
+~V^{- i,a}(x_2)\langle F^{+i,n}(x_2^-,x_{2_\perp})[x_2^-,-\infty]_{x_{2_\perp}+\delta^+}^{na}[-\infty,x_1^-]_{x_{1_\perp}+\delta^+}^{bc}\rangle_\cala^{\rm Fig.~\ref{fig:eik1}d-f}V^{-j,c}(x_1)U^{+b}_{~j}(x_1)\Big\}
\nonumber\\
&&\hspace{-1mm}
=~
\!\int\!\dhd{\alpha'_a}\dhd {k'}_{a_\perp}\dhd{\beta'_b}\dhd k_{b_\perp}\dhd\alpha_a\dhd {k'}_{a_\perp}\dhd\beta_b\dhd {k'}_{b_\perp}
~e^{-i{\alpha'_a}\vro x_2^- -i\alpha_a\vro x_1^-}e^{-i{\beta'_b}\vro x_2^+ -i\beta_b\vro x_1^+}
\nonumber\\
&&\hspace{4mm}
\times~e^{-i(k_a+k_a,x_2)_\perp-i(k'_a+k'_b,x_1)_\perp}
U^{+,b}_{~~i}({\alpha'_a},{k'}_{a_\perp})V^{-i,a}({\beta'_b},k_{b_\perp})
U^{+,b}_{~j}(\alpha_a,p'_{A_\perp})V^{-j,a}(\beta_b,k'_{b_\perp})
\nonumber\\
&&\hspace{9mm}
\times~\big[I^{\sigma_p,\sigma_t}_{\rm eik}(\alpha_a,\alpha'_a,{\beta_b},\beta'_b,k_{a_\perp}, k'_{a_\perp},k_{b_\perp}, k'_{b_\perp},x_{12_\perp})
+O(\lambda_p)+O(\lambda_t)\big]
\label{pizweikotvet}
\ega
with
\bega
&&\hspace{-5mm}
I^{\sigma_p,\sigma_t}_{\rm eik}(\alpha_a,\alpha'_a,{\beta_b},\beta'_b,k_{a_\perp}, k'_{a_\perp},k_{b_\perp}, k'_{b_\perp},x_{12_\perp})
\label{ieiksumma}\\
&&\hspace{-1mm}
=~-\half\ln^2\Big(-{i\over 4}({\alpha'_a}+\ie)\sigma_p s x_{12_\perp}^2e^\gamma\Big)
-\half\ln^2\Big(-{i\over 4}(\alpha_a+\ie)\sigma_p s x_{12_\perp}^2e^\gamma\Big)
\nonumber\\
&&\hspace{5mm}
-~
\half\ln^2\Big(-{i\over 4}({\beta'_b}+\ie)\sigma_t s x_{12_\perp}^2e^\gamma\Big)
-\half\ln^2\Big(-{i\over 4}(\beta_b+\ie)\sigma_t s x_{12_\perp}^2e^\gamma\Big)-\pi^2
\nonumber\\
&&\hspace{5mm}
+~I_K(-k_{a_\perp},x_{12_\perp})+I_K(k'_{a_\perp},x_{12_\perp})+I_K(-k_{b_\perp},x_{12_\perp})
+I_K(k'_{b_\perp},x_{12_\perp})
\nonumber
\ega
where  $I_K$ is defined in Eq. (\ref{ika}).  

Let us present also the derivative 
\footnote{It is worth
noting that handbag diagrams in Figs \ref{fig:eik2}i-k and \ref{fig:eik1}i-k do not contribute to evolution equation since they do not need a rapidity cutoff, see the discussion in  in Ref. \cite{Balitsky:2022vnb} and in Sect. \ref{sec:handbag}.}

\bega
&&\hspace{-1mm}
\sigma_t{d\over d\sigma_t}I^{\sigma_p,\sigma_t}_{\rm eik}({\alpha'_a},\alpha_a,{\beta'_b},\beta_b,k'_{a_\perp}, k_{a_\perp},k_{b_\perp}, k'_{b_\perp},x_{12_\perp})
\label{derieiksumma}\\
&&\hspace{5mm}
=~-\ln\Big(-{i\over 4}({\beta'_b}+\ie)\sigma_t s x_{12_\perp}^2e^\gamma\Big)
-\ln\Big(-{i\over 4}(\beta_b+\ie)\sigma_t s x_{12_\perp}^2e^\gamma\Big)
\nonumber
\ega
which translates into evolution equation \cite{Balitsky:2022vnb,Balitsky:2019ayf}
\bega
&&\hspace{-0mm}
\sigma_t{d\over d\sigma_t}\hacalo^{ij;\sigma_t}({\beta'_b},\beta_b,x_{2_\perp},x_{1_\perp})
\label{evoleqtar}\\
&&\hspace{0mm}
=~-{\alpha_sN_c\over 2\pi}\Big[2\ln{sx_{12_\perp}^2\over 4}+\ln(-i{\beta'_b}\sigma_t+\epsilon)+\ln(-i\beta_b\sigma_t+\epsilon)+2\gamma\Big]
\hacalo^{ij;\sigma_t}({\beta'_b},\beta_b,x_{2_\perp},x_{1_\perp})
 \nonumber
\ega
Similarly, one obtains
\bega
&&\hspace{-0mm}
\sigma_p{d\over d\sigma_p}\hacalo^{ij;\sigma_t}({\alpha'_a},\alpha_a,x_{2_\perp},x_{1_\perp})
\label{evoleqproj}\\
&&\hspace{0mm}
=~-{\alpha_sN_c\over 2\pi}\Big[2\ln{sx_{12_\perp}^2\over 4}+\ln(-i{\alpha'_a}\sigma_p+\epsilon)+\ln(-i\alpha_a\sigma_p+\epsilon)+2\gamma\Big]
\hacalo^{ij;\sigma_t}({\alpha'_a},\alpha_a,x_{2_\perp},x_{1_\perp})
 \nonumber
\ega
In the coordinate space,  the evolution equation (\ref{evoleqtar}) takes the form
\bega
&&\hspace{-1mm}
\sigma_t{d\over d\sigma_t}\hacalo^{ij;\sigma_t}(x_2^+,x_{2_\perp};x_1^+,x_{1_\perp})
\label{evoleqtarcoor}\\
&&\hspace{5mm}
=~-{\alpha_sN_c\over 2\pi}\Big\{2\ln{sx_{12_\perp}^2\over 4}\hacalo^{ij;\sigma_t}(x_2^+,x_{2_\perp};x_1^+,x_{1_\perp})
 \nonumber\\
&&\hspace{11mm}
-~
\int\! dz_2^+
 \Big[{\theta(x_2-z_2)^+\over (x_2-z_2)^+}-\delta(x_2-z_2)^+\!\int_0^{\sigma_t/\vro}{dz_2^+\over z_2^+} \Big]
 \hacalo_{ij}^{\sigma_p}(z_2^+,x_{2_\perp};x_1^+,x_{1_\perp})
  \nonumber\\
&&\hspace{11mm}
-~
\int\! dz_1^+ \Big[{\theta(x_1-z_1)^+\over (x_1-z_1)^+}-\delta(x_1-z_1)^+\!\int_0^{\sigma_t/\vro}{dz_1^+\over z_1^+} \Big]
 \hacalo_{ij}^{\sigma_p}(x_2^+,x_{2_\perp};z_1^+,x_{1_\perp})\Big\}
 \nonumber
\end{eqnarray}
where we used formula
\beq
\int\!\dhd\beta ~e^{-i\beta z}\big[\ln\big(-i\beta\sigma+\epsilon\big)+\gamma\big]
~=~-{\theta(z)\over z}+\delta(z)\!\int_0^\sigma\!{dz'\over z'}
\eeq
The evolution equation of $\hacalo^{ij;\sigma_p}(x_2^-,x_{2_\perp};x_1^-,x_{1_\perp})$ looks like Eq. (\ref{evoleqtarcoor}) with 
trivial replacements $x^+\rightarrow x^-$ and $\sigma_t\rightarrow\sigma_p$.

\subsection{Soft factor with rapidity-only cutoffs \label{app:softs}}
In this Section we demonstrate that the soft factor with rapidity-only cutoffs is a power correction. 
The soft factor is given by the correlation function of four Wilson lines 
\begin{eqnarray}
&&\hspace{-11mm}   
{1\over N_c^2-1}{\rm Tr}\langle\{x_2^-,-\infty^-\}_{x_{2_\perp}}\{-\infty^+,x_2^+\}_{x_{2_\perp}}
[x_1^+, -\infty^+]_{x_{1_\perp}}[-\infty^-,x_1^-]_{x_{1_\perp}}\rangle_\scras
\end{eqnarray}
The diagrams for the one-loop soft factor with rapidity-only regularization by ``point splitting'' is shown in Fig. \ref{fig:softfactor}
\begin{figure}[htb]
\begin{center}
\includegraphics[width=111mm]{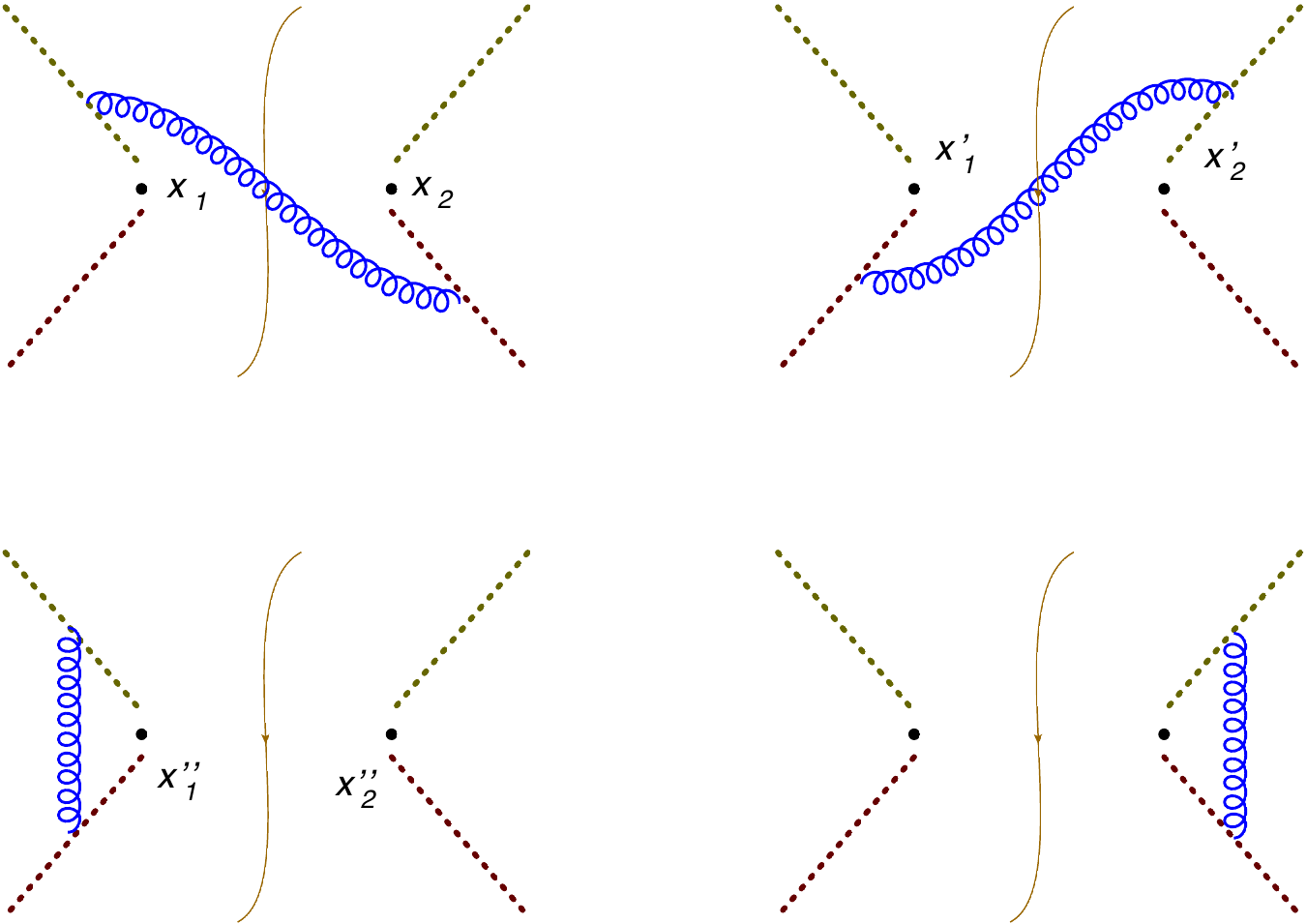}
\end{center}
\caption{First-order perturbative diagrams for the soft factor. 
The rapidity regularization is depicted by point splitting:  ``projectile'' Wilson lines start from  $x'_1=x_1+\delta^+$ 
and $x'_2=x_2+\delta^+$ while ``target'' Wilson lines from $x''_1=x_1+\delta^-$ and $x''_2=x_2+\delta^-$ \label{fig:softfactor}}
\end{figure}

%
\begin{eqnarray}
&&\hspace{-1mm}   
{{\rm Tr}\over N_c(N_c^2-1)}\langle\{x_2^-,-\infty^-\}_{x_{2_\perp}+\delta^+}\{-\infty^+,x_2^+\}_{x_{2_\perp}+\delta^-}
[x_1^+, -\infty^+]_{x_{1_\perp}+\delta^-}[-\infty^-,x_1^-]_{x_{1_\perp}+\delta^+}\rangle
\nonumber\\
&&\hspace{-1mm}
=~{1\over N_c(N_c^2-1)}{\rm Tr}\bigg[\!\int_{-\infty}^{x_2^-}\! d{x'_2}^-\!\int_{-\infty}^{x_2^+}\! d{x'_2}^+
~\tilA^+({x'_2}^- +\delta^+,x_{2_\perp})\tilA^-({x'_2}^+ +\delta^-,x_{2_\perp})
\nonumber\\
&&\hspace{25mm}
-~\!\int_{-\infty}^{x_2^-}\! d{x'_2}^-\!\int_{-\infty}^{x_1^+}\! d{x'_1}^+ 
\tilA^+({x'_2}^- +\delta^+,x_{2_\perp})A^-({x'_1}^+ +\delta^-,x_{1_\perp})
\nonumber\\
&&\hspace{25mm}
-~\!\int_{-\infty}^{x_2^+}\! d{x'_2}^+\!\int_{-\infty}^{x_1^-}\! d{x'_1}^-
~\tilA^-({x'_2}^++\delta^-,x_{2_\perp})A^+({x'_1}^- +\delta^+,x_{1_\perp})
\nonumber\\
&&\hspace{25mm}
+~\!\int_{-\infty}^{x_1^-}\! d{x'_1}^-\!\int_{-\infty}^{x_1^+}\! d{x'_1}^+
~A^+({x'_1}^- +\delta^+,x_{1_\perp})A^-({x'_1}^+ +\delta^-,x_{1_\perp})\bigg]
\nonumber\\
&&\hspace{-1mm}
=~\int\!\dhd^4p\bigg[\!\int_{-\infty}^{x_2^-}\! d{x'_2}^-\!\int_{-\infty}^{x_2^+}\! d{x'_2}^+
~e^{-i\alpha\vro ({x'_2}^- -\delta^-) +i\beta\vro({x'_2}^+ -\delta^+)}{i\over\alpha\beta s-p_\perp^2-\ie}
\nonumber\\
&&\hspace{12mm}
+~\!\int_{-\infty}^{x_2^-}\! d{x'_2}^-\!\int_{-\infty}^{x_1^+}\! d{x'_1}^+
~e^{-i\alpha\vro ({x'_2}^- -\delta^-)+i\beta\vro({x'_1}^+ -\delta^+)-i(p,x_{12})_\perp}
\delta_+(\alpha\beta s-p_\perp^2)
\nonumber\\
&&\hspace{12mm}
+~\!\int_{-\infty}^{x_2^+}\! d{x'_2}^+\!\int_{-\infty}^{x_1^-}\! d{x'_1}^-
~e^{i\alpha\vro({x'_1}^- -\delta^-)-i\beta\vro ({x'_2}^+ -\delta^+)-i(p,x_{12})_\perp}
\delta_+(\alpha\beta s-p_\perp^2)
\nonumber\\
&&\hspace{12mm}
-~\!\int_{-\infty}^{x_1^-}\! d{x'_1}^-\!\int_{-\infty}^{x_1^+}\! d{x'_1}^+
~e^{-i\alpha\vro ({x'_1}^- -\delta^-)+i\beta\vro ({x'_1}^+ -\delta^+)}{i\over\alpha\beta s-p_\perp^2+\ie}
\bigg]
\label{soft2}\\
&&\hspace{-1mm}
=~\int\!\dhd\alpha\dhd\beta\dhd p_\perp\Big\{{i\over (\alpha+\ie)(\beta-\ie)}
\Big[{e^{i\alpha\vro(\delta^- -x_2^-)-i\beta\vro(\delta^+-x_2^+)}\over\alpha\beta s-p_\perp^2-\ie}
-{e^{i\alpha\vro(\delta^- -x_1^-)-i\beta\vro (\delta^+ -x_1^+)}\over\alpha\beta s-p_\perp^2+\ie}\Big]
\nonumber\\
&&\hspace{11mm}
+~{s\over p_\perp^2}e^{-i(p,x_{12})_\perp}
\Big[e^{i\alpha\vro(\delta^- -x_2^-)-i\beta\vro(\delta^+-x_1^+)}+
e^{-i\alpha\vro(\delta^- -x_1^-)+i\beta\vro(\delta^+-x_2^+)}\Big]\delta_+(\alpha\beta s-p_\perp^2)
\Big\}
\nonumber
\end{eqnarray}
Since the $x_2^-,x_1^-\sim {1\over\vro{\alpha'_a}}\ll \delta^-$ 
and $x_2^+,x_1^+\sim {1\over\vro{\beta'_b}}\ll \delta^+$ we can neglect 
$x_2^-,x_1^-,x_2^+,x_1^+$ in the above integrals and get
\begin{eqnarray}
&&\hspace{-1mm}=~s\!\int\!\dhd\alpha\dhd\beta{\dhd p_\perp\over p_\perp^2}\Big\{ 
-e^{i\alpha\vro\delta^- -i\beta\vro\delta^+}
\tilde{\delta}(\alpha\beta s-p_\perp^2)
\nonumber\\
&&\hspace{11mm}
+~e^{-i(p,x_{12})_\perp}\big[e^{i\alpha\vro\delta^- -i\beta\vro\delta^+}+e^{-i\alpha\vro\delta^- +i\beta\vro\delta^+}\big]
\delta_+(\alpha\beta s-p_\perp^2)
\Big\}
\nonumber\\
&&\hspace{-1mm}
=~s\!\int\!\dhd\alpha\dhd\beta{\dhd p_\perp\over p_\perp^2}
\big[e^{i\alpha\vro\delta^- -i\beta\vro\delta^+}+e^{-i\alpha\vro\delta^- +i\beta\vro\delta^+}\big]
\delta_+(\alpha\beta s-p_\perp^2)
\big(e^{-i(p,x_{12})_\perp}-1\big)
\nonumber\\
&&\hspace{-1mm}
=~{1\over 2\pi^2}\int\!{dp_\perp^2\over p_\perp^2}[J_0(p_\perp\Delta_\perp)-1]K_0\Big(p_\perp\sqrt{2\delta^+\delta^-}\Big)
~=~{1\over 4\pi^2}{\rm Li}_2\Big(-{x_{12_\perp}^2\over 2\delta^+\delta^-}\Big)
\label{eikorr1}
\end{eqnarray}
Thus, we get the perturbative contribution to rapidity-regularized soft factor in the form
\begin{eqnarray}
&&\hspace{-11mm}   
\langle\{x_2^-,-\infty^-\}_{x_{2_\perp}+\delta^+}\{-\infty^+,x_2^+\}_{x_{2_\perp}+\delta^-}
[x_1^+, -\infty^+]_{x_{1_\perp}+\delta^-}[-\infty^-,x_1^-]_{x_{1_\perp}+\delta^+}\rangle
\nonumber\\
&&\hspace{-11mm}
=~{1\over 4\pi^2}{\rm Li}_2\Big(-{x_{12_\perp}^2\over 2\delta^+\delta^-}\Big)~\sim~O\Big({\Delta_\perp^2\over 2\delta^+\delta^-}\Big)~\sim~O\Big({\sigma_p\sigma_ts\over Q_\perp^2}\Big)
~=~O\Big({\mu_\sigma^2\over Q_\perp^2}\Big)~\sim\zeta^{-1/2}
\end{eqnarray}
which is a parametrically small power correction. Of course, there are  non-perturbative contributions to the soft factor - 
power corrections presumably  of order of $\Lambda_{\rm QCD}^2x_{12_\perp}^2$, but, as we mentioned above,  the lesson is that the soft factor with rapidity-only 
regularization does not have perturbative contributions which can mix with the TMD evolution, quite unlike the usual regularization of the soft factor with ``UV+rapidity'' cutoff.

\subsection{Approximation $x_{12}^\parallel=0$ for the calculation of coefficient function \label{sect:noparallel}}

In this Section we prove that for the calculation of the coefficient function $\frc_1$  in Eq. (\ref{facoord1}) one can set $x_{12}^\parallel=0$ with power accuracy. 
Let us start with the difference $I_{1a}-I_{1a}^{\rm eik}$ in Eq. (\ref{differenca}). Since vitual eikonals do not depend on $x$, it is sufficient to consider 
\beq
\hspace{-0mm}
I_{1a}({\alpha'_a}, k'_{a_\perp},{\beta'_b}, k_{b_\perp}, x_2,x_1)-I_{\rm Fig.~\ref{fig:eik2}a,b}^{\rm eik}({\alpha'_a}, {k'}_{a_\perp},{\beta'_b}, k_{b_\perp}, x_2,x_1)
~-~(x^\parallel\rightarrow 0)
\label{iaminuseik}
\eeq
The expression for $I_{1a}$ is given by Eq. (\ref{iodina})
\bega
&&\hspace{-1mm}
I_{1a}({\alpha'_a}, {k'}_{a_\perp},\beta_b, k_{b_\perp}, x_{2_\perp}, x_{1_\perp})~=~
8\pi^2\!\int_0^\infty\!{\dhd\alpha\over\alpha} \!\int\!\dhd p_\perp~
{e^{-i(p,x_{12})_\perp}\over \alpha\beta_bs+(p-k'_b)_\perp^2-p_\perp^2+\ie}
\nonumber\\
&&\hspace{5mm}
\times~e^{i\alpha \vro x_{12}^-}\bigg[
{p_\perp^2\over\alpha^2s\xi+p_\perp^2}
{(\alpha+{\alpha'_a})(\alpha\beta_bs-p_\perp^2)e^{i{p_\perp^2\over\alpha s}\vro x_{12}^+}
\over \big[(\alpha+{\alpha'_a})p_\perp^2-\alpha(p+k'_a)^2+\ie\big]}
\nonumber\\
&&\hspace{11mm}
+~{(p-k'_b)_\perp^2(\alpha+{\alpha'_a})e^{i{\beta_b}\vro x_{12}^+ +i{(p-k'_b)_\perp^2\over\alpha s}\vro x_{12}^+}
\over 
\alpha({\alpha'_a}+\alpha)\beta_bs+({\alpha'_a}+\alpha)(p-k_b)_\perp^2-\alpha(p+k'_a)_\perp^2+\ie({\alpha'_a}+\alpha)}
\bigg]
\label{iodinaee}
\end{eqnarray}
(cf. Eq.(\ref{iodinae}) at $x^\parallel=0$), and the expression for $I_{\rm Fig.~\ref{fig:eik2}a,b}^{\rm eik}$ can be taken as Eq. 
(\ref{ieik11ade}) without virtual term 
\bega
&&\hspace{0mm}
I^{\rm eik}_{\rm Fig.~\ref{fig:eik2}a,b}({\beta_b},k_{b_\perp},x_2,x_1)~=~8\pi^2
\!\int_0^\infty\!{\dhd\alpha\over\alpha}~e^{-i{\alpha\over\sigma_t}}\!\int\!\dhd p_\perp 
\bigg(
{\alpha{\beta_b}s-p_\perp^2\over  p_\perp^2}e^{-i{p_\perp^2\over\alpha s}\vro x_{12}^+}
\nonumber\\
&&\hspace{0mm}
+~{(p-k_b)_\perp^2e^{i({\beta_b}+{(p-k_b)_\perp^2\over\alpha s})\vro x_{12}^+}
\over [\alpha{\beta_b} s+(p-k_b)_\perp^2+\ie]}\bigg)
{e^{-i(p,x_{12})_\perp}\over  \alpha{\beta_b}s-p_\perp^2+(p-k_b)_\perp^2}
\label{ieik11ab}
\end{eqnarray}
For definiteness, let us take ${\beta_b}>0$
(the case of ${\beta_b}<0$ is similar).
The difference (\ref{iaminuseik}) can be represented as a sum of two contributions. The first one is 
the difference between first terms in Eqs. (\ref{iodinaee}) and (\ref{ieik11ab}) minus same difference at $x_{12}^\parallel=0$.
\begin{eqnarray}
&&\hspace{-1mm}
8\pi^2\!\int_0^\infty\!{\dhd\alpha\over\alpha} \!\int\!\dhd p_\perp~\bigg[
{{p_\perp^2\over\alpha^2s\xi+p_\perp^2}(\alpha+{\alpha'_a})(\alpha\beta_bs-p_\perp^2)
\Big(e^{i\alpha\vro x_{12}^- +i{p_\perp^2\over\alpha s}\vro x_{12}^+}-1\Big)
\over \big((\alpha+{\alpha'_a})p_\perp^2-\alpha(p+k'_a)^2+\ie\big)\big(\alpha\beta_bs+(p-k_b)_\perp^2-p_\perp^2+\ie\big)}
\nonumber\\
&&\hspace{11mm}
-~
\!\int_0^{\infty}\!{\dhd\alpha\over\alpha}~e^{-i{\alpha\over \sigma_t}}
{\Big(e^{i{p_\perp^2\over\alpha s}\vro x_{12}^+}-1\Big)(\alpha\beta_b s-p_\perp^2)
\over p_\perp^2[\alpha\beta_b s-p_\perp^2+(p-k_b)_\perp^2+\ie]}\bigg]e^{-i(p,x_{12})_\perp}
\nonumber\\
&&\hspace{-1mm}  
=~4\pi\!\int_0^\infty\!{dt\over t} \!\int\!\dhd p_\perp~\bigg\{  
{p_\perp^2\over{t^2\over {\alpha'_a}{\beta_b}s}|{{\alpha'_a}\over\beta_b}|\xi+p_\perp^2}
{(t+{\alpha'_a}\beta_bs)\big(e^{i{t{\alpha'_a}\vro x_{12}^-\over {\alpha'_a}\beta_bs}}-1\big)
e^{i{p_\perp^2\over t}{\beta'_b}\vro x_{12}^+}
\over (t+{\alpha'_a}\beta_bs)p_\perp^2-t(p+k'_a)^2+\ie}
\nonumber\\
&&\hspace{11mm}
+~
\bigg[
{p_\perp^2\over{t^2\over{\alpha'_a}\beta_bs}|{{\alpha'_a}\over\beta_b}|\xi+p_\perp^2}
{(t+{\alpha'_a}{\beta_b}s)
\over (t+{\alpha'_a}\beta_bs)p_\perp^2-t(p+k_a)^2+\ie}
-~{e^{-i{t\over \sigma_t\beta_bs}}
\over p_\perp^2}\bigg]
\nonumber\\
&&\hspace{22mm}
\times~\Big(e^{-i{p_\perp^2\over t}\beta_b\vro x_{12}^+}-1\Big)\bigg\}
{t-p_\perp^2\over t+(p-k_b)_\perp^2-p_\perp^2+\ie}
~=~O\Big({Q_\perp^2\over \sigma_t\beta_bs}\Big)
\label{idiffe1}
\end{eqnarray}
Indeed, integrals over $p_\perp$ and $t$ converge at $p_\perp\sim x_\perp^{-1}\sim Q_\perp$ and $t$ between $Q_\perp^2$ and 
${\alpha'_a}{\beta_b}s$.  
At $t\sim {\alpha'_a}{\beta_b}s$  the first term is $\sim{Q_\perp^2\over Q^2}=\lambda$ and the second 
$\sim {Q_\perp^2\over Q^2}{\beta_b}\vro x_{12}^+\sim\lambda$. At $t\sim Q_\perp^2$ 
the first term is $\sim{Q_\perp^2\over Q^2}{\alpha'_a}\vro x_{12}^-\sim{Q_\perp^2\over Q^2\lambda}$ while the second can be rewritten as
\beq
\hspace{-0mm} 
4\pi\!\int_0^\infty\!{dt\over t} \!\int\!{\dhd p_\perp\over p_\perp^2}~e^{-i(p,x_{12})_\perp}\Big(e^{i{p_\perp^2\over t}\beta_b\vro x_{12}^+}-1\Big)
{\Big(1-e^{-i{t\over \sigma_t\beta_bs}}\Big)(t-p_\perp^2)
\over [t-p_\perp^2+(p-k_b)_\perp^2+\ie]}~\sim~O(\lambda_t)
\label{fla839}
\eeq

The second contribution to the Eq.  (\ref{iaminuseik}) is 
the difference between second terms in Eqs. (\ref{iodinae}) and (\ref{ieik11ab})
\bega
&&\hspace{-1mm}
8\pi^2\!\int_0^\infty\!{\dhd\alpha\over\alpha} \!\int\!\dhd p_\perp~\bigg[
{(\alpha+{\alpha'_a})
\Big(e^{i\alpha \vro x_{12}^- +i{\beta'_b}\vro x_{12}^+ +i{(p-k_b)_\perp^2\over\alpha s}\vro x_{12}^+}-1\Big)
\over 
\alpha({\alpha'_a}+\alpha)\beta_bs+({\alpha'_a}+\alpha)(p-k_b)_\perp^2-\alpha(p+k'_a)_\perp^2+\ie({\alpha'_a}+\alpha)}
\nonumber\\
&&\hspace{22mm}
-~e^{-i{\alpha\over\sigma_t}} {e^{i({\beta'_b}+{(p-k'_b)_\perp^2\over\alpha s})\vro x_{12}^+}-1
\over [\alpha\beta_b s+(p-k_b)_\perp^2+\ie]}\bigg]
{(p-k'_b)_\perp^2e^{-i(p,x_{12})_\perp}\over  \alpha\beta_bs-p_\perp^2+(p-k_b)_\perp^2+\ie}
\label{idiffe2}
\end{eqnarray}
Again, at $\alpha\sim{\alpha'_a}$ this integral is of order of $\lambda$ whereas at small $\alpha\ll{\alpha'_a}$ 
it turns to 
\bega
&&\hspace{-1mm}
8\pi^2\!\int_0^\infty\!{\dhd\alpha\over\alpha} \!\int\!\dhd p_\perp~
\Big[e^{i\alpha \vro x_{12}^- +i{\beta_b}\vro x_{12}^+ +i{(p-k_b)_\perp^2\over\alpha s}\vro x_{12}^+}-1
-e^{-i{\alpha\over\sigma_t}}\Big(e^{i({\beta_b}+{(p-k_b)_\perp^2\over\alpha s})\vro x_{12}^+}-1\Big)\Big]
\nonumber\\
&&\hspace{22mm}
\times~
{(p-k_b)_\perp^2e^{-i(p,x_{12})_\perp}\over   [\alpha\beta_b s+(p-k_b)_\perp^2+\ie][\alpha\beta_bs-p_\perp^2+(p-k_b)_\perp^2+\ie]}
\nonumber\\
&&\hspace{-1mm}
=~
4\pi\!\int_0^\infty\!{dt\over t} \!\int\!\dhd p_\perp~
\Big[e^{i(1+{(p-k_b)_\perp^2\over t})\beta_b\vro x_{12}^+}\big[e^{i{t\over\beta_bs} \vro x_{12}^-}-e^{i{t\over\sigma_t\beta_bs}}\big]+
e^{-i{t\over\sigma_t{\beta_b}s}}-1\Big]
\nonumber\\
&&\hspace{22mm}
\times~
{(p-k_b)_\perp^2e^{i(p,x_{12})_\perp}\over   [t +(p-k_b)_\perp^2+\ie][t-p_\perp^2+(p-k_b)_\perp^2+\ie]}
~=~O(\lambda_t)
\label{idiffe3}
\end{eqnarray}
since the integral over $t$ converges at $t\sim Q_\perp^2$. 

Thus, 
\bega
&&\hspace{-0mm}
I_{1a}({\alpha'_a}, k'_{a_\perp},\beta_b, k_{b_\perp}, x_2,x_1)-I_{\rm Fig.~\ref{fig:eik2}a,b}^{\rm eik}({\alpha'_a}, k'_{a_\perp},\beta_b, k_{b_\perp}, x_1,x_2)
\label{iaminuseikotvet}\\
&&\hspace{5mm}
=~I_{1a}({\alpha'_a}, k'_{a_\perp},\beta_b, k_{b_\perp}, x_{12_\perp})-I_{\rm Fig.~\ref{fig:eik2}a,b}^{\rm eik}({\alpha'_a}, k'_{a_\perp},\beta_b, k_{b_\perp},x_{12_\perp})
~+~O(\sigma_t)
\nonumber
\ega
Similarly, one can demonstrate that
\bega
&&\hspace{-0mm}
I_{1b}({\alpha'_a}, k'_{a_\perp},\beta_b, k_{b_\perp}, x_1,x_2)
-I_{\rm Fig. \ref{fig:eik1}e,f}^{\rm eik}({\alpha'_a}, k'_{a_\perp},\beta_b, k_{b_\perp}, x_1,x_2)
\label{iaminuseikotvet1}\\
&&\hspace{5mm}
=~I_{1b}({\alpha'_a}, k'_{a_\perp},\beta_b, k_{b_\perp}, x_{12_\perp})
-I_{\rm Fig. \ref{fig:eik1}e,f}^{\rm eik}(\alpha'_a, k'_{a_\perp},\beta_b, k_{b_\perp}, x_{12_\perp})   
~+~O(\sigma_p)
\nonumber
\ega
so we get
\bega
&&\hspace{-1mm}   
I_1({\alpha'_a}, k'_{a_\perp},\beta_b, k_{b_\perp}, x_1,x_2)-[I_{\rm Fig.~\ref{fig:eik2}a,b}^{\rm eik}
+I_{\rm Fig. \ref{fig:eik1}e,f}^{\rm eik}]({\alpha'_a}, k'_{a_\perp},\beta_b, k_{b_\perp}, x_1,x_2)
~-~(x_{12}^\parallel\rightarrow 0)
\nonumber\\
&&\hspace{11mm}
=~O\Big({Q_\perp^2\over \sigma_t\beta_bs}\Big)+O\Big({Q_\perp^2\over \sigma_p{\alpha'_a}s}\Big)~\sim~O(\lambda_p)+O(\lambda_t)
\label{iodinapprox}
\ega

By projectile$\leftrightarrow$target replacement we get 
\bega
&&\hspace{-1mm}   
I_2(\alpha_a, k_{a_\perp},{\beta'_b}, k'_{b_\perp}, x_1,x_2)-[I_{\rm Fig.~\ref{fig:eik2}e,f}^{\rm eik}+I_{\rm Fig. \ref{fig:eik1}a,b}^{\rm eik}](\alpha_a, k_{a_\perp},{\beta'_b}, k'_{b_\perp}, x_1,x_2)]~-~(x_{12}^\parallel\rightarrow 0)
\nonumber\\
&&\hspace{11mm}
=~O\Big({Q_\perp^2\over \sigma_t{\beta'_b}s}\Big)+O\Big({Q_\perp^2\over \sigma_p\alpha_as}\Big)
~\sim~O(\lambda_p)+O(\lambda_t)
\label{idvapprox}
\ega
This justifies  the calculation of the coefficient function in Eq. (\ref{jotodin}) at $x_{12}^\parallel=0$.

\subsection{Diagrams with correction field $\barC$ \label{sect:corfildz}}

In this Section we demonstrate that diagrams in the the background field $\barC$ lead to power corrections. 
The correction fields $\barC_\mu$ are given by Eq. (\ref{corfildz}) and we will also use the expressions for field strengths
\footnote{Note that $\barC_{\mu\nu}~\neq~ \partial_\mu\barC_\nu-\partial_\nu\barC_\mu-ig[\barC_\mu,\barC_\nu]$
} from Ref. \cite{Balitsky:2017flc}
\begin{eqnarray}
&&\hspace{-1mm}
\barC_{\mu\nu}~\equiv~F_{\mu\nu}(\matA)-F_{\mu\nu}(\barA)-F_{\mu\nu}(\barB),
\label{corefs}\\
&&\hspace{-1mm}
g\barC^{- i}(x^+,x^-)~=~{i\over 2}\!\int_{-\infty}^{x^+}\!dx'^+\!\int_{-\infty}^{x^-}\!dx'^-(x-x')^- 
\nonumber\\
&&\hspace{-1mm}
\times~\Big(\partial^i[U_{~k}^{+}(x'^-),V^{- k}(x'^+)]-\partial_k[U^{+ k}(x'^-),V^{- i}(x'^+)]+\partial^k[U^{+ i}(x'^-),V^{- k}(x'^+)]\Big)~\sim~{Q_\perp^3\over\sqrt{s}},
\nonumber\\
&&\hspace{-1mm}
g\barC^{+ i}(x^+,x^-)~=~-{i\over 2}\!\int_{-\infty}^{x^-}\!dx'^-\!\int_{-\infty}^{x^+}\!dx'^+(x-x')^+
\nonumber\\
&&\hspace{-1mm}
\times~\Big(\partial^i[U_{~k}^{+}(x'^-),V^{- k}(x'^+)]+\partial_k[U^{+ k}(x'^-),V^{- i}(x'^+)]-\partial_k[U^{+ i}(x'^-),V^{- k}(x'^+)]\Big)
~\sim~{Q_\perp^3\over\sqrt{s}},
\nonumber\\
&&\hspace{-1mm}
g\barC^{+-}(x)~=~-i[U_j(x^-,x_\perp),V^j(x^+,x_\perp)]~\sim~Q_\perp^2,
\nonumber\\
&&\hspace{-1mm}
g\barC_{ik}(x)~=~U_{ik}(x^-,x_\perp)+V_{ik}(x^+,x_\perp)-i\big[U_i(x^-,x_\perp)V_k(x^+,x_\perp)-i\leftrightarrow k\big)
~\sim~Q_\perp^2
\nonumber
\end{eqnarray}

There are two types of  diagrams with background field $\barC$ shown in Fig. \ref{fig:corr1}. 
\begin{figure}[htb]
\begin{center}
\includegraphics[width=131mm]{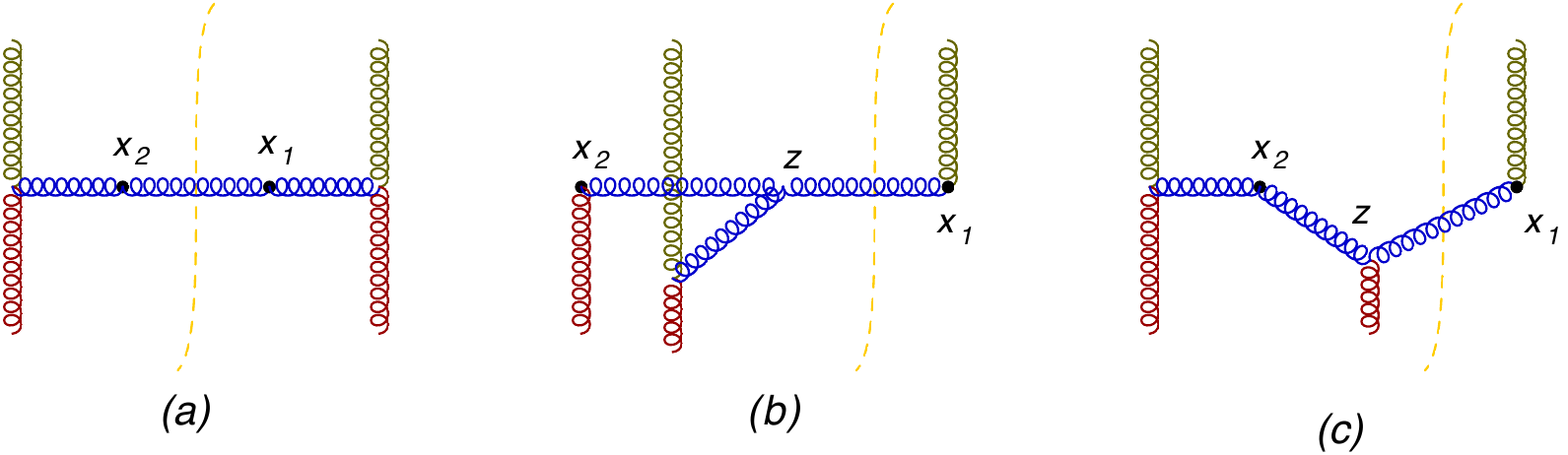}
\end{center}
\caption{Typical diagrams in the background correction field $\barC$.  \label{fig:corr1}    
}
\end{figure}

Let us start with the first one. We get
\begin{eqnarray}
&&\hspace{-1mm}
\barC^{-i}(x_2)\langle F^+_{~i}(x_2) F^-_{~j}(x_1)\rangle\barC^{+j}(x_1)
~=~{x_{12_\perp}^2g^{ij}+2x^i_{12}x^j_{12}\over\pi^2(-x_{12}^2-\ie x^0_{12})^3}\barC^-_{~i}(x_2)\barC^+_{~j}(x_1)
\label{corr1}\\
&&\hspace{-1mm}
\simeq~\Big(g^{ij}+2{x^i_{12}x^j_{12}\over x_{12_\perp}^2}\Big){1\over x_{12_\perp}^4}\barC^-_{~i}(x_2)\barC^+_{~j}(x_1)
~\sim~O\Big({Q_\perp^6\over s^3}\big)\times U^{- i}V^{+i}(x_2)U^{- j}V^{+j}(x_1)
\nonumber
\end{eqnarray}
since $\barC^{-i},\barC^{+j}\sim{Q_\perp^3\over\sqrt{s}}$, see Eq. (\ref{corefs}).

Next we consider diagram in Fig. \ref{fig:corr1}b where we   replaced Feynman propagator by 
the retarded one according to Eq. (\ref{gluprop2}).
\begin{eqnarray}
&&\hspace{-1mm}
V^{-i}(x_2)\langle (\cald^+ A_i-\cald_i A^+)(x_2)(\cald^- A_j-\cald_j A^-)(x_1)\rangle U^{+j}(x_1)~=~
\nonumber\\
&&\hspace{11mm}
=~-V^{-i}(x_2)(x_2|(p^+\delta_i^\alpha-p_ig^{+\alpha}){1\over p^2+\ie p_0}
\Big([\{p^\lambda,\barC_\lambda\}+\barC^2]g^{\alpha\beta}+2i\barC^{\alpha\beta}\Big)
\nonumber\\
&&\hspace{22mm}
\times~~\tilde\delta_+(p)
(p^-\delta^\beta_j-p_jg^{\beta -})|x_1) U^{+j}(x_1)
\label{vkladotfig7b}
\end{eqnarray}
Looking at power counting for the correction fields (\ref{corfildz}) and (\ref{corefs}) we see that the largest contribution to
the r.h.s. of Eq. (\ref{vkladotfig7b}) comes from the term
\begin{eqnarray}
&&\hspace{-1mm}
V^{-i}(x_2)\langle (\cald^+ A_i-\cald_i A^+)(x_2)(\cald^- A_j-\cald_j A^-)(x_1)\rangle U^{+j}(x_1)~=~
\nonumber\\
&&\hspace{-1mm}
=~-V^{-i}(x_2|p^+{1\over p^2+\ie p_0}
\big(\{p^\lambda,\barC_\lambda\}g^{ij}+\barC^2g^{ij}+2i\barC^{ij}\big)\tilde\delta_+(p)p^-|x_1) U^{+j}(x_1)
\label{vkladotfig7be}
\end{eqnarray}
Let us estimate the term with $\barC^{ij}$. It is similar to Eq. (\ref{fla74}), only instead of $U^{+i}{1\over p^2}V^{-j}\geq m_\perp^2$ we have 
here $\barC_{ij}\sim{m_\perp^4\over s}$ or  $\barC^+\barC^-\sim{m_\perp^4\over s}$,  see Eq. (\ref{corfildz}). Next, let us consider term with $\{p^\lambda,\barC_\lambda\}
=\{p^+ ,C^-\}+\{p^- ,C^+\}+\{p_i ,C^i\}$. From Eq. (\ref{corfildz}) we see that the last term is $O\big({m_\perp^2\over s}\big)$ with respect to the first two terms so we get
\footnote{For the estimate of power corrections the exact form of singularity in the gluon propagator is not important}
\begin{eqnarray}
&&\hspace{-1mm}
(x_2|p^+{1\over p^2+\ie p_0}(\{p^+ ,C^-\}+\{p^- ,C^+\}) \tilde\delta_+(p)p^-|x_1)^{ab}~
\label{fla1142}\\
&&\hspace{-1mm}
=~(x_2|{1\over p^2}\Big(\{p^+ ,i\partial^+C^-\}+\{p^- ,i\partial^+C^+\}) p^-
+(\{p^+ ,C^-\}+\{p^- ,C^+\}) {p_\perp^2\over 2}\Big)\tilde\delta_+(p)|x_1)^{ab}
\nonumber
\end{eqnarray}
Since $\partial^+,\partial^-\sim \sqrt{s}$ the second term is $O\big({p_\perp^2\over s}\sim {m_\perp^2\over s}\big)$ in comparison to the first one,
the r.h.s. of Eq. (\ref{fla1142}) reduces to
\begin{eqnarray}
&&\hspace{-1mm}
(x_2|{1\over p^2}\Big(\{p^+ ,i\partial^+C^-\}+\{p^- ,i\partial^+C^+\}) p^-\tilde\delta_+(p)|x_1)^{ab}
\label{fla1143}\\
&&\hspace{-1mm}
=~{1\over 2}\!\int\! \dhd{\alpha'_a}\dhd k'_{a_\perp}\dhd{\beta_b}\dhd k_{b_\perp}
{e^{-ik'_ax_2-ik_bx_1}\over{\alpha'_a}{\beta'_b}}[U^+_{~i}({\alpha'_a}, k'_{a_\perp}),V^{-i}({\beta_b},k_{b_\perp})]^{ab}
\!\int\!\dhd\alpha\dhd\beta\dhd p_\perp~
\nonumber\\
&&\hspace{11mm}
\times~e^{i\alpha\vro x_{12}^-+i\beta\vro x_{12}^+ -i(p,x_{12})_\perp}{\theta(\beta-{\beta_b})(-{\alpha'_a})\over(\alpha+{\alpha'_a})\beta s-(p+k_a)_\perp^2}\Big[1+{\alpha\over{\alpha'_a}}-{\beta\over{\beta_b}}\Big]
\tilde{\delta}\Big(\alpha-{(p-k'_a)_\perp\over(\beta-{\beta_b})s}\Big)
\nonumber\\
&&\hspace{-1mm}
=~-\!\int\! \dhd{\alpha'_a}\dhd  k'_{a_\perp}\dhd{\beta_b}\dhd k_{b_\perp}
{e^{-ik'_ax_2-ik_bx_1}\over 2{\beta_b}}[U^+_{~i}({\alpha'_a},k'_{a_\perp}),V^{-i}({\beta_b},k_{b_\perp})]^{ab}
\!\int\!\dhd\beta\dhd p_\perp~
\nonumber\\
&&\hspace{11mm}
\times~
{\theta(\beta)\beta e^{i{(p-k_b)_\perp^2\over\beta s}\vro x_{12}^- +i(\beta+{\beta_b})\vro x_{12}^+ -i(p,x_{12})_\perp}\over {\alpha'_a}\beta(\beta+{\beta_b}) s+(p-k_b)_\perp^2(\beta+{\beta_b})-(p+k'_a)_\perp^2\beta}
\Big[{\beta\over{\beta_b}}-{(p-k_b)_\perp^2\over{\alpha'_a}\beta s}\Big]
\nonumber\\
&&\hspace{-1mm}
=~\!\int\! \dhd{\alpha'_a}\dhd k'_{a_\perp}\dhd{\beta_b}\dhd k_{b_\perp}
[U^+_{~i}({\alpha'_a},k'_{a_\perp}),V^{-i}({\beta_b},k_{b_\perp})]^{ab}
e^{-i{\alpha'_a}\vro x_2^- +i(k'_a,x_2)_\perp-i{\beta_b}\vro x_1^- +i(k_b,x_1)_\perp}
\nonumber\\
&&\hspace{11mm}
\times~{1\over4\pi  }
\!\int\!\dhd p_\perp\!\int_0^\infty\! dt{[t^2-Q_{ab}^2(p-k_b)_\perp^2] 
e^{it^{-1}(p-k_b)_\perp^2{\alpha'_a}\vro x_{12}^- +i\big(Q_{a'b}^{-2}t+1\big){\beta_b}\vro x_{12}^+ -i(p,x_{12})_\perp}
\over Q_{a'b}^4[t(t+Q_{a'b}^2)-(p-k _b)_\perp^2(t+Q_{a'b}^2)+(p+k'_a)_\perp^2t]}
\nonumber
\end{eqnarray}
This should be compared to the leading order contribution (\ref{calcula1})\\
$\sim U^+_{~i}({\alpha'_a},k'_{a_\perp})V^{-j}({\beta_b},k_{b_\perp})\times$logs. It is clear that contribution from 
$t\lesssim Q_{a'b}^2$ to the last line in Eq. (\ref{fla1143}) is $O\big({Q_\perp^2\over Q_{a'b}^2}\big)$, and if $t\gg Q_{a'b}^2$ the last line in the above equation reduces to
\bega
&&\hspace{-1mm}
{1\over 4\pi Q_{a'b}^4}
\!\int\!\dhd p_\perp\!\int_0^\infty\! dt~
e^{it^{-1}(p-k_b)_\perp^2{\alpha'_a}\vro x_{12}^- +iQ_{a'b}^{-2}t{\beta_b}\vro x_{12}^+ -i(p,x_{12})_\perp}
\\
&&\hspace{-1mm}
=~-{ie^{-i(k_b,x_{12})_\perp}\over 16\pi^2 Q_{a'b}^4}
\!\int_0^\infty\! dt~{t\over{\alpha'_a}\vro x_{12}^-}
e^{iQ_{a'b}^{-2}t{\beta_b}\vro x_{12}^+ -it{x_{12_\perp}^2\over 4{\alpha'_a}\vro x_{12}^-}}
~=~-{i{\alpha'_a}\vro x_{12}^- e^{-i(k_b,x_{12})_\perp}\over \pi^2 Q_{a'b}^4(x_{12_\perp}^2-x_{12_\parallel}^2)^2}
\nonumber
\end{eqnarray}
which is a power correction $\sim {Q_\perp^4\over Q_{ab}^4}$ since ${\alpha'_a}\vro x_{12}^-\sim 1$. Thus, the contribution of the diagram in Fig. \ref{fig:corr1}b is a power correction.

Finally, let us consider diagram in Fig. \ref{fig:corr1}c.
We get
\begin{eqnarray}
&&\hspace{-1mm}
\barC^{-i}(x_2)\langle (\cald^+ A_i-\cald_i A^+)(x_2)(\cald^- A_j-\cald_j A^-)(x_1)\rangle_\barB U^{+j}(x_1)~=~
\nonumber\\
&&\hspace{-1mm}
=~-\barC^{-i}(x_2)(x_2|(p^+\delta_i^\alpha-p_ig^{+\alpha}){1\over p^2+\ie p_0}\barB_{\alpha\beta}2i\tilde\delta_+(p)
(p^-\delta^\beta_j-p_jg^{\beta -})|x_1)U^{+j}(x_1)
\nonumber\\
&&\hspace{-1mm}
=~-2i\barC^{-i}(x_2)(x_2|{p^+\over p^2+i\epsilon p_0}V_{-i }\tilde\delta_+(p)p^j|x_1)U^+_{~j}(x_1)
\nonumber\\
&&\hspace{-1mm}
=~-i\barC^{-;a}_{~~i}(x_2)U^{+;b}_{~j}(x_1)\!\int\! \dhd{\beta_b}\dhd k_{b_\perp}
V^{-i;ab}({\beta_b},k_{b_\perp})e^{i{\beta_b}\vro x_1^+ -i(k_b,x_1)_\perp}
\nonumber\\
&&\hspace{-1mm}
\times~\!\int\!\dhd\alpha\dhd\beta\dhd p_\perp~e^{i\alpha\vro x_{12}^- +i(\beta+{\beta_b})\vro x_{12}^+ -i(p+k_b,x_{12})_\perp}
{\vro(\beta+{\beta_b})(p+k_b)^j\over\alpha(\beta+{\beta_b}) s-(p+k_b)_\perp^2}{\theta(\beta)\over \beta}
\tilde{\delta}\Big(\alpha-{p_\perp^2\over \beta s}\Big)
\nonumber\\
&&\hspace{-1mm}
=~-{i\over 2\pi}\vro\barC^{-;a}_{~~i}(x_2)U^{+j;b}(x_1)\!\int\! \dhd{\beta_b}\dhd k_{b_\perp}
V^{-i;ab}({\beta_b},k_{b_\perp})e^{i{\beta_b}\vro x_1^+ -i(k_b,x_1)_\perp}
\nonumber\\
&&\hspace{6mm}
\times~\!\int_0^\infty\! d\beta\!\int\!\dhd p_\perp~e^{i{p_\perp^2\over\beta s}\vro x_{12}^- +i(\beta+{\beta_b})\vro x_{12}^+ -i(p+k_b,x_{12})_\perp}
{(\beta+{\beta_b})(p+k_b)^j\over p_\perp^2(\beta+{\beta_b}) -(p+k_b)_\perp^2\beta}
\label{fig14c}
\end{eqnarray}
Since $\vro \barC^{-i}\sim m_\perp^3$ in comparison to $U^{+i}V^{-k}\sim sm_\perp^2$ in the leading term,
we need an extra $s/m_\perp$ from the last line.
At finite $\beta\sim{\beta'_b}$  the integral in the last line is $\sim k_a^j\sim m_\perp$ so we need to check this
integral at very small or very large $\beta$. It is convenient to make change of variables $p_\perp=k_\perp\sqrt{b}$:
\bega
&&\hspace{-1mm}
\!\int_0^\infty\! d\beta\!\int\!\dhd p_\perp~e^{i{p_\perp^2\over\beta s}\vro x_{12}^- +i(\beta+{\beta_b})\vro x_{12}^+ -i(p+k_b,x_{12})_\perp}
{(\beta+{\beta_b})(p+k_b)^j\over p_\perp^2(\beta+{\beta_b}) -(p+k_b)_\perp^2\beta}
\\
&&\hspace{-1mm}
=~\!\int_0^\infty\! d\beta\!\int\!\dhd k_\perp~e^{i{k_\perp^2\over s}\vro x_{12}^- +i(\beta+{\beta_b})\vro x_{12}^+ -i(k\sqrt{\beta}+k_b,x_{12})_\perp}
{(\beta+{\beta_b})(k\sqrt{\beta}+k_b)^j\over k_\perp^2(\beta+{\beta_b}) -(k\sqrt{\beta}+k_b)_\perp^2}
\nonumber
\ega
As $\beta\rightarrow 0$ we get
\bega
&&\hspace{-2mm}
e^{i{\beta_b}\vro x_{12}^+ -i(k_b,x_{12})_\perp}
\!\int_0^\infty\!\! d\beta~e^{i\beta\vro x_{12}^-}\!\int\!\dhd k_\perp~
{k_b^je^{i{k_\perp^2\over s}\vro x_{12}^-}\over k_\perp^2 -{k_{b_\perp}^2\over{\beta_b}}}
~\sim {{\beta_b}k_b^j\over {\beta_b}\vro x_{12}^+}\ln{Q_{ab}^2\over k_{b_\perp}^2{\alpha'_a}\vro x_{12}^-}
~\sim~{\beta_b} m_\perp
\nonumber\\
\end{eqnarray}
since ${\beta_b}\vro x_{12}^+\sim{\alpha_a}\vro x_{12}^-\sim 1$.  Conversely, as $\beta\rightarrow\infty$ one obtains
\bega
&&\hspace{-1mm}
\!\int_0^\infty\! d\beta\!\int\!\dhd p_\perp~e^{i{p_\perp^2\over\beta s}\vro x_{12}^- +i\beta\vro x_{12}^+ -i(p+k_b,x_{12})_\perp}
{(p+k_b)^j\over k_{b_\perp}^2 +2(p,k_b)_\perp}
\\
&&\hspace{-1mm}
=~2e^{-i(k_b,x_{12})_\perp}
\!\int\!\dhd p_\perp~\sqrt{p_\perp^2x_{12}^-\over sx_{12}^+}K_1\Big(p_\perp\sqrt{-x_{12_\parallel}^2}\Big)
{(p+k_b)^je^{-i(p,x_{12})_\perp}\over k_{b_\perp}^2 +2(p,k_b)_\perp}~\sim~{m_\perp\over\rho x_{12}^+}
~\sim~{\beta_b} m_\perp
\nonumber
\ega
Thus,  we got  $m_\perp$ instead of an extra $s/m_\perp$ needed 
to compensate the smallness in Eq. (\ref{fig14c}) so the contribution of the diagram in Fig. \ref{fig:corr1}c
 is a power correction $O\big({m_\perp^2\over s}\big)$ in comparison to the leading term  (\ref{calcula1}).

Summarizing, we demonstrated that the diagrams with correction fields  (\ref{corfildz}) are power corrections 
$\sim O\big({1\over\zeta}\big)$.

\subsection{Necessary integrals \label{sect:integrals}}

\subsubsection{Integrals for virtual diagrams\label{app:vints}}
Master integral for virtual diagrams can be taken from integrals (11) - (18) of Ref. \cite{Usyukina:1993ch}.  At $k_1^2,k_2^2<0$ and 
$(k_1+k_2)^2>0$ it reads
\begin{eqnarray}
&&\hspace{-1mm}
\!\int\! {\dhd^4p\over i}{1\over[(p+k_1)^2+\ie][(p-k_2)^2+\ie](p^2+\ie)}~=~
\label{ussyu}\\
&&\hspace{11mm}
=~{1\over 16\pi^2\kappa}\Bigg[{\rm Li}_2({-k_1^2\over (k_1,k_2)+\kappa})+{\rm Li}_2({-k_2^2\over (k_1,k_2)+\kappa})
+{1\over 2}\ln{k_2^2\over k_1^2}\ln{k_2^2+k_1\cdot k_2+\kappa\over k_1^2+k_1\cdot k_2+\kappa}
\nonumber\\
&&\hspace{24mm}
+~{1\over 2}\ln\Big({k_1^2\over (k_1,k_2)+\kappa}+\ie\Big)\ln\Big({k_2^2\over (k_1,k_2)+\kappa}+\ie\Big)+{\pi^2\over 6}\Bigg]
\nonumber
\end{eqnarray}
where $\kappa\equiv\sqrt{(k_1\cdot k_2)^2-k_1^2k_2^2}$. 
In our kinematics $k_1\cdot k_2\gg k_1^2,k_2^2$ so
\begin{eqnarray}
&&\hspace{-1mm}
\!\int\! {\dhd^4p\over i}{1\over[(p+k_1)^2+\ie][(p-k_2)^2+\ie](p^2+\ie)}
\label{integral1}\\
&&\hspace{11mm}
=~{1\over 32\pi^2(k_1\cdot k_2)}\ln{-2(k_1\cdot k_2)-\ie\over -k_1^2}\ln{-2(k_1\cdot k_2)-\ie\over -k_2^2}~+~O\Big({k_1^2,k_2^2\over (k_1\cdot k_2)}\Big)
\nonumber
\end{eqnarray}
We will also need similar integral with cut propagators.  The standard calculation yields
\begin{eqnarray}
&&\hspace{-1mm}
\int\! \dhd^4p~\tilde{\delta}(p+k_1)^2(p+k_1)_0{1\over p^2}\tilde{\delta}(p-k_2)^2\theta(k_2-p)_0
\label{imtegral1}\\
&&\hspace{-1mm}
=~-{\theta(k_1+k_2)^2\theta(k_1+k_2)_0\over 32\pi\kappa}\ln{k_1\cdot k_2+\kappa\over k_1\cdot k_2-\kappa}~\simeq~
-{\theta(k_1+k_2)^2\theta(k_1+k_2)_0\over 32\pi k_1\cdot k_2}\ln{4(k_1\cdot k_2)^2\over k_1^2k_2^2}
\nonumber
\end{eqnarray}
in agreement with Eq. (\ref{integral1}).
Note that at $k_1^2,k_2^2<0$ the denominator $1/p^2$ in the l.h.s. of this equation is not singular.

Using Eqs. (\ref{integral1}) and (\ref{imtegral1}) it is easy to obtain
\begin{eqnarray}
&&\hspace{-2mm}
\int\!\dhd^4p{{\alpha_a}\over (p+k_a)^2+\ie}{s\over p^2+\ie}{{\beta_b}\over (p-k_b)^2+\ie}
~=~{i\over 16\pi^2}\Big[\ln{-{\alpha_a}{\beta_b}s-\ie\over k_{a_\perp}^2}\ln{-\alpha_a\beta_bs-\ie\over k_{b_\perp}^2}+{\pi^2\over 3}\Big]
\nonumber\\
&&\hspace{-1mm}
\!\int\!\dhd^4p~\tilde{\delta}(p+k_a)^2\theta(-\beta){{\alpha_a}{\beta_b}s\over p^2-\ie}\tilde{\delta}(p-k_b)^2\theta(\alpha)
~=~-{\theta(-{\alpha_a})\theta(-{\beta_b})\over 8\pi}\ln{({\alpha_a}{\beta_b}s)^2\over k_{a_\perp}^2k_{b_\perp}^2}
\label{integrali}
\end{eqnarray}
To compare to the calculation of the ``production'' diagrams in Sect. \ref{sect:real}, 
we need also an explicit calculation of the sum of the ``virtual'' integrals in Eq. (\ref{virtsum}).
Performing the integration over $\alpha$,  we get
\begin{eqnarray}
&&\hspace{-1mm}
16\pi^2\!\int{\dhd^4p\over i}\Big\{{\alpha_a\beta_bs\over [(\alpha+\alpha_a)\beta s-(p+k_a)_\perp^2+\ie](\alpha\beta s-p_\perp^2+\ie)
[\alpha(\beta-\beta_b)s-(p-k_b)_\perp^2+\ie]}
\nonumber\\
&&\hspace{11mm}
+~\tilde{\delta}[(\alpha_a+\alpha)\beta s-(p+k_a)^2]\theta(-\beta){\alpha_a\beta_bs\over \alpha\beta s-p_\perp^2-\ie}
\tilde{\delta}[\alpha(\beta-\beta_b)s-(p-k_b)_\perp^2]\theta(\alpha)\Big\}
\nonumber\\
&&\hspace{11mm}
=~-4\pi\!\int\!\dhd p_\perp\!\int_0^1\! du{\baru Q_{ab}^2\over [\baru p_\perp^2+u(p-k_b)^2]
[\baru(p+k_a)_\perp^2+u(p-k_b)_\perp^2
-Q_{ab}^2\baru u]}
\nonumber\\
&&\hspace{11mm}
=~\ln{-Q_{ab}^2\over k_{a_\perp}^2}\ln{-Q_{ab}^2\over k_{b_\perp}^2}+{\pi^2\over 3}~+~O(\lambda)
\label{virtegralo}
\ega
where $Q_{ab}^2$ is defined in Eq. (\ref{kuab}). To get the last line in the above equation, 
we performed the integration over $p_\perp$ 
\begin{eqnarray}
&&\hspace{-5mm}
{\rm Eq.~(\ref{virtegralo})}~=~\!\int_0^1\! dudv{ -Q_{ab}^2\over k_{a_\perp}^2v(1-\baru v)+k_{b_\perp}^2u-
(Q_{ab}^2-2(k_a,k_b)_\perp)uv}
\nonumber\\
&&\hspace{-1mm}
=~\!\int_0^1\! dudv{1\over av(1-\baru v)+bu+uv}~+~O(a,b)
\nonumber\\
&&\hspace{-1mm}
=~\!\int_0^1\! dudv\Big[{1\over av+bu+uv}
+{a\baru v^2\over [av+bu+uv][av(1-\baru v)+bu+uv]}\Big]
\nonumber\\
&&\hspace{-1mm}
=~\!\int_0^1\! dudv\Big[{1\over av+bu+uv}
+{a\baru v^2\over [av+bu+uv][av(1-\baru v)+bu+uv]}\Big]~+~O(a,b)
\nonumber\\
&&\hspace{-1mm}
=~\!\int_0^1\! dudv\Big[{1\over av+bu+uv}
+{a\over (a+u)(a\barv+u)}\Big]~+~O(a,b)~=~\ln a\ln b+{\pi^2\over 3}~+~O(a,b)
\nonumber
\end{eqnarray}
where $a= -k_{a_\perp}^2/Q_{ab}^2$ and $b= -k_{b_\perp}^2/Q_{ab}^2$.
\subsubsection{Integrals for ``production'' diagrams\label{app:rints}}
To calculate the integral (\ref{fla721}) we will represent it as follows
\bega
&&\hspace{-1mm}
4\pi\!\int\!\dhd p_\perp~
{e^{-i(p,x)_\perp}Q_{ab}^2\over Q_{ab}^2p_\perp^2+(p+k_a)_\perp^2(p-k_b)_\perp^2}
\ln{-Q_{ab}^2p_\perp^2\over (p+k_a)_\perp^2(p-k_b)_\perp^2}
\nonumber\\
&&\hspace{1mm}
=~4\pi\!\int\!\dhd p_\perp~
{e^{-i(p,x)_\perp}Q_{ab}^2\over Q_{ab}^2p_\perp^2+k_{a_\perp}^2k_{b_\perp}^2}
\ln{-Q_{ab}^2p_\perp^2\over (p+k_a)_\perp^2(p-k_b)_\perp^2}
\nonumber\\
&&\hspace{11mm}
+~{4\pi\over Q_{ab}^2}\!\int\!\dhd p_\perp~
{e^{-i(p,x)_\perp}[k_{a_\perp}^2k_{b_\perp}^2-(p+k_a)_\perp^2(p-k_b)_\perp^2]
\over \Big[p_\perp^2+{k_{a_\perp}^2k_{b_\perp}^2\over Q_{ab}^2}\big]
\big[p_\perp^2+{(p+k_a)_\perp^2(p-k_b)_\perp^2\over Q_{ab}^2}\big]}
\ln{-Q_{ab}^2p_\perp^2\over (p+k_a)_\perp^2(p-k_b)_\perp^2}
\nonumber\\
&&\hspace{1mm}
=~4\pi\!\int\!\dhd p_\perp~
{e^{-i(p,x)_\perp}Q_{ab}^2\over Q_{ab}^2p_\perp^2+k_{a_\perp}^2k_{b_\perp}^2}
\ln{-Q_{ab}^2p_\perp^2\over (p+k_a)_\perp^2(p-k_b)_\perp^2}~+~O(\lambda)
\ega
To get the last line we note that the integral in the third line is at best logarithmic at small $p_\perp$. Next, we split 
the integral in the last line in three parts
\bega
&&\hspace{-1mm}
8\pi\!\int\!\dhd p_\perp~
{e^{-i(p,x)_\perp}\over p_\perp^2+{k_{a_\perp}^2k_{b_\perp}^2\over Q_{ab}^2}}
\Big[\ln{-Q_{ab}^2p_\perp^2\over k_{a_\perp}^2k_{b_\perp}^2}-\ln{(p+k_a)_\perp^2\over k_{a_\perp}^2}
-\ln{(p-k_b)_\perp^2\over k_{b_\perp}^2}\Big],
\ega
use integrals ($m^2=-{k_{a_\perp}^2k_{b_\perp}^2\over Q_{ab}^2}$)
\begin{eqnarray}
&&\hspace{-1mm}
4\pi\!\int\!\dhd^2 p{e^{-i(p,x)}\over p^2-m^2}\ln{p^2\over m^2}~=~\half\Big(\ln{m^2x^2\over 4}+2\gamma\Big)^2
+{\pi^2\over 3}~+~O(m^2x^2),
\nonumber\\
&&\hspace{-1mm}
4\pi\!\int\!\dhd^2 p{e^{-i(p,x)}\over p^2}\ln{(p+k)^2\over k^2}~=~
\half\big(\ln{k^2x^2\over 4}+2\gamma\big)^2-I_K(k,x)
\ega
where  $I_K$ is defined in Eq. (\ref{ika}), 
and get
\bega
&&\hspace{-1mm}
4\pi\!\int\!\dhd p_\perp~
{e^{-i(p,x)_\perp}Q_{ab}^2\over Q_{ab}^2p_\perp^2+(p+k_a)_\perp^2(p-k_b)_\perp^2}
\ln{-Q_{ab}^2p_\perp^2\over (p+k_a)_\perp^2(p-k_b)_\perp^2}
\nonumber\\
&&\hspace{11mm}
=~\ln{-Q_{ab}^2\over k_{a_\perp}^2}\ln{-Q_{ab}^2\over k_{b_\perp}^2}-\half\Big(\ln{-Q_{ab}^2x_\perp^2\over 4}+2\gamma\Big)^2
\nonumber\\
&&\hspace{22mm}
+~{\pi^2\over 3}+I_K(k_{a_\perp},x_\perp)+I_K(-k_{b_\perp},x_\perp)~+~O(\lambda)
\label{mintegral1}
\ega

 For the calculation of the integral (\ref{ieiktarget1}) we need also
\beq
 4\pi\!\int\!{\dhd^2 p\over p^2}\big(e^{-ipx}-1\big)\ln{(p+q)_\perp^2\over p^2}
 ~=~\bigg\{4\pi\!\int\!\dhd^2 p\big(e^{-ipx}-1\big){1\over (p^2)^{1+\delta}[(q+p)^2]^{-\delta}}\bigg\}_{(\delta)}
\eeq
 where $\big\{..\big\}_{(\delta)}$ denotes the first term of the expansion in power of $\delta$.
After some algebra, one obtains
\begin{eqnarray}
&&\hspace{-5mm}
 4\pi\!\int\!{\dhd^2 p\over p^2}\big(e^{-ipx}-1\big)\ln{(p+q)_\perp^2\over p^2}
 \nonumber\\
&&\hspace{-5mm}
=~
\bigg\{4\pi\!\int\!\dhd p\big(e^{-ipx}-1\big){1\over (p^2)^{1+\delta}[(q+p)^2]^{-\delta}}\bigg\}_{(\delta)}~
=~2\!\int_0^1\! {du\over u}~\big(1-e^{i(qx)u}\big)K_0(qx\sqrt{\baru u})
\nonumber\\
&&\hspace{-1mm}
+~\bigg[{\delta\over \Gamma(1-\delta)\Gamma(1+\delta)}\!\int_0^1\! du~u^{-\delta-1}\baru^\delta \big\{-(\ln q^2x^2\baru u-\ln 4+2\gamma)
-2K_0(qx\sqrt{\baru u})\big\}\bigg]_\delta  
\nonumber\\
&&\hspace{-1mm}
=~2\!\int_0^1\! {du\over u}~\big(1-e^{i(qx)u}\big)K_0(qx\sqrt{\baru u})-\!\int_0^1\! {du\over u}\big[2K_0(qx\sqrt{\baru u})-\big(\ln{q^2x^2\baru u\over 4}+2\gamma\big)\big]
\nonumber\\
&&\hspace{-1mm}
=~-\!\int_0^1\! {du\over u}\big[\ln{q^2x^2\baru u\over 4}+2\gamma+2e^{i(q,x)u}K_0(qx\sqrt{\baru u})\big]
~=~-I_K(q_\perp,x_\perp)
\label{eikfla}
\end{eqnarray}
where  $I_K$ is defined in Eq. (\ref{ika}).

\subsection{Coefficient function from the calculation with background gluons on the mass shell\label{onmashell}}
In this Section we double-check the calculation of the coefficient function (\ref{masteresult}) using background fields 
on the mass shell  (\ref{bezmassi}).
For the background fields on the mass shell the hadronic tensor (\ref{pizw}) is parametrized as follows
\begin{eqnarray}
&&\hspace{-1mm}
\pizw(x_2,x_1)-\pizw^{\sigma_p,\sigma_t}_{\rm eik}(x_2,x_1)~=~
\!\int\!\dhd{\alpha'_a}\dhd{\beta'_b}\dhd\alpha_a\dhd\beta_b
e^{-i{\alpha'_a}\vro x_2^- -i\alpha_a\vro x_1^-}e^{-i{\beta'_b}\vro x_2^+ -i\beta_b\vro x_1^+}
\nonumber\\
&&\hspace{5mm}
\times~U^{+,b}_{~~i}({\alpha'_a})V^{-i,a}({\beta'_b})
U^{+,b}_{~j}(\alpha_a)V^{-j,a}(\beta_b)
[I-I^{\sigma_p,\sigma_t}_{\rm eik}]({\alpha'_a},\alpha_a,{\beta'_b},\beta_b,  x_2,x_1)
\label{paramashell}
\ega
As demonstrated below (see Sect. \ref{sec:sGmasshell}),
for the massless background fields there are no soft contributions, and Glauber gluons
cancel as usual.

\subsubsection{Virtual contributions}
Let us again start from the virtual diagram and take $\beta_b<0$. From Eq. (\ref{virtsG}) we get
\begin{eqnarray}
&&\hspace{-1mm}
I^{\rm virt~msh}_{\rm Fig. \ref{fig:nlovirt}}~=~-16\pi^2\!\int{\dhd p\over i}
\Big\{{\alpha_a\beta_bs
(\alpha\beta s-p_\perp^2+\ie)^{-1}\over [(\alpha+\alpha_a)\beta s-p_\perp^2+\ie]
[\alpha(\beta-\beta_b)s-p_\perp^2+\ie]}
\nonumber\\
&&\hspace{-1mm}
+~\tilde{\delta}[(\alpha_a+\alpha)\beta s-p^2]\theta(-\beta){\alpha_a\beta_bs\over \alpha\beta s-p_\perp^2-\ie}
\tilde{\delta}[\alpha(\beta-\beta_b)s-p_\perp^2]\theta(\alpha)\Big\}
\nonumber\\
&&\hspace{-1mm}
=~-4\pi\!\int_0^1\! du\!\int\! {\dhd p_\perp\over p_\perp^2}{\baru\alpha_a|\beta_b|s\over \baru u\alpha_a|\beta_b|s+p_\perp^2+\ie}
=~-{1\over\ve^2}\big({(\alpha_a+\ie)|\beta_b|s\over 4\pi}\big)^\ve{\Gamma(1-\ve)\Gamma^2(1+\ve)\over \Gamma(1+2\ve)}
\nonumber\\
&&\hspace{-1mm}
=~-{1\over\ve^2}-{1\over\ve}\Big(\ln{(\alpha_a+\ie)|\beta_b|s\over 4\pi}+\gamma\Big)
-\half\Big(\ln{(\alpha_a+\ie)|\beta_b|s\over 4\pi}+\gamma\Big)^2+{\pi^2\over 12}+{\pi^2\over 6}~+~O(\ve)
\nonumber\\
\ega
where $\ve={d_\perp\over 2}-1={d\over 2}-2$. 
It is easy to see that  for $\beta_b>0$ the singularity is changed to $\ln{-(\alpha_a+\ie)\beta_bs\over 4\pi}$. Adding the similar contribution 
of Fig. \ref{fig:nlovirtleft} diagram, we obtain
\begin{eqnarray}
&&\hspace{-1mm}
I^{\rm virt}_{\rm mass~shell}~=~\label{ivirtmsh}\\
&&\hspace{-1mm}
=~-{2\over\ve^2}-{1\over\ve}\Big(\ln{-(\alpha_a+\ie)(\beta_b+\ie)s\over 4\pi}+\gamma\Big)
-\half\Big(\ln{-(\alpha_a+\ie)(\beta_b+\ie)s\over 4\pi}+\gamma\Big)^2~
\nonumber\\
&&\hspace{-1mm}
-~{1\over\ve}\Big(\ln{-(\alpha'_a+\ie)(\beta'_b+\ie)s\over 4\pi}+\gamma\Big)
-\half\Big(\ln{-(\alpha'_a+\ie)(\beta'_b+\ie)s\over 4\pi}+\gamma\Big)^2~+~O(\ve)
\nonumber
\ega

Let us consider now virtual eikonals. From Eq. (\ref{ieiktarget1}) we get
\begin{eqnarray}
&&\hspace{-0mm}
I^{\rm eik}_{\rm Fig.~\ref{fig:eik2}c,d}(\beta_b,k_b)\Big|_{k_{b_\perp}=0}
\nonumber\\
&&\hspace{-0mm}
=~-8\pi^2\!\int_0^\infty\!\dhd\alpha~e^{-i{\alpha\over\sigma_t}}\!\int\!{\dhd p_\perp\over  p_\perp^2}
{\beta_bs\over  \alpha(\beta_b+\ie)s+p_\perp^2}
=~-{2\over\ve^2}\Big({-i\beta_b\sigma_ts\over 4\pi}\Big)^\ve\Gamma(1-\ve)\Gamma(1+\ve)
\nonumber\\
&&\hspace{-0mm}
=~-{1\over\ve^2}-{1\over\ve}[\ln(-i\beta_b\sigma_ts)-\ln4\pi]-\half[\ln(-i\beta_b\sigma_ts)-\ln4\pi]^2-{\pi^2\over 6}~+~O(\ve)
\label{ieiktarget1bezmass}
\end{eqnarray}
Similarly,
\begin{eqnarray}
&&\hspace{-1mm}    
I^{\rm eik}_{\rm Fig.~\ref{fig:eik2}g,h}({\beta'_b},k'_b)\Big|_{k_{b_\perp}=0}~=~-8\pi^2
\!\int_0^{\infty}\!\dhd\alpha\!\int\!{\dhd p_\perp\over p_\perp^2}~e^{i{\alpha\over \sigma_t}}
{{\beta'_b} s\over \alpha({\beta'_b}+\ie)s-p_\perp^2}
\nonumber\\
&&\hspace{-0mm}
=~-{1\over\ve^2}-{1\over\ve}[\ln(-i{\beta'_b}\sigma_ts)-\ln4\pi]-\half[\ln(-i{\beta'_b}\sigma_ts)-\ln4\pi]^2-{\pi^2\over 6}~+~O(\ve)
\nonumber
\end{eqnarray}
where $-i\beta'\equiv -i\beta'+\epsilon$ etc.

Next, we can get contributions of virtual diagrams in Fig. diagrams in Fig. \ref{fig:eik2}c,d and Fig. \ref{fig:eik2}g,h 
by usual projectile$\leftrightarrow$target replacements (\ref{projtarepl}) so
the final result for virtual eikonal TMD contributions on the mass shell reads
\begin{eqnarray}
&&\hspace{-0mm}
I^{\rm virt~eik}_{\rm mass~shell}~=~I^{\rm eik}_{\rm Fig.~\ref{fig:eik2}c,d}(\beta_b)+I^{\rm eik}_{\rm Fig.~\ref{fig:eik2}g,h}({\beta'_b})+I^{\rm eik}_{\rm Fig.~\ref{fig:eik1}c,d}(\alpha_a)
+I^{\rm eik}_{\rm Fig.~\ref{fig:eik1}g,h}({\alpha'_a})
\nonumber\\
&&\hspace{-0mm}
=~-{4\over\ve^2}-{1\over \ve}\Big[\ln{-i{\alpha'_a}\sigma_ps\over 4\pi}+\ln{-i\alpha_a\sigma_ps\over 4\pi}+\ln{-i{\beta'_b}\sigma_ts\over 4\pi}+\ln{-i\beta_b\sigma_ts\over 4\pi}
\Big]
\label{virteikmashell}\\
&&\hspace{-0mm}
-~\half\Big[\ln^2{-i{\alpha'_a}\sigma_ps\over 4\pi}+\ln^2{-i\alpha_a\sigma_ps\over 4\pi}+\ln^2{-i{\beta'_b}\sigma_ts\over 4\pi}+\ln^2{-i\beta_b\sigma_ts\over 4\pi}\Big]
-{2\pi^2\over 3}~+~O(\ve)
\nonumber
\end{eqnarray}
%
\subsubsection{Production terms minus TMD matrix elements}

Next we will calculate the difference $J_1({\alpha'_a},k'_{a_\perp},\beta_b, k_{b_\perp}=0, x_1,x_2)$ from Eq. (\ref{jotodin}) 
at $ k'_{a_\perp}=0$ and $k_{b_\perp}=0$. To avoid confusing singularities, we keep  $x_{12}^\parallel\neq 0$ in this calculation. 
\begin{eqnarray}
&&\hspace{-1mm}
J_1({\alpha'_a}, k'_{a_\perp},\beta_b,k_{b_\perp}, x_1,x_2)\Big|_{k'_{a_\perp}= k_{b_\perp}=0}
~=~
\Big(I_1({\alpha'_a}, k'_{a_\perp},\beta_b, k_{b_\perp}, x_1,x_2)
\nonumber\\
&&\hspace{1mm}
-~I_{\rm Fig.  \ref{fig:eik1}e,f}^{\rm eik}({\alpha'_a}, k'_{a_\perp}, x_1,x_2)
-I_{\rm Fig.  \ref{fig:eik2}a,b}^{\rm eik}(\beta_b, k_{b_\perp}, x_1,x_2)\Big)_{ {k'}_{a_\perp}= {k'}_{b_\perp}=0}
\nonumber\\
&&\hspace{-1mm}
=~8\pi^2\!\int\!\dhd p_\perp~e^{-i(p,x_{12})_\perp}\bigg\{\!\int\!\dhd\beta
\bigg[
{\theta(\beta)\over\beta^2s}
 {(\beta_b-\beta)({\alpha'_a}\beta s+p_\perp^2)\over ({\alpha'_a}+\ie)(\beta_b+\ie)p_\perp^2}
e^{i\beta\vro x_{12}^+ +i{p_\perp^2\over\beta s}\vro x_{12}^-} 
\nonumber\\
&&\hspace{11mm}
+~
{\theta(\beta)\over\beta^2s}{(\beta_b-\beta)p_\perp^2
\over ({\alpha'_a}+\ie)
\big[({\alpha'_a}+\ie)(\beta_b-\beta)\beta s-\beta_bp_\perp^2\big]}
e^{i\beta\vro x_{12}^+-i{\alpha'_a}\vro x_{12}^- +i{p_\perp^2\over\beta s}\vro x_{12}^-} 
\nonumber\\
&&\hspace{11mm}
+~{\theta(\beta-\beta_b)(\beta-\beta_b)[{\alpha'_a}(\beta-\beta_b)s+p_\perp^2]
\over p_\perp^2\big[({\alpha'_a}+\ie)\beta(\beta-\beta_b)s+\beta_b p_\perp^2\big]
(\beta_b+\ie)}  
e^{i\beta\vro x_{12}^+ +i{p_\perp^2\over(\beta-\beta_b) s}\vro x_{12}^-} 
\bigg]
\nonumber\\
&&\hspace{11mm}
-~{1\over p_\perp^2}
\!\int_0^{\infty}\!\dhd\beta~e^{-i{\beta\over \sigma_p}}
{e^{i{p_\perp^2\over\beta s}\vro x_{12}^-}\over \beta^2({\alpha'_a}+\ie)s}
\Big[
{\alpha'_a}\beta s+p_\perp^2+~{p_\perp^4e^{-i{\alpha'_a}\vro x_{12}^-}\over {\alpha'_a}\beta s-p_\perp^2+\ie}
\Big]
\nonumber\\
&&\hspace{11mm}
-~{1\over p_\perp^2}
\!\int_0^{\infty}\!\dhd\alpha~e^{i{\alpha\over \sigma_t}}{e^{i{p_\perp^2\over\alpha s}\vro x_{12}^+}
\over \alpha^2(\beta_b+\ie)s}
\Big[
\alpha\beta_b s-p_\perp^2
+{p_\perp^4e^{i\beta_b\vro x_{12}^+}
\over\alpha\beta_b s+p_\perp^2+\ie}
\Big]\bigg\}
\label{masshell1}
\ega

It is convenient to rearrange terms in the r.h.s. as follows
\bega
&&\hspace{-1mm}
J_1({\alpha'_a}, k'_{a_\perp},\beta_b, k_{b_\perp}, x_1,x_2)\Big|_{ k'_{a_\perp}= k_{b_\perp}=0}
~=~J_1^{(1)}+J_1^{(2)}+J_1^{(3)}+J_1^{(4)}+J_1^{(5)}
\label{jots}
\ega
where
\bega
&&\hspace{-3mm}
J_1^{(1)}({\alpha'_a}, \beta_b,  x_1,x_2)
\label{jeiodin}\\
&&\hspace{-3mm}
=~~8\pi^2\!\int\!{\dhd p_\perp\over p_\perp^2}~e^{-i(p,x_{12})_\perp}\bigg[\!\int_0^\infty\!{\dhd\beta\over\beta}
e^{i{p_\perp^2\over\beta s}\vro x_{12}^-} \big(e^{i\beta\vro x_{12}^+}-e^{i{\beta\over \sigma_p}}\big)
-\!\int_0^{\infty}\!{\dhd\alpha\over\alpha}e^{i{\alpha\over \sigma_t}}e^{i{p_\perp^2\over\alpha s}\vro x_{12}^+}
\bigg],
\nonumber
\ega
\bega
&&\hspace{-8mm}
J_1^{(2)}({\alpha'_a},\beta_b, x_1,x_2)
\label{jeivan}\\
&&\hspace{-8mm}
=~8\pi^2\!\int\!\dhd p_\perp~e^{-i(p,x_{12})_\perp}
\bigg[\!\int_0^{\infty}\!\dhd\alpha~e^{i{\alpha\over \sigma_t}}
{e^{i{p_\perp^2\over\alpha s}\vro x_{12}^+}\over \alpha^2s(\beta_b+\ie)}
-~\!\int_0^\infty\!\dhd\beta~{e^{i\beta\vro x_{12}^+ +i{p_\perp^2\over\beta s}\vro x_{12}^-}\over (\beta_b+\ie)p_\perp^2}
\bigg],
\nonumber
\ega
\bega
&&\hspace{-1mm}
J_1^{(3)}({\alpha'_a}, \beta_b, x_1,x_2)
\label{jeitu}\\
&&\hspace{-1mm}
=~8\pi^2\!\int\!\dhd p_\perp~e^{-i(p,x_{12})_\perp}\!\int_0^\infty\!\dhd\beta\bigg[{e^{i{p_\perp^2\over\beta s}\vro x_{12}^-}\over ({\alpha'_a}+\ie)\beta^2s}
 \big(e^{i\beta\vro x_{12}^+}-e^{-i{\beta\over \sigma_p}}\big)
-{e^{i\beta\vro x_{12}^+ +i{p_\perp^2\over\beta s}\vro x_{12}^-} \over \beta s(\beta_b+\ie)({\alpha'_a}+\ie)} 
\bigg], 
\nonumber
\ega
\bega
&&\hspace{-1mm}
J_1^{(4)}({\alpha'_a}, \beta_b, x_1,x_2)
~=~8\pi^2\!\int\!\dhd p_\perp~e^{-i(p,x_{12})_\perp}\bigg[
\!\int\!\dhd\beta ~e^{i\beta\vro x_{12}^+ +i{p_\perp^2\over(\beta-\beta_b) s}\vro x_{12}^-} 
{\theta(\beta-\beta_b)\over p_\perp^2(\beta_b+\ie)}
\nonumber\\
&&\hspace{-1mm}
\times~
{(\beta-\beta_b)[{\alpha'_a}(\beta-\beta_b)s+p_\perp^2]
\over ({\alpha'_a}+\ie)\beta(\beta-\beta_b)s+\beta_b p_\perp^2}  
-~\!\int_0^{\infty}\!{\dhd\alpha\over\alpha}~e^{-i{\alpha\over \sigma_t}}{p_\perp^2e^{i{p_\perp^2\over\alpha s}\vro x_{12}^+  +i\beta_b\vro x_{12}^+}
\over (\alpha\beta_b s+\ie)(\alpha\beta_b s+p_\perp^2+\ie)}\bigg]
\label{jeisri}
\ega
and
\bega
&&\hspace{-1mm}
J_1^{(5)}({\alpha'_a}, \beta_b,  x_1,x_2)
~=~{8\pi^2\over{\alpha'_a}+\ie}\!\int\!\dhd p_\perp\!\int\! \dhd\beta
\bigg[\big(e^{i\beta\vro x_{12}^+}-e^{i{\beta\over \sigma_p}}\big)
{p_\perp^2e^{i{p_\perp^2\over\beta s}\vro x_{12}^- -i{\alpha'_a}\vro x_{12}^-}
\over\beta^2s[{\alpha'_a}\beta s-p_\perp^2+\ie]}
\nonumber\\
&&\hspace{11mm}
+~
{p_\perp^4e^{i\beta\vro x_{12}^+ -i{\alpha'_a}\vro x_{12}^- +i{p_\perp^2\over\beta s}\vro x_{12}^-} 
\over \beta s
\big[({\alpha'_a}+\ie)(\beta_b-\beta)\beta s-\beta_bp_\perp^2\big]\big[({\alpha'_a}+\ie)\beta s-p_\perp^2\big]}
\bigg]~e^{-i(p,x_{12})_\perp}
\label{jeifo}
\ega

Let us start with $J_1^{(1)}$ term. After changing $\alpha={p_\perp^2\over\beta s}$ in the last term it takes the form
\begin{eqnarray}
&&\hspace{-1mm}
J_1^{(1)}({\alpha'_a}, \beta_b,  x_2,x_1)
\label{jeiodin1}\\
&&\hspace{-1mm}
=~
4\pi\!\int\!{\dhd p_\perp\over p_\perp^2}~e^{-i(p,x_{12})_\perp}\!\int_0^\infty\!{d\beta\over\beta}\Big[
e^{i{p_\perp^2\over\beta s}\vro x_{12}^- +i\beta\vro x_{12}^+}-e^{i{p_\perp^2\over\beta s}\vro x_{12}^- -i\beta\vro \delta^+}
-e^{i{p_\perp^2\over\beta s}\vro \delta^- +i\beta\vro x_{12}^+}\Big]
\nonumber
\end{eqnarray}
Using integral
\begin{eqnarray}
&&\hspace{-1mm}
4\pi\!\int\!{\dhd p_\perp\over p_\perp^2}~e^{-i(p,x_{12})_\perp}\!\int_0^\infty\!{d\beta\over\beta}
e^{i{p_\perp^2\over\beta s}\vro x_{12}^- +i\beta\vro x_{12}^+}~
=~(\pi x_{12_\perp}^2)^{-\epsilon}\!\int_0^1\! du~{\Gamma(\epsilon)u^{\epsilon-1}
\over \big(u+{2(ix_{12}^+ -\epsilon)(ix_{12}^- -\epsilon)\over x_{12_\perp}^2}\big)^{\epsilon}}
\nonumber\\
&&\hspace{1mm}
=~{1\over\epsilon^2}-{\ln[2\pi (ix_{12}^+ +\epsilon)(ix_{12}^- +\epsilon)]+\gamma\over\epsilon}+\half
\big[\ln(\pi x_{12_\perp}^2)+\gamma\big]
\nonumber\\
&&\hspace{1mm}
\times~\Big[\ln(\pi x_{12_\perp}^2)+\gamma+2\ln{2\pi (ix_{12}^+ -\epsilon)(ix_{12}^- -\epsilon)\over x_{12_\perp}^2}\Big]
-~{\pi^2\over 12}~+~O\big({x_{12}^+x_{12}^-\over x_{12_\perp}^2}\big)~+~O(\epsilon)
\end{eqnarray}
we get
\begin{eqnarray}
&&\hspace{-1mm}
J_1^{(1)}({\alpha'_a}, \beta_b,  x_2,x_1)~=~-{1\over\epsilon^2}+{1\over\epsilon}\big[\ln 2\pi\delta^+\delta^-+\gamma\big]
\label{jotidinotvet}\\
&&\hspace{-1mm}
-~\half\big[\ln(\pi x_{12_\perp}^2)+\gamma\big]\Big[\ln(\pi x_{12_\perp}^2)+\gamma+2\ln{2\pi\delta^+\delta^-\over x_{12_\perp}^2}\Big]
+{\pi^2\over 12}~+~O(\epsilon)+O(\lambda_p)+O(\lambda_t)
\nonumber
\end{eqnarray}

In the next Section we will prove now that all $J_1^{(i)}$ except $J_1^{(1)}$ are power corrections so the result for ``production - eikonal'' terms is
\begin{eqnarray}
&&\hspace{-1mm}
J_1({\alpha'_a}, \beta_b,  x_2,x_1)+J_2(\alpha_a, {\beta'_b},  x_2,x_1)~=~-{2\over\epsilon^2}+{2\over\epsilon}\big[\ln 2\pi\delta^+\delta^-+\gamma\big]
\label{jotvet}\\
&&\hspace{-1mm}
-~\big[\ln(\pi x_{12_\perp}^2)+\gamma\big]\Big[\ln(\pi x_{12_\perp}^2)+\gamma+2\ln{2\pi\delta^+\delta^-\over x_{12_\perp}^2}\Big]
+{\pi^2\over 6}~+~O(\epsilon)+O(\lambda_p)+O(\lambda_t)
\nonumber
\end{eqnarray}
where we have added second term obtained by projectile$\leftrightarrow$target replacements.

Now we are in a position to assemble the result for the coefficient function obtained by on-the-mass-shell calculation. We get
\begin{eqnarray}
&&\hspace{-1mm}
J_1({\alpha'_a}, \beta_b,  x_2,x_1)+J_2(\alpha_a, {\beta'_b},  x_2,x_1)
+I^{\rm virt}_{\rm mass~shell}-I^{\rm virt~eik}_{\rm mass~shell}
\label{mashellotvet}\\
&&\hspace{-1mm}
=~-{2\over\epsilon^2}+{2\over\epsilon}\big[\ln 2\pi\delta^+\delta^-+\gamma\big]
+\big[\ln(\pi x_{12_\perp}^2)+\gamma\big]\Big[\ln x_{12_\perp}^2+2\ln{\sigma_p\sigma_ts\over 4}-3\ln\pi -\gamma\Big]
+{\pi^2\over 6}
\nonumber\\
&&\hspace{5mm}
-~{2\over\ve^2}-{1\over\ve}\Big(\ln{-({\alpha'_a}+\ie)({\beta'_b}+\ie)s\over 4\pi}+\gamma\Big)
-\half\Big(\ln{-({\alpha'_a}+\ie)({\beta'_b}+\ie)s\over 4\pi}+\gamma\Big)^2~
\nonumber\\
&&\hspace{5mm}
-~{1\over\ve}\Big(\ln{-(\alpha_a+\ie)(\beta_b+\ie)s\over 4\pi}+\gamma\Big)
-\half\Big(\ln{-(\alpha_a+\ie)(\beta_b+\ie)s\over 4\pi}+\gamma\Big)^2+{\pi^2\over 6}
\nonumber\\
&&\hspace{5mm}
+~{4\over\ve^2}+{1\over \ve}\Big[\ln{-i{\alpha'_a}\sigma_ps\over 4\pi}+\ln{-i\alpha_a\sigma_ps\over 4\pi}+\ln{-i{\beta'_b}\sigma_ts\over 4\pi}+\ln{-i\beta_b\sigma_ts\over 4\pi}
\Big]
\nonumber\\
&&\hspace{5mm}
+~\half\Big[\ln^2{-i{\alpha'_a}\sigma_ps\over 4\pi}+\ln^2{-i\alpha_a\sigma_ps\over 4\pi}+\ln^2{-i{\beta'_b}\sigma_ts\over 4\pi}+\ln^2{-i\beta_b\sigma_ts\over 4\pi}\Big]
+{2\pi^2\over 3}~+~O(\ve)
\nonumber\\
&&\hspace{-1mm}
=
~\ln^2{x_{12_\perp}^2s\sigma_p\sigma_t\over 4}
-\ln{(-i{\alpha'_a})e^{\gamma}\over\sigma_t}\ln{(-i{\beta'_b})e^{\gamma}\over\sigma_p}
-\ln{(-i\alpha_a)e^{\gamma}\over\sigma_t}\ln{(-i\beta_b)e^{\gamma}\over\sigma_p}+\pi^2
\nonumber
\end{eqnarray}
which agrees with Eq. (\ref{masteresult}). 

\subsubsection{$J_1^{(i>1)}$ are power corrections}
Here we prove that all terms in the r.h.s. of Eq. (\ref{jots}) except for $J_1^{(1)}$ are power corrections.
Let us start from  $J_1^{(2)}$ which can be rewritten as
\bega
&&\hspace{-1mm}
J_1^{(2)}~=~{4\pi\over\beta_b}\!\int\!\dhd p_\perp~e^{-i(p,x_{12})_\perp}\!\int_0^\infty\!{d\alpha\over\alpha^2s}
e^{i{p_\perp^2\over\alpha s}\vro x_{12}^+}
\big[e^{i{\alpha\over \sigma_t}}-e^{i\alpha\vro x_{12}^-}\big] 
\label{oxraterms}\\
&&\hspace{-1mm}
=~-{i\over\vro x_{12}^+}\!\int_0^\infty\!{d\alpha\over\alpha}e^{-i{x_{12_\perp}^2\over 4\vro x_{12}^+}\alpha s}
\big[e^{i\alpha\vro \delta^-}-e^{i\alpha\vro x_{12}^-}\big]
\nonumber\\
&&\hspace{-1mm}
=~{i\over\beta_b\vro x_{12}^+}\ln{x_{12_\perp}^2-2x_{12}^-x_{12}^+
\over x_{12_\perp}^2+2 \delta^-x_{12}^+}
~=~2i{{\alpha'_a}(\vro \delta^--\vro x_{12}^- )\over Q_{a'b}^2x_{12_\perp}^2}~\simeq~O(\lambda_t)
\nonumber
\end{eqnarray}
Next, after the integration over $p_\perp$ the term $J_1^{(3)}$ turns to
\begin{eqnarray}
&&\hspace{-7mm}
J_1^{(3)}~
=~{i\over\vro x_{12}^-}\!\int_0^\infty\!d\beta\Big[
{1\over {\alpha'_a}\beta}
e^{-i{x_{12_\perp}^2\over 4\vro x_{12}^-}\beta s} \big(e^{i\beta\vro x_{12}^+}-e^{-i{\beta\over \sigma_p}}\big)}
-~{1 \over {\alpha'_a}\beta_b}e^{i\beta\vro x_{12}^+ -i{x_{12_\perp}^2\over4\vro x_{12}^-}\beta s
\Big]
\nonumber\\
&&\hspace{1mm}
=~{i\over{\alpha'_a}\vro x_{12}^-}\ln{x_{12_\perp}^2-2x_{12}^+ x_{12}^-\over x_{12_\perp}^2+2\delta^+ x_{12}^-}
+{4\over Q_{ab'}^2}{1\over x_{12_\perp}^2-x_{12_\parallel}^2}
~\simeq ~O(\lambda_p) 
\label{blueterms}
\end{eqnarray}

To calculate  $J_1^{(4)}$ term, it is convenient to change variable $\beta=\beta_b+{p_\perp^2\over\alpha s}$ in the first integral. 
For simplicity,  we take $\beta_b>0$ and get
\bega
&&\hspace{-5mm}
J_1^{(4)}~=~{8\pi^2\over\beta_b}\!\int\!\dhd p_\perp~e^{-i(p,x_{12})_\perp}\!\int_0^\infty\!\dhd\alpha \Big[{p_\perp^2\over \alpha^2s}
\big(e^{i\alpha \vro x_{12}^-}-e^{i{\alpha\over \sigma_t}}\big)
{e^{i{p_\perp^2\over\alpha s}\vro x_{12}^++i\beta_b\vro x_{12}^+}\over\alpha\beta_b s+p_\perp^2+\ie}
\nonumber\\
&&\hspace{11mm}
+~{p_\perp^4\over \alpha s}{e^{i\alpha\vro x_{12}^- +i{p_\perp^2\over\alpha s}\vro x_{12}^+ +i\beta_b\vro x_{12}^+}
\over[\alpha\beta_b s+p_\perp^2+\ie][({\alpha'_a}+\alpha)\alpha\beta_bs+{\alpha'_a} p_\perp^2]}\Big]
\nonumber\\
&&\hspace{4mm}
=~\!\int_0^\infty\! dp_\perp^2~J_0(px_{12_\perp})\!\int_0^\infty\! dt~\Big\{{t\over t^2(t+p_\perp^2)}
e^{i{t+p_\perp^2\over t}\beta_b\vro x_{12}^+}
\nonumber\\
&&\hspace{11mm}
\times~\Big(e^{i{t\over Q_{a'b}^2}{\alpha'_a} \vro x_{12}^-}-e^{i{t\over \beta_b\sigma_ts}}\Big)
+~{p_\perp^4\over t}{e^{i{t\over Q_{a'b}^2}{\alpha'_a}\vro x_{12}^- 
+i{p_\perp^2\over t}\beta_b\vro x_{12}^+ +i\beta_b\vro x_{12}^+}
\over(t+p_\perp^2)[(Q_{a'b}^2+t)t+Q_{a'b}^2 p_\perp^2]}\Big\}
\label{greenterms1}
\end{eqnarray}
The first term in the r.h.s. can be represented as 
\begin{eqnarray}
&&\hspace{-11mm}
\!\int_0^\infty\! dp_\perp^2~J_0(px_{12_\perp})\!\int_0^\infty\!dt {p_\perp^2\over t^2(t+p_\perp^2)}e^{i{t+p_\perp^2\over t}\beta_b\vro x_{12}^+}
\Big[e^{i{t\over Q_{a'b}^2}{\alpha'_a} \vro x_{12}^-}-e^{i{t\over Q_{a'b}^2}{\alpha'_a} \vro \delta^-}\Big]
\label{greenterms2}\\
&&\hspace{-11mm}
=~2\!\int_0^\infty\! {dv\over v}~J_0(v)\!\int_0^\infty\!dt {1\over t^2(t+1)}e^{i{t+1\over t}\beta_b\vro x_{12}^+}
\Big[e^{i{t\over Q_{a'b}^2x_{12_\perp}^2}{\alpha'_a} \vro x_{12}^- v^2}-~(x_{12}^-\rightarrow \delta^-)\Big]
\nonumber
\ega
We need to check two regions: $t\ll 1$ and $t\gg 1$. Assuming essential $t's$ are small, we get
\bega
&&\hspace{-1mm}
2\!\int_0^\infty\! {dv\over v}~J_0(v)\!\int_0^\infty\!dt {1\over t^2}e^{i{t+1\over t}\beta_b\vro x_{12}^+}
\Big[e^{i{t\over Q_{a'b}^2x_{12_\perp}^2}{\alpha'_a} \vro x_{12}^- v^2}-~(x_{12}^-\rightarrow \delta^-)\Big]
\nonumber\\
&&\hspace{-1mm}
=~2i{e^{i\beta_b\vro x_{12}^+}\over \beta_b\vro x_{12}^+}
\!\int_0^\infty\! dv~J_0(v)\Big[\sqrt{-2x_{12}^+x_{12}^-\over x_{12_\perp}^2}K_1\Big(v\sqrt{-2x_{12}^+x_{12}^-/x_{12_\perp}^2}\Big)
~-~(x_{12}^-\rightarrow \delta^-)\Big]
\nonumber\\
&&\hspace{-1mm}
=~i{e^{i\beta_b\vro x_{12}^+}\over \beta_b\vro x_{12}^+}
\ln{x_{12_\perp}^2-x_{12}^+\delta^-\over x_{12_\perp}^2-x_{12}^+x_{12}^-}~\simeq~{i\delta^-\over \beta_b\vro x_{12_\perp}^2}
~=~O(\lambda_t)
\label{smaltchek}
\ega
which is a power correction. However, quick look at the integral over $t$ shows that 
$t\sim vQ_{a'b}x_{12_\perp}\sqrt{{\alpha'_a}\delta^-\over \beta_bx_{12}^+}\sim v{\sigma_t\beta_bs\over Q_\perp^2}$ so
 $t\ll 1$ corresponds to $v\ll \lambda_t$ which gives negligible contribution to the integral (\ref{smaltchek}).
 Thus, characteristic $t$'s in the integral (\ref{greenterms2}) are not small. Let us assume they are large and check 
 it {\it a posteriori}. At $t\gg 1$ we get
\bega
&&\hspace{-1mm}
{\rm Eq.~(\ref{greenterms2})}~=~2\!\int_0^\infty\! {dv\over v}~J_0(v)\!\int_0^\infty\!dt {1\over t^3}
e^{i{t+1\over t}\beta_b\vro x_{12}^+}
\Big[e^{i{t\over Q_{ab'}^2x_{12_\perp}^2}{\alpha'_a} \vro x_{12}^- v^2}-~(x_{12}^-\rightarrow \delta^-)\Big]
\nonumber\\
&&\hspace{-1mm}
=~-2{e^{i\beta_b\vro x_{12}^+}\over (\beta_b\vro x_{12}^+)^2}
\!\int_0^\infty\! dv~vJ_0(v)\Big[\Big(-{2x_{12}^+x_{12}^-\over x_{12_\perp}^2}\Big)
K_2\Big(v\sqrt{-2x_{12}^+x_{12}^-/x_{12_\perp}^2}\Big)
~-~(x_{12}^-\rightarrow \delta^-)\Big]
\nonumber\\
&&\hspace{11mm}
=~-2{e^{i\beta_b\vro x_{12}^+}\over (\beta_b\vro x_{12}^+)^2}
\Big[\ln{x_{12_\perp}^2-x_{12}^+\delta^-\over x_{12_\perp}^2-x_{12}^+x_{12}^-}
+{x_{12_\perp}^2\over x_{12_\perp}^2-x_{12}^+x_{12}^-}-{x_{12_\perp}^2\over x_{12_\perp}^2-x_{12}^+\delta^-}\Big]
\nonumber\\
&&\hspace{11mm}
\simeq~-~2{e^{i\beta_b\vro x_{12}^+}\over s{\beta_b}^2}{(\delta^-)^2 -(x_{12}^-)^2\over x_{12_\perp}^4}
~=~O(\lambda_t^2)
\end{eqnarray}
which is smaller than a typical power correction.  Also, in this integral $v\sim 1$ and 
$t\sim vQ_{a'b}x_{12_\perp}\sqrt{{\alpha'_a}\delta^-\over \beta_bx_{12}^+}\sim v{\sigma_t\beta_bs\over Q_\perp^2}\sim {1\over\lambda_t}$ so we justified large-$t$ approximation.

The last term in the r.h.s. of Eq. (\ref{greenterms1}) is even smaller
\bega
&&\hspace{-1mm}
\!\int_0^\infty\! dp_\perp^2~J_0(px_{12_\perp})\!\int_0^\infty\! dv~
{e^{iv{p_\perp^2\over Q_{a'b}^2}{\alpha'_a}\vro x_{12}^- +i{1\over v}\beta_b\vro x_{12}^+ +i\beta_b\vro x_{12}^+}
\over v(v+1)[Q_{a'b}^2(v+1)+v^2p_\perp^2]}
\nonumber\\
&&\hspace{-1mm}
\stackrel{p_\perp x_{12_\perp}\sim 1}\simeq~
\!\int_0^\infty\! dp_\perp^2~J_0(px_{12_\perp})\!\int_0^\infty\! dv~
{e^{i{1\over v}\beta_b\vro x_{12}^+ +i\beta_b\vro x_{12}^+}
\over v(v+1)[Q_{a'b}^2(v+1)+v^2p_\perp^2]}
\nonumber\\
&&\hspace{-1mm}
=~2\!\int_0^\infty\! dv~{v^2\over v+1}e^{i(v+1){\beta_b}\vro x_{12}^+}K_0\Big(Q_{a'b}x_{12_\perp}\sqrt{v(v+1)}\Big)~\sim~{Q_\perp^6\over Q^6}
\label{greenterm3}
\end{eqnarray}
Thus, $J_1^{(4)}~=~O(\lambda_t^2)$ and can be neglected.

Next, it is easy to see that $J_1^{(5)}$ of Eq. (\ref{jeifo}) differs from the first two lines in Eq. (\ref{greenterms1}) for $J_1^{(4)}$ 
by projectile$\leftrightarrow$ target replacements ${\alpha'_a}\leftrightarrow\beta_b, \sigma_p\leftrightarrow\sigma_t, x_{12}^+\leftrightarrow x_{12}^-$ so we have $J_1^{(5)}~=~O(\lambda_p^2)$. 

\subsubsection{sG-contribution \label{sec:sGmasshell}}
The soft-Glauber contribution to virtual diagram in Fig. (\ref{fig:nlovirt}) can be easily obtained from Eq. (\ref{sgintegral}) by setting ${k'}_{a_\perp}=k_{b_\perp}=0$
\bega
&&\hspace{-1mm}
I^{\rm virt~sG ~ms}_{\rm Fig. \ref{fig:nlovirt}}~=~
-{\mu_\sigma^{2\ve}\over (4\pi)^\ve}\!\int_0^\infty\!{dl^2\over l^2}l^{2\ve}{e^{-il^2}\over 1-{l^2\sigma_t\over {\alpha'_a}}+\ie}
 \int_0^\infty\!{dt^2\over t^2}t^{2\ve}{e^{it^2}\over 1-{t^2\sigma_p\over |{\beta'_b}|}-\ie}
 \nonumber\\
&&\hspace{-1mm}
=~-{1\over\ve^2}+{1\over\ve}\Big(2\gamma-\ln{\mu_\sigma^2\over 4\pi i}\Big)-\half\ln^2{\mu_\sigma^2\over 4\pi i}
-2\gamma\ln{\mu_\sigma^2\over 4\pi i}-\gamma^2~+~O(\lambda_t,\lambda_p)
\label{sgintegralmashell}
\ega
where we used integral
\beq
\int_0^\infty\! dv~v^{\ve-1}{e^{-v}\over 1+\lambda v}~=~{1\over\ve}-\gamma-\lambda+\ve\big[\gamma\lambda
+{\gamma^2\over 2}+{\pi^2\over 12}\big]+O(\lambda^2)+O(\ve^2)
\eeq
The expression for $I^{\rm virt~sG~ms}_{\rm Fig. \ref{fig:nlovirtleft}}$ will be the same as seen from Eq. (\ref{virtsG6}).
Moreover, it is easy to see  from Eqs. (\ref{iodinsgotvet})-(\ref{idvasgotvet}) that  the sG-contribution to ``production'' diagrams 
differs in sign from virtual contribution (\ref{sgintegralmashell})
\beq
I^{\rm sG~ms}_{\rm Fig.\ref{fig:real1}}~=~I^{\rm sG~ms}_{\rm Fig.\ref{fig:real2}}~=~-I^{\rm virt~sG ~ms}_{\rm Fig. \ref{fig:nlovirt}}
~=~-I^{\rm virt~sG ~ms}_{\rm Fig. \ref{fig:nlovirtleft}}
\eeq
and therefore the total sG-contribution to the sum of the diagrams with background fields on the mass shell is a power correction.

\bibliography{tmdconfb.bib}

\end{document}